\documentclass[a4paper,fleqn,usenatbib]{mnras}

\usepackage{newtxtext,newtxmath}
\usepackage[T1]{fontenc}
\usepackage{ae,aecompl}

\usepackage{graphicx}	% Including figure files
\usepackage{amsmath}	% Advanced maths commands

\newcommand{\bars}{\mathrm{bar}}
\newcommand{\dd}{\mathrm{d}}
\title[Globular clusters and bar]{Globular clusters and bar: captured or not captured?}

\author[A. A. Smirnov et al.]{
Anton~A.~Smirnov$^{1}$,
Anisa~T.~Bajkova$^{1}$,
Vadim~V.~Bobylev$^{1}$\\
$^{1}$Central (Pulkovo) Astronomical Observatory, Russian Academy of Sciences, Pulkovskoye chaussee 65/1, St. Petersburg 196140, Russia\\
}

\date{Accepted XXX. Received YYY; in original form ZZZ}

\pubyear{2021}

\begin{document}
\label{firstpage}
\pagerange{\pageref{firstpage}--\pageref{lastpage}}
\maketitle

\begin{abstract}
  Studies of the dynamics of globular clusters assume different values of bar parameters (mass, velocity, size) and analyse the results of orbit classifications over the range of the chosen values. It is also a usual thing that a spherical bulge component is converted into the bar to obtain a non-axisymmetric potential from an axisymmetric one. The choice of bar parameters and the way the bar is converted from the bulge introduce systematics into the orbit classifications that we explore in the present study. We integrate orbits of 30 bulge globular clusters residing in the inner area of the Galaxy ($R \lesssim 5$ kpc) backwards in time for three different potentials, two of which are obtained by fitting the rotation curve, and one is taken from the surrogate $N$-body model representing our Galaxy. We analyse each orbit in terms of dominant frequencies obtained from its coordinate spectra. We find that the bar pattern speed is a key factor in orbital classification. With an increase of it, frequencies deviate more and more from the ``bar'' frequency ratio 2:1. The \textcolor{black}{bar-to-bulge mass ratio (assuming the total mass of the bar plus the bulge is fixed)} and size of the bar play a smaller role.  We also find that, in the $N$-body potential, the fraction of orbits that follow the bar is higher than in those obtained from fitting the rotation curve.
\end{abstract}

\begin{keywords}
(Galaxy:) globular clusters: general -- Galaxy: kinematics and dynamics -- Galaxy: bulge
\end{keywords}

%%%%%%%%%%%%%%%%%%%%%%%%%%%%%%%%%%%%%%%%%%%%%%%%%%

%%%%%%%%%%%%%%%%% BODY OF PAPER %%%%%%%%%%%%%%%%%%

%%%%%%%%%%%%%%%%%%%%%%%%%%%%%%%%%%%%%%%%%%%%%%%%%%%%%%%
%%%%%%%%%%%%%%%%%%%%%%%%%%%%%%%%%%%%%%%%%%%%%%%%%%%%%%%
\section{Introduction}
%%%%%%%%%%%%%%%%%%%%%%%%%%%%%%%%%%%%%%%%%%%%%%%%%%%%%%%
%%%%%%%%%%%%%%%%%%%%%%%%%%%%%%%%%%%%%%%%%%%%%%%%%%%%%%%
Several physical components co-exist within the area of about 5 kiloparsecs from the centre of our Galaxy. These components are a bar, its vertically thick part, which is usually referred to as the boxy/peanut-shaped (B/P) bulge~\citep{McWilliam2010, Nataf2010, Wegg2013, Mosenkov2021}, and possibly an another bulge, commonly referred to as the classical one. The existence of the latter has came under the question in the past few years due to various indicators pointing out that bulge stars exhibit cylindrical rotation~\textcolor{black}{\citep{2012AJ....143...57K, 2013MNRAS.430..836N,2016ApJ...819....2N}}, i.~e. support the B/P bulge rather than the classical one, although there are some exceptions (\textcolor{black}{\citealt{2016ApJ...821L..25K}}, also see the review by \citealt{BG2016}). We do not mention here the most inner part subsystems, such as the nuclear disc and the nuclear star cluster~\citep{Becklin1968}, since they are not relevant to the present work and are important for considering on much smaller spatial scales than those considered here.
\par 
Globular clusters (GCs) are tracers of the secular evolution of bar and bulge components, since GCs include a large bulk of stars that reflect how these components form/evolve in their metallicity and stellar populations. However, the question of whether a particular GC belongs to a certain component (e.g. a bulge, a bar, a disc, or a halo) is not easy to answer. On the contrary, determining the origin of a globular cluster is a rather difficult task, which requires reliable knowledge of the clusters' proper motions, their radial velocities, positions, and metallicity ~\citep{Cote1999,Bica2016,2019A&A...630L...4M,Pasquato2019,Ortolani2019a, Ortolani2019b, PV2020, BB2020, BB2021, Sun2022}. For example,~\cite{Ortolani2019a} recently found that the CGs  Terzan 10  and Djorgovski 1 have typical halo orbits, while their orbits are contained within the bulge volume. Another illustrative example is that \cite{PV2020} and~\cite{Ortolani2019b} showed that several GCs, while do not belong to either the disc or the halo and appear to belong to the bulge, nevertheless do not follow the bar. Meaning that these GCs move either faster or slower than the bar, but not synchronously with it. The ambiguity in the classification of GCs stems from the fact that several physical components of the Galaxy mentioned above overlap in physical space  and, at the same time, the observations of the inner part of the Galaxy are affected by heavy extinction and crowding~\citep{BG2016}. %For example, in () author obtained that ... . While in ... . In recent work by ...
%classified 78 GCs employed the unsupervised machine algorithm taking into account two orbital parameters (the apogalactic distance and maximum height from the disc plane) obtained from a sample of calculated orbits. It is also importan And the latter was   .
\par 
An additional problem, which especially concerns the dynamics of the GCs of the inner Galaxy and the classifications based on it, is that the parameters of the bar itself are also not set in stone. Bar pattern speed estimates range from about 30 km/s/kpc to 40 km/s/kpc~\citep{Portail2017,Bovy2019,Sanders2019, Binney2020,Asano2020, Kawata2021, Chiba2021,Li2022, Clarke2022}, while some authors provide an even higher value of about 50 km/s/kpc~\citep{Minchev2007, Antoja2014}. Naturally, the centrifugal force that influence the motion of GCs depends on the bar pattern speed. \textcolor{black}{It is also important that the changes in the pattern speed force the resonances to move, and, thus, orbits will differ depending on how close the GC to a particular resonance.} Therefore, the classifications of the orbits should differ depending on the bar pattern speed, and one should consider a set of bar pattern speed values, as it was done, for example, in ~\cite{Ortolani2019b} and~\cite{PV2020}. In \cite{PV2020}, the authors calculated the probability that an orbit belongs to one or another component separately for each of the pattern speeds considered there.
\par 
The uncertainty in the existence of classical bulge mentioned above can also implicitly affect the results of GCs' classification. One of the approaches to modelling of GCs' orbits is to transform the spherical central component into a bar. This means the the central spherical bulge in the originally axisymmetric model of the Milky Way is replaced by an elongated bar with exactly the same mass as the bulge. This approach has been used in recent studies by~\cite{Ortolani2019a, Ortolani2019b, PV2020} and many previous ones. At the same time, various $N$-body studies showed that the inclusion of even a small classic bulge component can drastically change the overall evolution of the model, leading to the formation of the so-called barlenses~\citep{Salo_Laurikainen2017_v2,Smirnov_Sotnikova2018} or preventing the bar buckling~\citep{Smirnov_Sotnikova2018} altogether. 
\par 
In the present work, we want to address the mentioned issues in the context of the capturing of CGs by the bar. We want to explore how the choice of the bar parameters (pattern speed, mass, size) affects the state of the CGs relative to the bar, i.~e. is there any systematics in the frequency rations of GCs orbits $f_\mathrm{R}/f_x$ (see definition is Section~\ref{sec:sim}) depending on the bar parameters. %The frequency ratio is commonly used indicator of  %Specifically, we check can be captured depending on the choice of the bar parameters within the reasonable limits,
\par 
To this aim, we study the motion of GCs in three different instances of the Milky Way potential. Two of them are based on observational data from~\cite{Bajkova_2016, Bajkova_2017} and~\cite{McMillan_2017} and one is based on the $N$-body model from~\cite{2021arXiv211105466T}, which was specifically prepared to represent the mass distribution of the Milky Way and has a spatial resolution of about 30 pc. This $N$-body model also contains a classical bulge and a naturally formed bar, thus providing an opportunity to study the GCs kinematics in case of a self-consistent model, obtained without transforming one component into another.
%Additionally, we want to compare the obtained results if integrate the GCs in the $N$-body potential, given that some authors provide $N$-body version of Milky Way, suggesting that  %In particular, we consider our recent model by~ and one, which is commonly used too, by~... . 
%Additionally, we want to check how stable the results of classification with changes in the bar parameters such as mass, pattern speed, and size.
\par 
The article is structured as follows. In Section~\ref{sec:data}, we describe our sample of GCs. In Section~\ref{sec:sim}, we provide details on the potentials considered in the present work and how the classification and integration of the orbits backwards in time was carried out. In Section~\ref{sec:systematics}, we analyse the systematics in the classification of orbits introduced by changing bar parameters using one GC, NGC 6266, as an example. Section~\ref{sec:all_orbits} presents the results of the classification for all GCs in the sample. We compare our results with those of previous works in Section~\ref{sec:comp}. In Section~\ref{sec:conc}, we give our conclusions.

%In Section~\ref{sec:methods}, we describe spectral dynamics approach to identify orbit types.

\section{Data}
\label{sec:data}

\begin{table*}
\begin{tabular}{ c| c | c | c | c | c | c | c | c | c | c |}
\hline
%ID & R & $V_\mathrm{R}$ & $V_\mathrm{T}$ & $z$ & $V_z$& $\phi$\\
ID & $\alpha$, deg & $\delta$, deg & $D$, kpc & $E(D)$, kpc & $V_r$, km/s &  $E(V_r)$, km/s & $\mu_{\alpha}^*$, mas/yr &  $\mu_\delta$, mas/yr & $E(\mu_\alpha)$, mas/yr & $E(\mu_\delta$), mas/yr\\
\hline 
\hline
BH 229& 262.772& -29.982& 6.995& 0.140& 40.61& 1.29& 2.511& -10.105& 0.036& 0.034\\ 
ESO 452-11& 249.856& -28.399& 7.389& 0.200& 16.27& 0.48& -1.424& -6.470& 0.032& 0.030\\ 
Liller 1& 263.352& -33.389& 8.061& 0.350& 58.2& 2.2& -5.398& -7.475& 0.130& 0.099\\ 
NGC 6144& 246.808& -26.023& 8.151& 0.130& 195.74& 0.74& -1.744& -2.607& 0.026& 0.026\\ 
NGC 6266& 255.303& -30.114& 6.412& 0.090& -73.49& 0.70& -4.979& -2.948& 0.026& 0.026\\ 
NGC 6273& 255.657& -26.268& 8.343& 0.160& 145.54& 0.59& -3.248& 1.661& 0.025& 0.025\\ 
NGC 6293& 257.543& -26.582& 9.192& 0.280& -143.66& 0.39& 0.871& -4.326& 0.028& 0.028\\ 
NGC 6342& 260.292& -19.587& 8.013& 0.230& 116.56& 0.74& -2.904& -7.115& 0.027& 0.026\\ 
NGC 6355& 260.994& -26.353& 8.655& 0.220& -194.13& 0.83& -4.740& -0.573& 0.030& 0.029\\ 
NGC 6380& 263.617& -39.069& 9.607& 0.300& -6.54& 1.48& -2.173& -3.224& 0.031& 0.030\\ 
NGC 6401& 264.652& -23.910& 8.064& 0.220& -99.26& 3.18& -2.747& 1.445& 0.035& 0.034\\ 
NGC 6440& 267.220& -20.360& 8.248& 0.240& -69.39& 0.93& -1.180& -4.004& 0.037& 0.036\\ 
NGC 6453& 267.715& -34.599& 10.070& 0.220& -91.16& 3.08& 0.205& -5.940& 0.035& 0.036\\ 
NGC 6522& 270.892& -30.034& 7.295& 0.210& -13.90& 0.71& 2.565& -6.438& 0.037& 0.035\\ 
NGC 6528& 271.207& -30.056& 7.829& 0.240& 210.31& 0.75& -2.154& -5.651& 0.040& 0.036\\ 
NGC 6558& 272.573& -31.764& 7.474& 0.180& -195.70& 0.70& -1.725& -4.149& 0.035& 0.033\\ 
NGC 6624& 275.919& -30.361& 8.019& 0.110& 54.26& 0.45& 0.121& -6.935& 0.029& 0.029\\ 
NGC 6626& 276.137& -24.870& 5.368& 0.100& 11.11& 0.60& -0.277& -8.922& 0.029& 0.028\\ 
NGC 6638& 277.734& -25.497& 9.775& 0.340& 8.63& 2.00& -2.520& -4.078& 0.029& 0.029\\ 
NGC 6637& 277.846& -32.348& 8.900& 0.110& 46.63& 1.45& -5.034& -5.831& 0.027& 0.027\\ 
NGC 6642& 277.975& -23.475& 8.049& 0.200& -33.23& 1.13& -0.172& -3.893& 0.031& 0.030\\ 
NGC 6717& 283.775& -22.701& 7.524& 0.130& 32.45& 1.44& -3.124& -5.009& 0.027& 0.026\\ 
NGC 6723& 284.888& -36.632& 8.267& 0.100& -94.18& 0.26& 1.030& -2.418& 0.026& 0.026\\ 
Pal 6& 265.926& -26.223& 7.047& 0.450& 176.28& 1.53& -9.200& -5.317& 0.036& 0.033\\ 
Terzan 1& 263.949& -30.481& 5.673& 0.170& 57.55& 1.61& -2.695& -4.883& 0.064& 0.058\\ 
Terzan 2& 261.888& -30.802& 7.753& 0.330& 128.96& 1.18& -2.166& -6.245& 0.042& 0.039\\ 
Terzan 4& 262.663& -31.596& 7.591& 0.310& -39.93& 3.76& -5.495& -3.624& 0.089& 0.065\\ 
Terzan 5& 267.020& -24.779& 6.617& 0.150& -81.40& 1.36& -1.904& -5.276& 0.080& 0.075\\ 
Terzan 6& 267.693& -31.275& 7.271& 0.350& 137.15& 1.70& -4.980& -7.437& 0.046& 0.039\\ 
Terzan 9& 270.412& -26.840& 5.770& 0.340& 29.31& 2.96& -2.109& -7.783& 0.065& 0.047\\ 
\hline
\end{tabular}
    \caption{Observational parameters of globular clusters considered in the present work. \textcolor{black}{Note that $\mu_{\alpha}^*$ is the corrected value of proper motion, 	$\mu_{\alpha}^* = \mu_{\alpha} \cos \delta$.}}
    \label{tab:CG_prop_obs}
\end{table*}
%%%%%%%%%%%%%%%%%%%%%%%%%%%%%%%%%%%%%%%%%%%%%%%%%%%%%%%
%%%%%%%%%%%%%%%%%%%%%%%%%%%%%%%%%%%%%%%%%%%%%%%%%%%%%%%

\begin{table*}
\begin{tabular}{ c | c | c | c | c | c | c |}
\hline
%ID & R & $V_\mathrm{R}$ & $V_\mathrm{T}$ & $z$ & $V_z$& $\phi$\\
ID & $X$, kpc & $Y$, kpc & $Z$, kpc & $V_\mathrm{R}$, km/s & $V_\mathrm{T}$, km/s& $V_z$, km/s\\
\hline 
\hline
BH 229 &1.33${\displaystyle \pm}0.34$ &0.27${\displaystyle \pm}0.02$ &0.28${\displaystyle \pm}0.01$ & 50${\displaystyle \pm}  1$ & 20${\displaystyle \pm} 11$ &-244${\displaystyle \pm} 13$ \\ 
ESO 452-11 &1.47${\displaystyle \pm}0.32$ &-0.44${\displaystyle \pm}0.05$ &1.56${\displaystyle \pm}0.08$ & 20${\displaystyle \pm}  1$ & 48${\displaystyle \pm} 10$ &-95${\displaystyle \pm}  5$ \\ 
Liller 1 &0.55${\displaystyle \pm}0.40$ &-0.54${\displaystyle \pm}0.04$ &-0.01${\displaystyle \pm}0.01$ & 37${\displaystyle \pm}  3$ &-100${\displaystyle \pm} 19$ & 25${\displaystyle \pm}  5$ \\ 
NGC 6144 &0.95${\displaystyle \pm}0.38$ &-0.77${\displaystyle \pm}0.05$ &2.22${\displaystyle \pm}0.11$ &185${\displaystyle \pm}  1$ &110${\displaystyle \pm}  6$ & 44${\displaystyle \pm}  1$ \\ 
NGC 6266 &2.10${\displaystyle \pm}0.32$ &0.19${\displaystyle \pm}0.04$ &0.83${\displaystyle \pm}0.04$ &-88${\displaystyle \pm}  1$ &103${\displaystyle \pm}  8$ & 63${\displaystyle \pm}  3$ \\ 
NGC 6273 &0.26${\displaystyle \pm}0.38$ &-0.37${\displaystyle \pm}0.02$ &1.38${\displaystyle \pm}0.07$ &130${\displaystyle \pm}  1$ &226${\displaystyle \pm}  2$ &171${\displaystyle \pm}  7$ \\ 
NGC 6293 &-0.56${\displaystyle \pm}0.46$ &-0.68${\displaystyle \pm}0.02$ &1.27${\displaystyle \pm}0.06$ &-117${\displaystyle \pm}  1$ &130${\displaystyle \pm}  7$ &-152${\displaystyle \pm}  7$ \\ 
NGC 6342 &0.11${\displaystyle \pm}0.38$ &0.79${\displaystyle \pm}0.03$ &1.37${\displaystyle \pm}0.07$ &160${\displaystyle \pm}  2$ &-18${\displaystyle \pm} 13$ &-29${\displaystyle \pm}  3$ \\ 
NGC 6355 &-0.26${\displaystyle \pm}0.44$ &-0.19${\displaystyle \pm}0.01$ &0.84${\displaystyle \pm}0.04$ &-197${\displaystyle \pm}  1$ &130${\displaystyle \pm}  7$ &136${\displaystyle \pm}  8$ \\ 
NGC 6380 &-0.35${\displaystyle \pm}0.46$ &-1.97${\displaystyle \pm}0.08$ &-0.56${\displaystyle \pm}0.03$ &-25${\displaystyle \pm}  2$ & 83${\displaystyle \pm}  9$ & 12${\displaystyle \pm}  1$ \\ 
NGC 6401 &0.04${\displaystyle \pm}0.40$ &0.55${\displaystyle \pm}0.02$ &0.58${\displaystyle \pm}0.03$ &-95${\displaystyle \pm}  3$ &241${\displaystyle \pm}  1$ &118${\displaystyle \pm}  6$ \\ 
NGC 6440 &-0.34${\displaystyle \pm}0.39$ &1.06${\displaystyle \pm}0.05$ &0.56${\displaystyle \pm}0.03$ &-34${\displaystyle \pm}  1$ & 91${\displaystyle \pm}  7$ &-38${\displaystyle \pm}  2$ \\ 
NGC 6453 &-1.24${\displaystyle \pm}0.50$ &-1.41${\displaystyle \pm}0.04$ &-0.66${\displaystyle \pm}0.03$ &-108${\displaystyle \pm}  3$ & 25${\displaystyle \pm} 12$ &-139${\displaystyle \pm}  7$ \\ 
NGC 6522 &0.87${\displaystyle \pm}0.34$ &0.55${\displaystyle \pm}0.01$ &-0.48${\displaystyle \pm}0.02$ &-13${\displaystyle \pm}  1$ &105${\displaystyle \pm}  7$ &-177${\displaystyle \pm}  9$ \\ 
NGC 6528 &0.38${\displaystyle \pm}0.40$ &0.35${\displaystyle \pm}0.01$ &-0.55${\displaystyle \pm}0.03$ &223${\displaystyle \pm}  1$ & 38${\displaystyle \pm} 11$ &-40${\displaystyle \pm}  2$ \\ 
NGC 6558 &0.77${\displaystyle \pm}0.36$ &0.39${\displaystyle \pm}0.01$ &-0.77${\displaystyle \pm}0.04$ &-185${\displaystyle \pm}  1$ & 97${\displaystyle \pm}  8$ & 12${\displaystyle \pm}  1$ \\ 
NGC 6624 &0.17${\displaystyle \pm}0.40$ &0.50${\displaystyle \pm}0.02$ &-1.09${\displaystyle \pm}0.06$ & 59${\displaystyle \pm}  1$ & 25${\displaystyle \pm} 11$ &-122${\displaystyle \pm}  6$ \\ 
NGC 6626 &2.42${\displaystyle \pm}0.29$ &1.93${\displaystyle \pm}0.04$ &-0.51${\displaystyle \pm}0.03$ & 40${\displaystyle \pm}  1$ & 53${\displaystyle \pm} 10$ &-91${\displaystyle \pm}  5$ \\ 
NGC 6638 &-1.75${\displaystyle \pm}0.48$ &0.66${\displaystyle \pm}0.07$ &-1.20${\displaystyle \pm}0.06$ & 52${\displaystyle \pm}  2$ & 39${\displaystyle \pm} 11$ & 25${\displaystyle \pm}  2$ \\ 
NGC 6637 &-0.52${\displaystyle \pm}0.45$ &0.05${\displaystyle \pm}0.01$ &-1.57${\displaystyle \pm}0.08$ & 81${\displaystyle \pm}  2$ &-56${\displaystyle \pm} 16$ & 82${\displaystyle \pm}  4$ \\ 
NGC 6642 &-0.20${\displaystyle \pm}0.38$ &1.41${\displaystyle \pm}0.07$ &-0.89${\displaystyle \pm}0.04$ & -5${\displaystyle \pm}  1$ &116${\displaystyle \pm}  7$ &-50${\displaystyle \pm}  3$ \\ 
NGC 6717 &0.30${\displaystyle \pm}0.35$ &1.96${\displaystyle \pm}0.08$ &-1.41${\displaystyle \pm}0.07$ & 93${\displaystyle \pm}  3$ & 61${\displaystyle \pm} 10$ & 26${\displaystyle \pm}  2$ \\ 
NGC 6723 &0.36${\displaystyle \pm}0.38$ &0.18${\displaystyle \pm}0.01$ &-2.44${\displaystyle \pm}0.12$ &-100${\displaystyle \pm}  1$ &182${\displaystyle \pm}  4$ &-34${\displaystyle \pm}  4$ \\ 
Pal 6 &1.03${\displaystyle \pm}0.34$ &0.77${\displaystyle \pm}0.01$ &0.24${\displaystyle \pm}0.01$ &193${\displaystyle \pm}  2$ &-50${\displaystyle \pm} 15$ &181${\displaystyle \pm}  8$ \\ 
Terzan 1 &2.49${\displaystyle \pm}0.29$ &0.89${\displaystyle \pm}0.01$ &0.12${\displaystyle \pm}0.01$ & 62${\displaystyle \pm}  2$ &104${\displaystyle \pm}  8$ & -2${\displaystyle \pm}  2$ \\ 
Terzan 2 &0.73${\displaystyle \pm}0.38$ &-0.21${\displaystyle \pm}0.02$ &0.33${\displaystyle \pm}0.01$ &127${\displaystyle \pm}  1$ & 13${\displaystyle \pm} 11$ &-49${\displaystyle \pm}  3$ \\ 
Terzan 4 &0.88${\displaystyle \pm}0.37$ &-0.17${\displaystyle \pm}0.03$ &0.19${\displaystyle \pm}0.01$ &-46${\displaystyle \pm}  4$ & 42${\displaystyle \pm} 11$ &100${\displaystyle \pm}  6$ \\ 
Terzan 5 &1.35${\displaystyle \pm}0.34$ &1.12${\displaystyle \pm}0.02$ &0.21${\displaystyle \pm}0.01$ &-58${\displaystyle \pm}  2$ & 79${\displaystyle \pm} 10$ &-29${\displaystyle \pm}  3$ \\ 
Terzan 6 &1.02${\displaystyle \pm}0.36$ &0.27${\displaystyle \pm}0.01$ &-0.26${\displaystyle \pm}0.01$ &141${\displaystyle \pm}  2$ &-55${\displaystyle \pm} 15$ & 19${\displaystyle \pm}  2$ \\ 
Terzan 9 &2.15${\displaystyle \pm}0.28$ &1.40${\displaystyle \pm}0.02$ &-0.18${\displaystyle \pm}0.01$ & 52${\displaystyle \pm}  3$ & 45${\displaystyle \pm} 10$ &-49${\displaystyle \pm}  3$ \\ 
\hline
\end{tabular}
    \caption{Cartesian coordinates and velocities of globular clusters considered in the present work.}
    \label{tab:CG_prop}
\end{table*}
%%%%%%%%%%%%%%%%%%%%%%%%%%%%%%%%%%%%%%%%%%%%%%%%%%%%%%%
%%%%%%%%%%%%%%%%%%%%%%%%%%%%%%%%%%%%%%%%%%%%%%%%%%%%%%%
To study the kinematics of GCs in different barred potentials, we first selected 30 CGs, which were previously identified in~\cite{BB2020} as those that belong to the bar/bulge. \textcolor{black}{These GCs were selected from a catalogue of 152 GCs from~\cite{BB2021} based on the following criteria. First, a geometric criterion was applied to retain only those GCs whose apocentric distance $r_\mathrm{apo}$ is less than 3.5 kpc~\citep{2019A&A...630L...4M, BB2020}. This reduces the sample to 39 members. Then, nine GCs were found to belong to the disc based on the angular momentum and eccentricity of the corresponding orbits (see details in~\citealt{BB2020}) and, thus, were removed from the sample.} 
\textcolor{black}{Table~\ref{tab:CG_prop_obs} and Table~\ref{tab:CG_prop} list the chosen GCs, as well as their observational parameters and Cartesian coordinates and velocities, used below to integrate orbits backwards in time \textcolor{black}{for 5 Gyr}.}  Coordinates and velocities are obtained from equatorial coordinates $(\alpha_{J2000},\delta_{J2000})$, line-of-sight velocities from the catalogue of~\cite{Vasiliev2019}, distances from~\cite{Baumgardt2021}, and proper motions from~\cite{Vasiliev2021}. The catalogue of~\cite{Vasiliev2019} is compiled based on the Gaia DR2 data, while the catalogues of \cite{Vasiliev2021, Baumgardt2021} contain new proper motions and refined distances based on Gaia EDR3 data, Hubble Space Telescope (HST) data, and some literature estimates. The transformation from angular coordinates and velocities is performed~\textcolor{black}{using the values obtained by~\cite{Bajkova_2016,Bajkova_2017} from rotation curve fitting, i.e.} under the assumption that the distance from the Galaxy centre to the Sun $R_{\sun}=8.3$ kpc, the height of the Sun above the disc plane $Z_{\sun}=17$ pc~\citep{2016AstL...42....1B}, the velocity of local standard of rest (LSR) $V_{\sun}=244$ km/s. The peculiar velocity  of the Sun relative to LSR $(u_{\sun},v_{\sun},w_{\sun})=(-11.1,12.2,7.3)$ km/s is taken from~\cite{Sch2010}. \textcolor{black}{For the bar viewing angle, the value 23 deg was taken from~\cite{Mosenkov2021}, where it was estimated from fitting the boxy/peanut bulge intensity profile for different viewing angles.}
\par

%%%%%%%%%%%%%%%%%%%%%%%%%%%%%%%%%%%%%%%%%%%%%%%%%%%%%%%
%%%%%%%%%%%%%%%%%%%%%%%%%%%%%%%%%%%%%%%%%%%%%%%%%%%%%%%
\section{Simulations}
\label{sec:sim}
%%%%%%%%%%%%%%%%%%%%%%%%%%%%%%%%%%%%%%%%%%%%%%%%%%%%%%%
%%%%%%%%%%%%%%%%%%%%%%%%%%%%%%%%%%%%%%%%%%%%%%%%%%%%%%%

\subsection{Mass models}
%%%%%%%%%%%%%%%%%%%%%%%%%%%%%%%%%%%%%%%%%%%%%%%%%%%%%%%%%%%%%%%%%%%%%%%%%%%%%%%%%%%%%%%%%%%%%%%%%%%%%%%
\begin{table}
\centering
\caption{Description of the model parameters. \textcolor{black}{For bar parameters, a range of values is indicated (symbol ``$\div$'' ), which is considered in the course of this work.}}
\begin{tabular}{ c | c | c | c | c | c | c |}
\hline
Parameter & Meaning & Value \\
\hline 
\hline
$M_\mathrm{b}$ & Bulge mass & $1.0 \cdot 10^{10}$ $M_{\sun}$\\
$r_\mathrm{b}$ & Bulge scale length &  $0.2672$ kpc \\
$M_\mathrm{d}$ & Disc mass &  $6.5 \cdot 10^{10}$ $M_{\sun}$ \\
$a_\mathrm{d}$ & Disc scale length &  $4.4$ kpc\\
$b_\mathrm{d}$ & Disc scale height &  $0.3$ kpc \\
$M_\mathrm{h}$ & Halo mass &  $2.9 \cdot 10^{11}$ $M_{\sun}$ \\
$a_\mathrm{h}$ & Halo scale length &  $7.7$ kpc \\
$M_\bars$ & Bar mass &  $(0 \div 0.95) M_\mathrm{b}$ \\
$a_\bars$ & Bar major hemi-axis &  $2.5\div5.0$ kpc \\
$p$ & Ratio of the bar axes in the disc plane &  $2.0\div4.0$ \\
$q$ & Flattening of the bar &  $3.0\div4.0$ \\
$\alpha$ & Bar viewing angle & 23 deg \\
%%%%%%%%%%%%%%%%%%%%%%%%%%%%%%%%%%%%%%%%%%%%%%%%%%%%%%%%%%%%%%%%%%%%%%%%%%%%%%%%%%%%%%%%%%%%%%%%%%%%%%%
%\multicolumn{3}{p{0.4\textwidth}}
%{\footnotesize{\textit{Notes}: }}
\end{tabular}
\label{tab:models1_pars}
\end{table} 
%%%%%%%%%%%%%%%%%%%%%%%%%%%%%%%%%%%%%%%%%%%%%%%%%%%%%%%%%%%%%%%%%%%%%%%%%%%%%%%%%%%%%%%%%%%%%%%%%%%%%%%

In the present work, we consider several types of mass models of the Milky Way. 
The first one was obtained by~\cite{Bajkova_2016,Bajkova_2017} (hereinafter, BB2016) via fitting the rotation curve to the kinematic data of a set of different objects with distances up 200 taken from~\cite{2014ApJ...785...63B}. The mass model consists of three distinct components, namely the bulge~\citep{1911MNRAS..71..460P}, the disc~\citep{1975PASJ...27..533M}, and the halo~\citep{NFW}:
\begin{equation}
    \Phi_\mathrm{bulge}(r)= - \displaystyle \frac{M_\mathrm{b}}{(r^2 + b_\mathrm{b}^2)^{1/2}},
\end{equation}
\begin{equation}
   \Phi_\mathrm{disc} (R, Z)= - \displaystyle \frac{M_\mathrm{d}}{\left[R^2 + \left(a_\mathrm{d} + \sqrt{Z^2 + b_\mathrm{d}^2}\right)^2\right]^{1/2}}, 
\end{equation}
\begin{equation}
    \Phi_\mathrm{halo}(r) = - \frac{M_\mathrm{h}}{r} \ln{\left(1 + \frac{r}{a_\mathrm{h}}\right)}.
    \label{eq:NFW}
\end{equation}
Description of the parameters and their respective values are given in Table~\ref{tab:models1_pars}.
%We introduce the bar component in this model by reducing the mass of the central component (bulge) for a particular value and assigning the bar mass to this value. Below, we consider a set of of values fore the bar mass. 
%The bar potential itself is described by a triaxial rotating ellipsoid model:
%\begin{equation}
%    \Phi_\mathrm{bar} = \frac{M_\bars}{\sqrt{(a_\bars^2 + \tilde{r}^2)}}, 
%    \label{eq:bar}
%\end{equation}
%where $\tilde{r}=\sqrt{x^2 + (y/p)^2 + (z/q)^2}$ is the elliptical radius and $p$ and $q$ characterise the flattening of the bar in disc plane and along the vertical axis, respectively.

\par 
The second model is taken from~\cite{McMillan_2017} (hereinafter, MC2017) and consists of six different components, namely thin and thick stellar discs, dark matter halo, and \begin{small}H\,I\end{small} and molecular discs. In this model, the dark halo is also described by a Navarro-Frank-White profile, given in eq.~(\ref{eq:NFW}). The stellar discs are exponential both in the plane and the vertical direction:
\begin{equation}
\rho_\dd(R, z) =  \frac{\Sigma_0}{2 z_\dd} \exp{\left(-\frac{|z|}{z_\dd} - \frac{R}{R_\dd}\right)},
\end{equation}
while gaseous discs are exponential in the plane and isothermal in the vertical direction and have a hole in the centre with the scale of $R_\mathrm{m}$:
\begin{equation}
\rho_\dd(R, z) = \frac{\Sigma_0}{4z_\dd}\exp{\left(-\frac{R_\mathrm{m}}{R} - \frac{R}{R_\dd}\right)} \mathrm{sech}^2\left[z/(2z_\dd)\right]
\label{eq:disc}
\end{equation}
The central component (bulge) is implemented via the following parametric model:
\begin{equation}
    \rho_\mathrm{b} = \frac{\rho_\mathrm{0,b}}{(1+r'/r_0)^\alpha} \exp{\left[-(r/r_\mathrm{cut})^2\right]},
\end{equation}
where $r'=\sqrt{R^2 + (z/q_\mathrm{bulge})^2}$. To avoid repeatance, we refer the reader to~\cite{McMillan_2017} for a description of the parameters and their values.\par  
In both models, we introduce a bar component by decreasing the mass of the central component (bulge) by a certain value and then assigning the bar mass to this value. \textcolor{black}{Essentially, this means that, for all models considered below~(except the $N$-body one), the total mass of a spherical bulge and the bar is fixed:
\begin{equation}
     M_\mathrm{bar} + M_\mathrm{b} = M_\mathrm{b,0},
\end{equation}
where $M_\mathrm{b,0}$ is the initial bulge mass of the axisymmetric model and $M_\mathrm{b}$ is the residue mass of the bulge. Hereinafter, we refer to  $M_\mathrm{b,0}$ simply as $M_\mathrm{b}$, since we do not consider the residue bulge mass as an independent parameter at any part of this work. Below, we consider a set of bar mass values, or, more precisely, a number of bar-to-bulge mass ratios $M_\mathrm{bar}/M_\mathrm{b}$ (see Table~\ref{tab:models1_pars}). \cite{Ortolani2019a,Ortolani2019b,PV2020} assigned all the bulge mass to the bar component in their models. Here, we introduce the ratio of the bulge and bar masses as a free parameter to investigate how uncertainty in the classical bulge parameters possibly existing in our Galaxy can affect the results of orbital classification.} 
 For the bar density profile, we take a Ferrers profile:
\begin{equation}
    \rho = \displaystyle\frac{105M_\mathrm{bar}}{32\pi p q a^3} \left[ 1-\left(\frac{\tilde{r}}{a_\bars}\right)^2\right]^2,
\end{equation}
where $M_\bars$ is the bar mass, $a_\bars$ is the bar major axis, $\tilde{r}=\sqrt{x^2 + (y/p)^2 + (z/q)^2}$ is the elliptical radius and $p$ and $q$ characterise the flattening of the bar in disc plane and along the vertical axis, respectively. 
%, $p$, and $q$ have the same meaning as in eq.~(\ref{eq:bar}). 
The bar parameters and their description are given in Table~\ref{tab:models1_pars}.
\par 
The third type of potential is taken from a recent work by~\cite{2021arXiv211105466T} (hereinafter, TG2021), where a surrogate Milky Way $N$-body model was presented. Time snapshots of the model were made publicly available by the authors. At start of the simulations, the model consisted of two spherical components, NWF-like halo~\citep{NFW} and a stellar bulge~\citep{Hernquist1990}, and an exponential disc isothermal in the vertical direction~(similar to eq.~(\ref{eq:disc}), but without the hole). The evolution of the model was followed up to about 4.3 Gyr. There is no need to insert the bar component separately or transform the bulge as the bar in this model is formed naturally (see Fig.~\ref{fig:nbody_model}). For simplicity, we consider here only the last snapshot of ~\cite{2021arXiv211105466T}'s simulations, neglecting the time evolution of the bar properties. We leave this for future studies. 
\par 
\textcolor{black}{For the selected time moment, the $N$-body bar has the size $a_\mathrm{bar}$ about 4.5 kpc and the pattern speed $\Omega_\mathrm{p}\approx39$ km/s/kpc. The mass of the bar was not estimated directly in~\cite{2021arXiv211105466T}, but the authors provided an overall estimate $M_\mathrm{bar}+M_\mathrm{disc}+M_\mathrm{bulge}=3.5\times10^{10} M_{\sun}$ of stellar mass inside the area of $R<5$ kpc (bar region), where $M_\mathrm{disc}$ is the mass associated with the inner area of the disc and $M_\mathrm{bulge}$ is the mass of the classic bulge originally included in the model.}
\par 
The number of particles in the $N$-body model is about $7\cdot 10^{7}$. To avoid very time-consuming calculations of gravitational force at each time-step when integrating the orbits, we prepared a multipole expansion of the potential using the convenient \texttt{Multipole} subroutine from \texttt{AGAMA} software package~\citep{agama}:
%%%%%%%%%%%%%%%%%%%%%%%%%%%%%%%%%%%%%%%%%%%%%%%%%%%%%%%
%%%%%%%%%%%%%%%%%%%%%%%%%%%%%%%%%%%%%%%%%%%%%%%%%%%%%%%
\begin{equation}
\Phi(r, \theta, \varphi)=\Sigma_{l,m} \Phi_{l,m}(r)Y_l^m(\theta, \varphi) \,,
\label{eq:pot}
\end{equation}
 %%%%%%%%%%%%%%%%%%%%%%%%%%%%%%%%%%%%%%%%%%%%%%%%%%%%%%%
%%%%%%%%%%%%%%%%%%%%%%%%%%%%%%%%%%%%%%%%%%%%%%%%%%%%%%%
 where $Y_{l}^m$ are spherical functions of degree $l$ and order $m$. We truncate the series at $l_\mathrm{max}=6$ and $m_\mathrm{max}=6$ and impose a triaxial type of symmetry (only even harmonics are calculated). Isolines of the potential approximations are shown in the right panel of Fig.~\ref{fig:nbody_model}. Note that the potential isolines are rounder than the density isolines, as they should be~(\citealt{Binney_Tremaine2008}, Chapter 2), but still showing the flattening in the bar area. In the very central part, the classic bulge overweighs other components and, thus, the isolines are circular here. 
\par
\textcolor{black}{For potentials of BB2016 and MC2017, we consider a range of  bar pattern speeds ($\Omega_\mathrm{p}$) and sizes ($a_\mathrm{bar}$, the half length of the bar major axis), \textcolor{black}{from 30 km/s/kpc to 60 km/s/kpc and from 5.0 kpc to 2.5 kpc, respectively.} Fig.~\ref{fig:cor} shows how the mentioned limits correspond to the main resonances in the potentials of BB2016 and MC2017. The dynamics of the bars is usually characterised by the rotation rate parameter $\mathcal{R}=R_\mathrm{CR}/a_\mathrm{bar}$ \citep{Binney_Tremaine2008}. As can be seen from the figure, we consider both slow ($\mathcal{R}\gg 1$) and fast bars ($\mathcal{R}\lesssim 1$) here. For other galaxies, $\mathcal{R}$ spans the range from almost 0 to about 4~\citep{2009ApJS..182..559B, 2019A&A...632A..51C, 2019MNRAS.482.1733G, 2020MNRAS.491.3655G, 2022MNRAS.517.5660G}. For our bars, $\mathcal{R}$ is from about 0.5 for $\Omega_\mathrm{p}=60$ km/s/kpc and $a_\mathrm{bar}=5$ kpc to about 10 for $\Omega_\mathrm{p}=10$ km/s/kpc and $a_\mathrm{bar}=2.5$ kpc. \textcolor{black}{For generality, we consider bars with major axes and pattern speeds having the values from the suggested ranges indisciminately, although longer bars tend to have lower pattern speeds (see figure 15 in~\citealt{2022MNRAS.517.5660G}).}}

\subsection{Orbit integration and classification}
\par
For each of the described potentials we add the rotation in accordance with the chosen value of the pattern speed $\Omega_\mathrm{p}$ and integrate orbits of GCs backwards in time for a time period of $5$ Gyr. \textcolor{black}{In the present work, we are interested in the orbital families which an orbit can belong to potentially. This ``property'' should not depend on the type of the integration (forward or backaward) \textcolor{black}{for regular orbits}\footnote{\textcolor{black}{For chaotic orbits, forward and backward integration can produce different results in terms of frequencies, but we are mostly concerned with the regular orbits captured or oscillating around the bar in the present work.}}, since orbital frequencies are ``integral'' properties of the orbit. Here, we consider backward integration, because it is in line with our previous studies, e. g.~\cite{2019MNRAS.488.3474B,2020ApJ...895...69B}.} Integration is carried out using the~{\texttt{AGAMA}} software package. {\texttt{AGAMA}} performs integration via 8th order Runge–Kutta scheme with an adaptive time-step. We choose an output time step $\Delta t=1$ Myr. The latter is unrelated to the actual time-step of integration, which is determined internally in ODE solver based on the imposed value of the relative accuracy, $10^{-8}$ in our case. \textcolor{black}{For each orbit, we traced the evolution of the Jacobi energy as an indicator of the accuracy of our calculations. A typical example is shown in Fig.~\ref{fig:JE}. In short, the energy is well conserved during integration (up to six decimal places).}
%\subsection{Potential from~\cite{Bajkova_2016,Bajkova_2017}}
%\subsection{\cite{McMillan_2017} potential}
%\subsection{$N$-body potentail}

\begin{figure}
	\centering\includegraphics[width=0.33\textwidth]{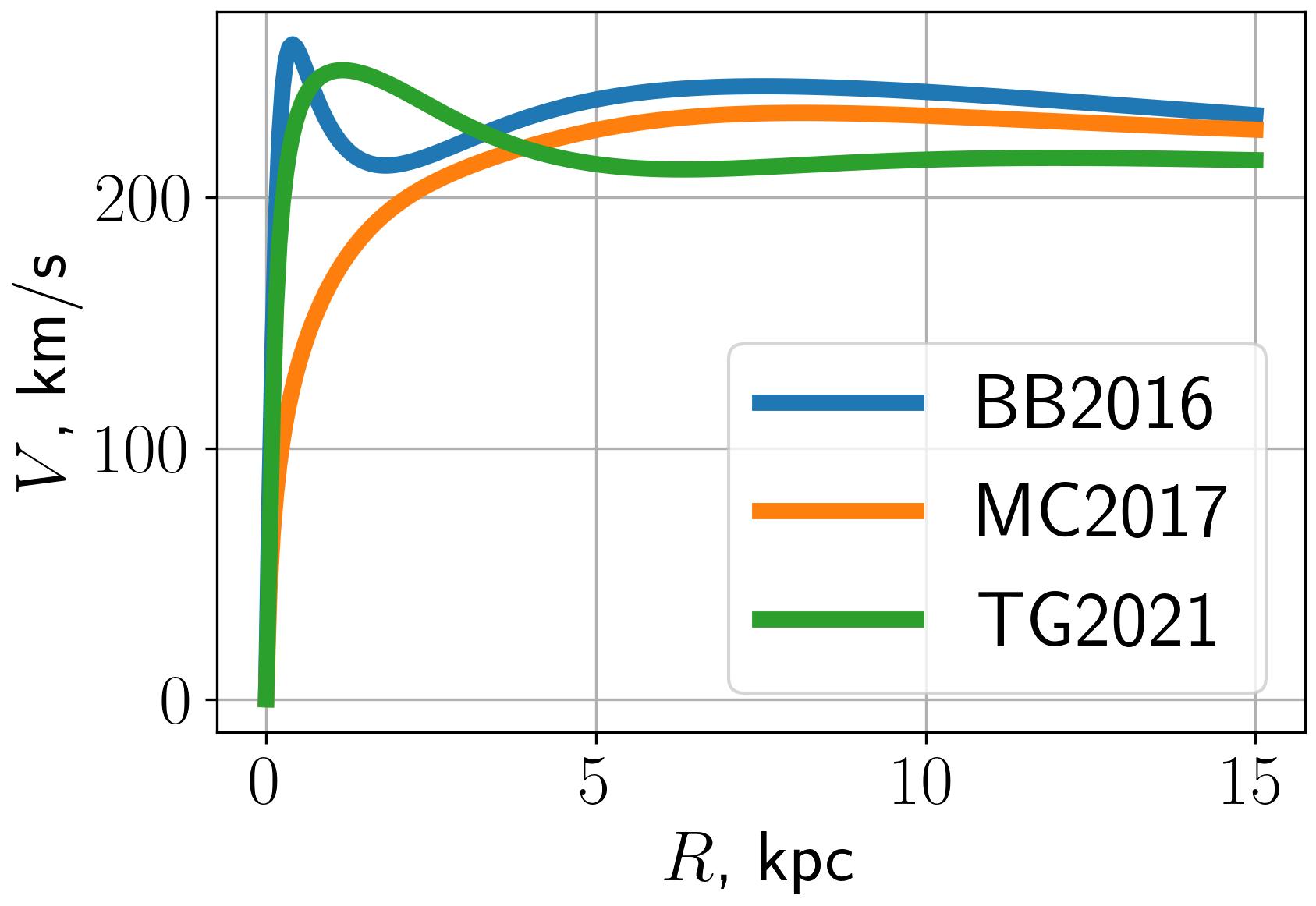}% 
    \caption{Rotation curves for the three potentials considered in the present work. Note that the velocity is calculated for azimuthally averaged potential in case of TG2021.}
    \label{fig:pot_vel}
\end{figure}

\begin{figure}
	\centering\includegraphics[width=0.33\textwidth]{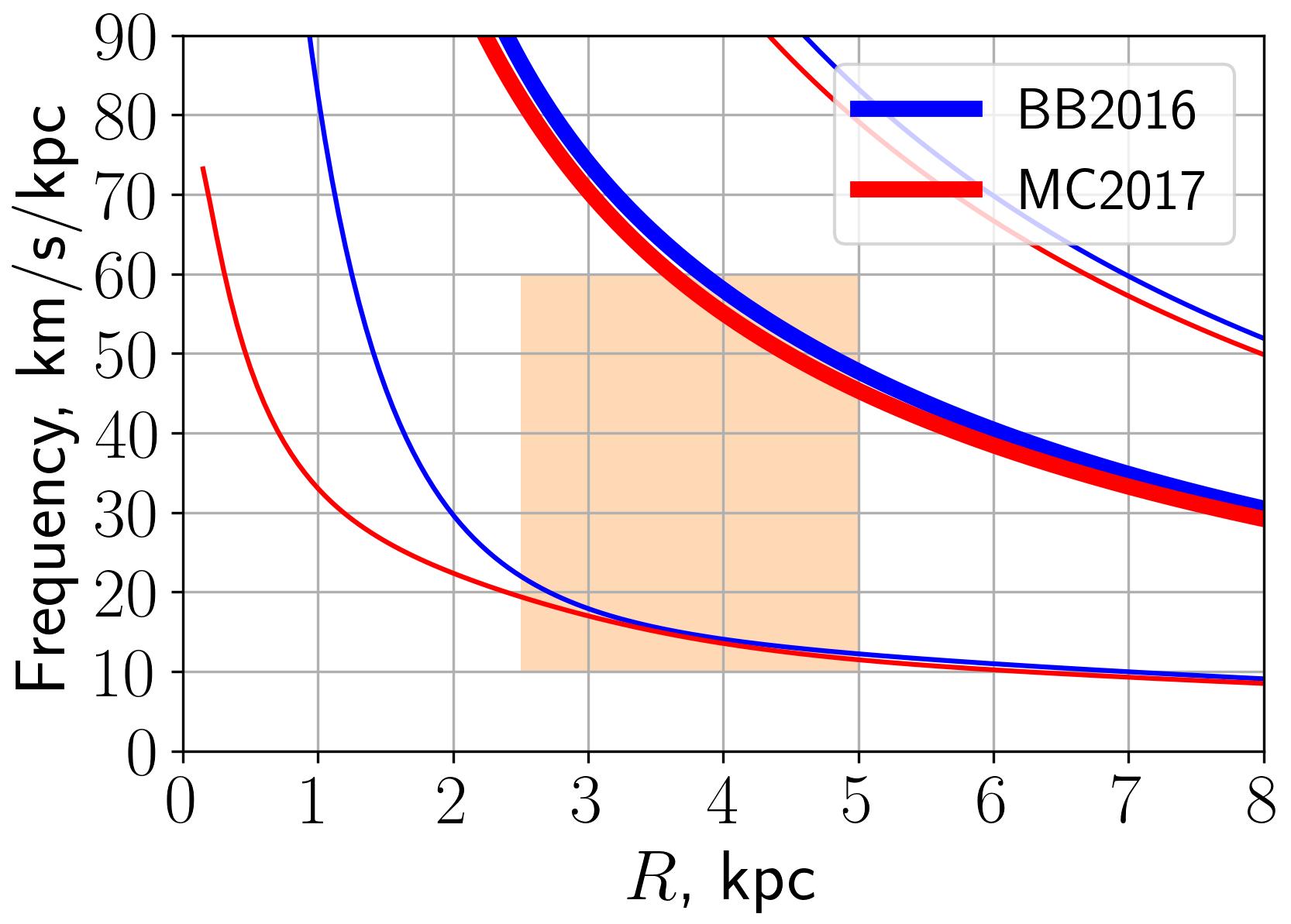}% 
    \caption{\textcolor{black}{The co-rotation (thick lines), inner Linblad resonance (thin lines below the co-rotation), and outer Lindblad resonance (thin lines above the co-rotation). The shaded area indicates the range of pattern speeds and bar sizes considered in the present work.}}
    \label{fig:cor}
\end{figure}

\begin{figure}
\begin{minipage}[c]{0.222\textwidth}
	\includegraphics[width=0.9\textwidth]{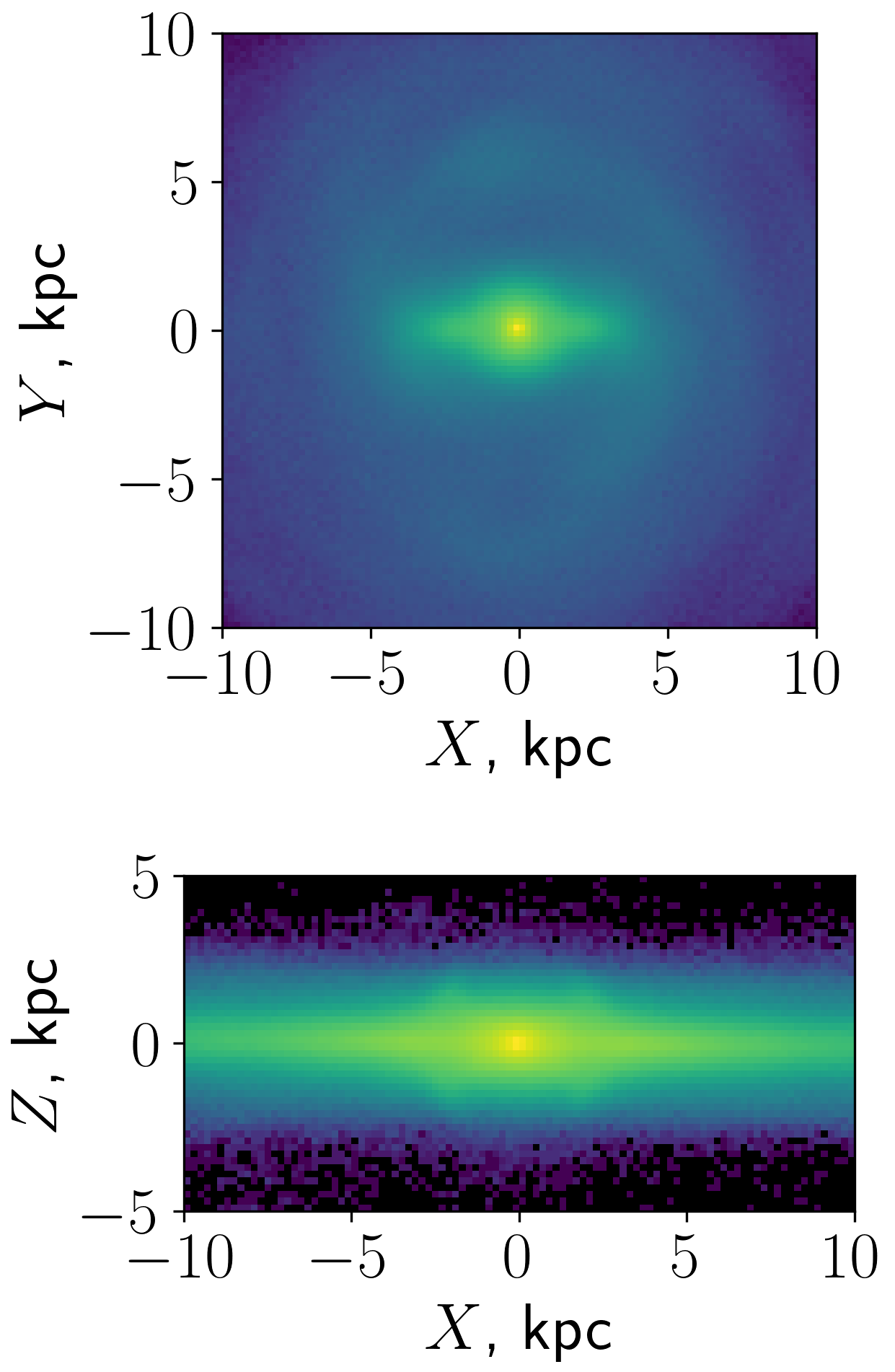}% 
\end{minipage}%
\begin{minipage}[c]{0.222\textwidth}
	\includegraphics[width=0.9\textwidth]{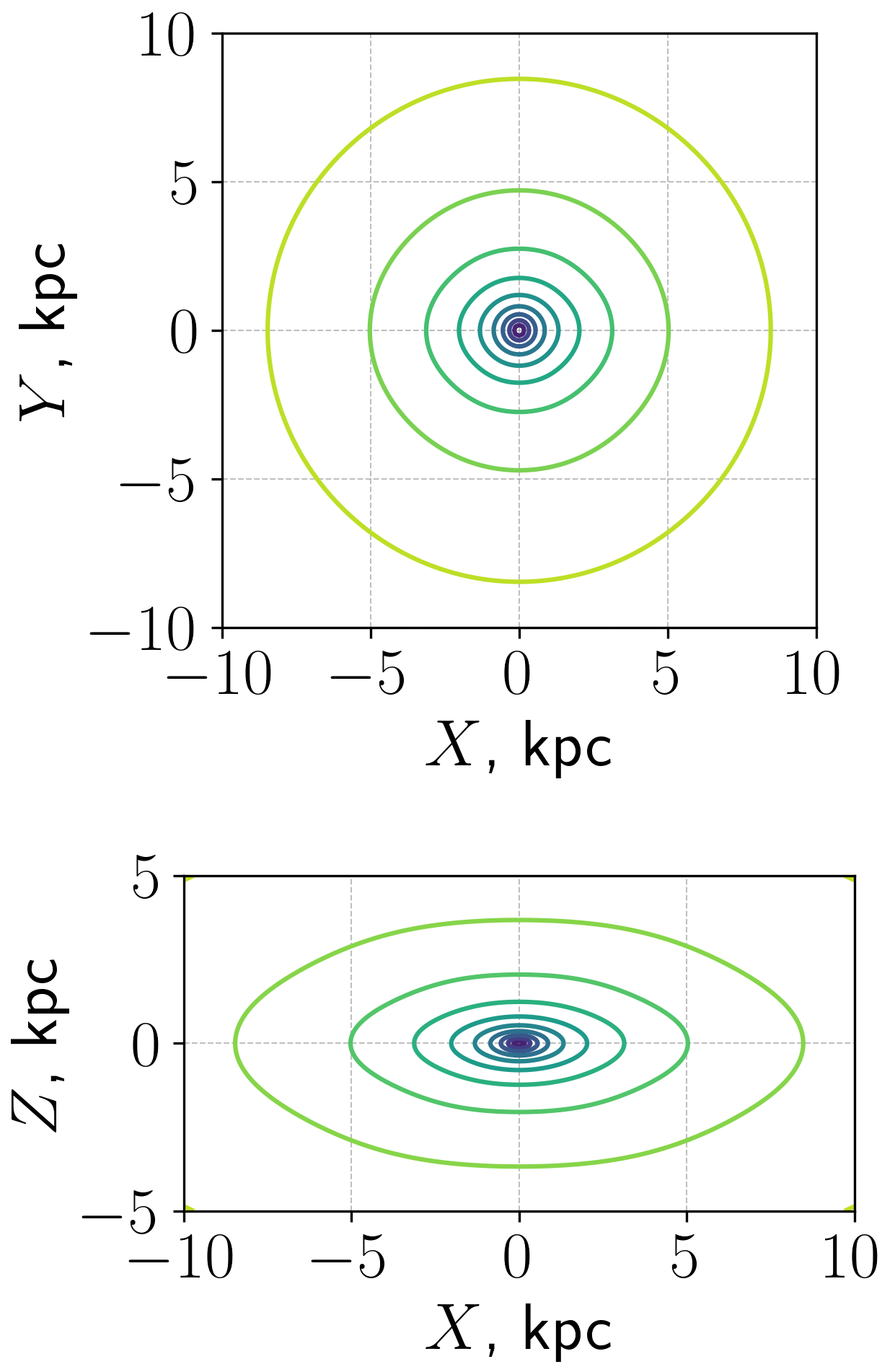}%
\end{minipage}

    \caption{Face-on (\textit{top}) and \textcolor{black}{edge-on (\textit{bottom}) views} of the density distribution in the $N$-body model (\textit{left}) and isolines of the potential approximation obtained from eq.~\ref{eq:pot} (\textit{right}).}
    \label{fig:nbody_model}
\end{figure}

\begin{figure}
	    %\vspace*{-0.1cm}
		\center{\includegraphics[width=0.45\textwidth]{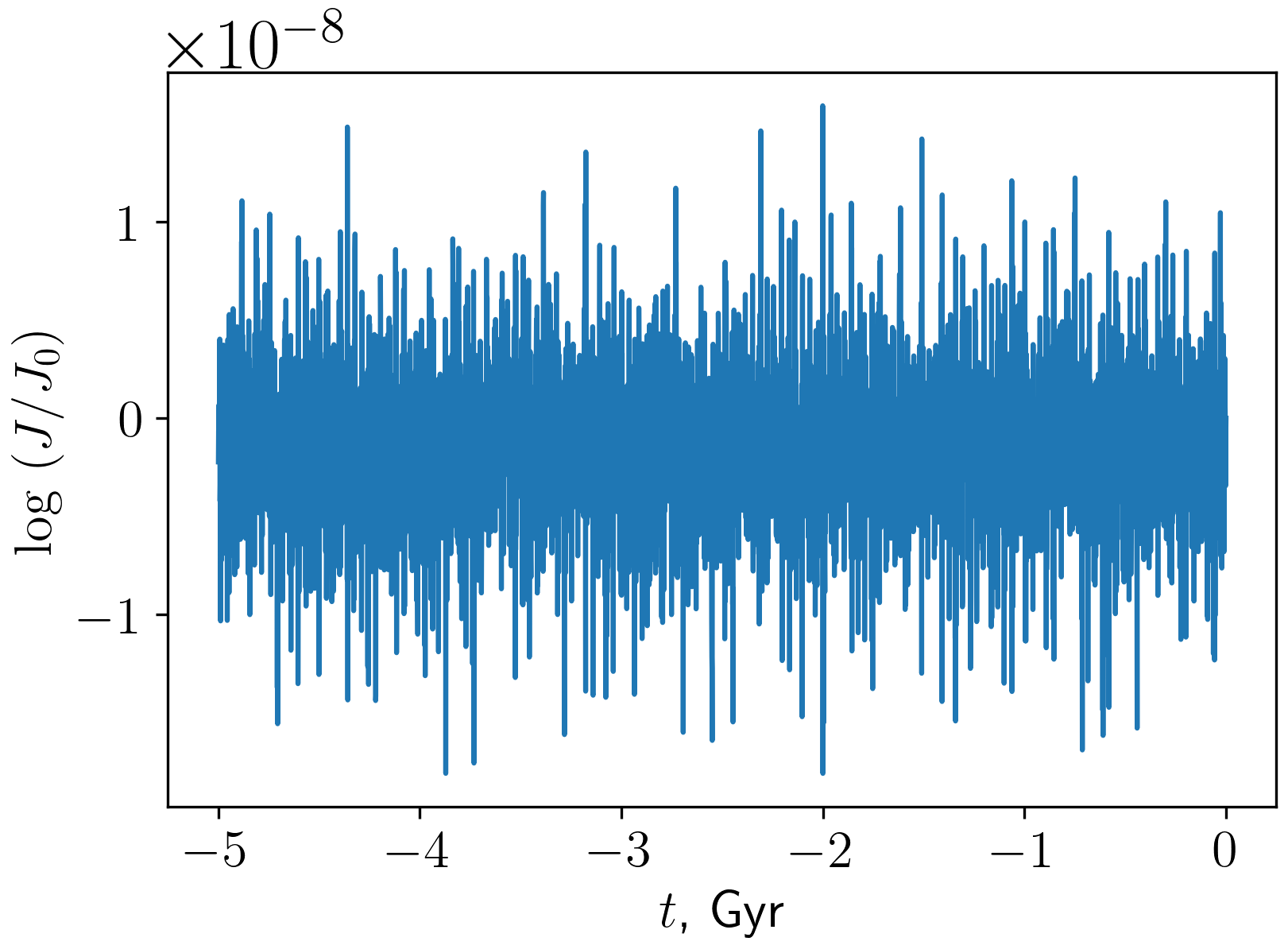}}
  \caption{\textcolor{black}{The decimal logarithm of the Jacobi energy normalised to its value at the beginning of integration for a typical orbit studied in the present work.}}
  \label{fig:JE}
\end{figure}%

    \begin{figure*}
    \centering
    \begin{minipage}[c]{0.4\textwidth}
	    %\vspace*{-0.1cm}
		\includegraphics[width=0.99\textwidth]{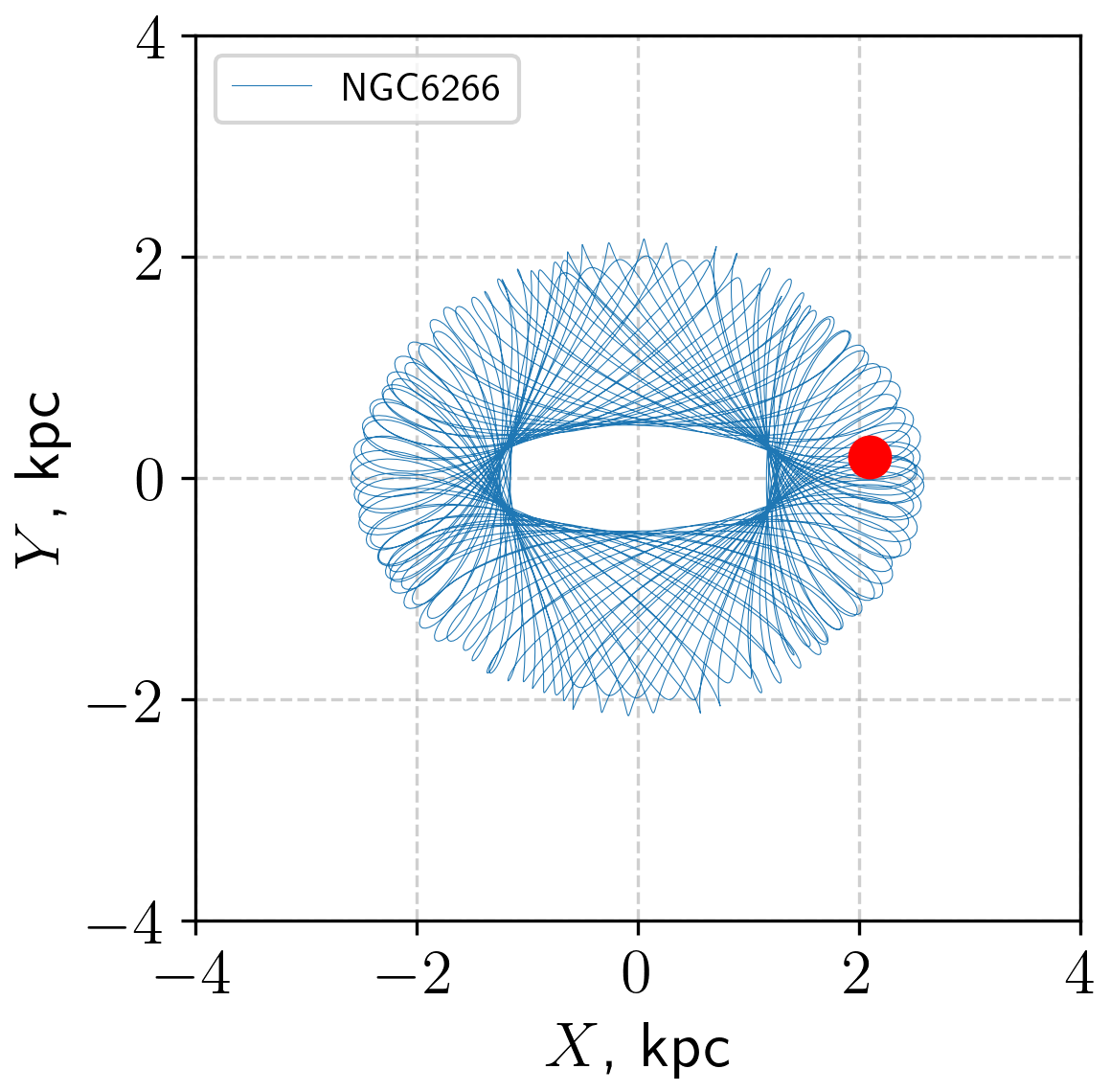} 
	\end{minipage}% 
    \begin{minipage}[t]{0.66\textwidth}
        \begin{minipage}[t]{0.49\textwidth}
	    %\vspace*{-0.1cm}
		\includegraphics[width=0.85\textwidth]{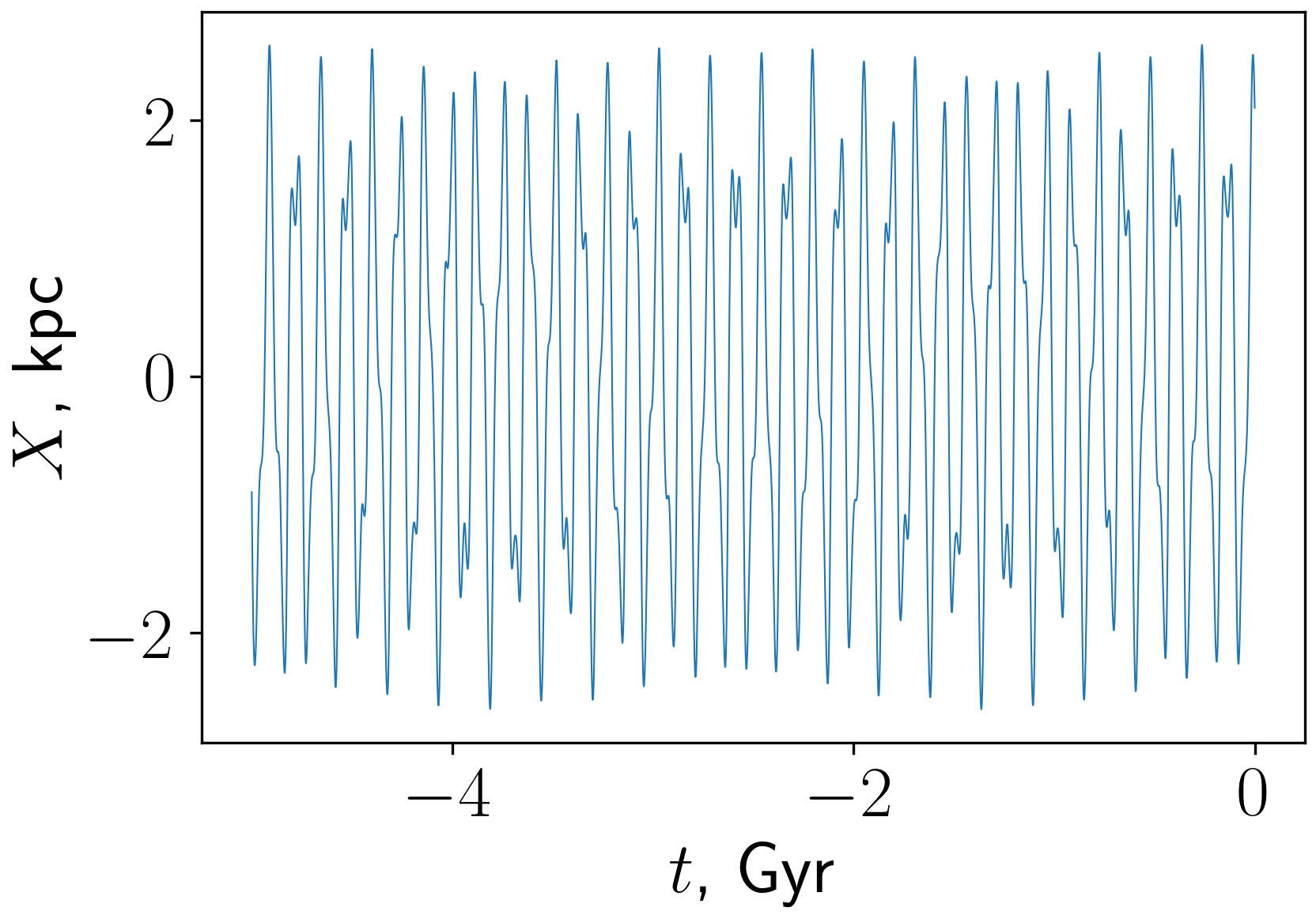} 
	\end{minipage}%
     \begin{minipage}[t]{0.49\textwidth}
	    %\vspace*{-0.1cm}
		\includegraphics[width=0.85\textwidth]{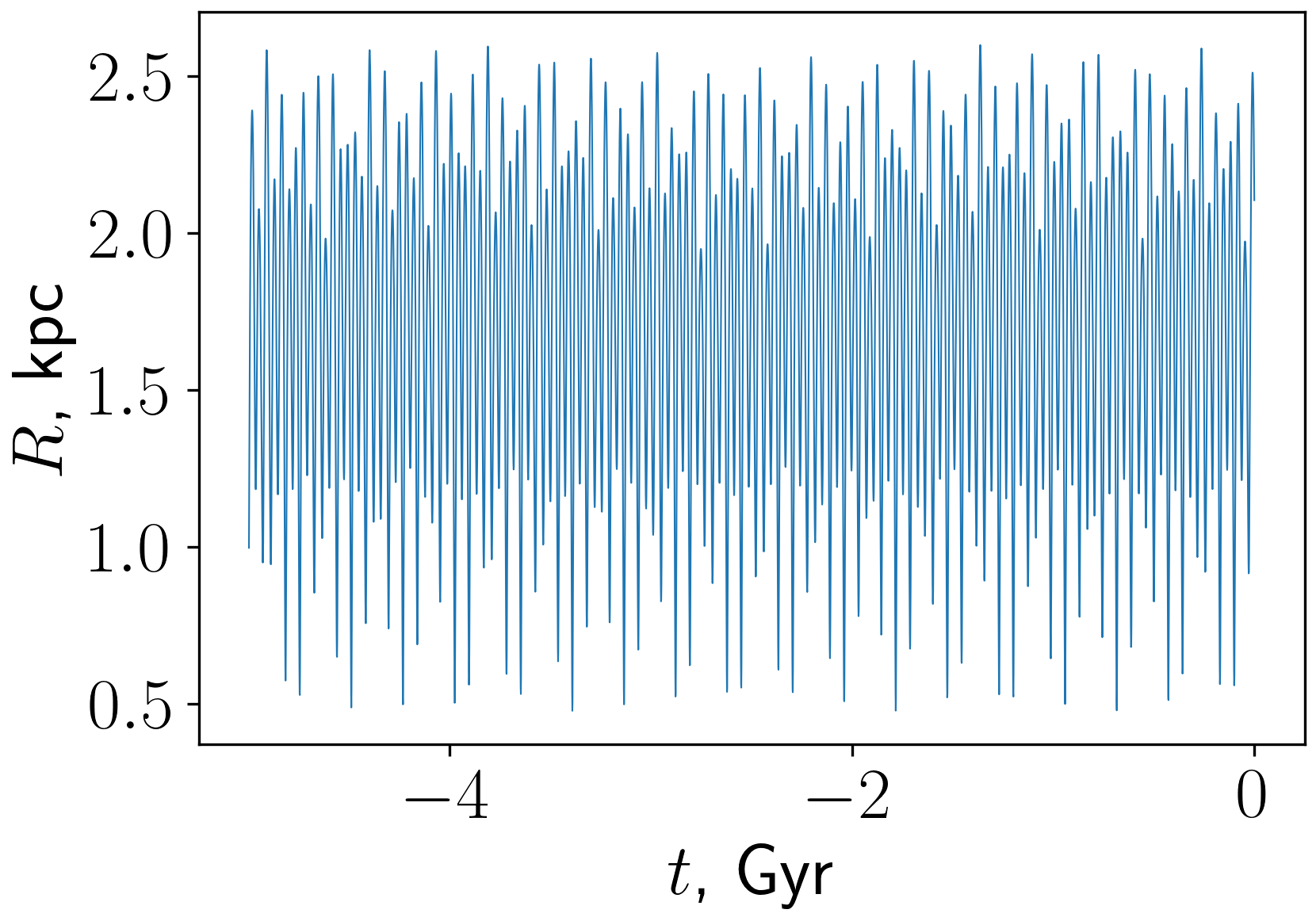} 
	\end{minipage}%
    \\
     \begin{minipage}[t]{0.49\textwidth}
	    %\vspace*{-0.1cm}
		\includegraphics[width=0.85\textwidth]{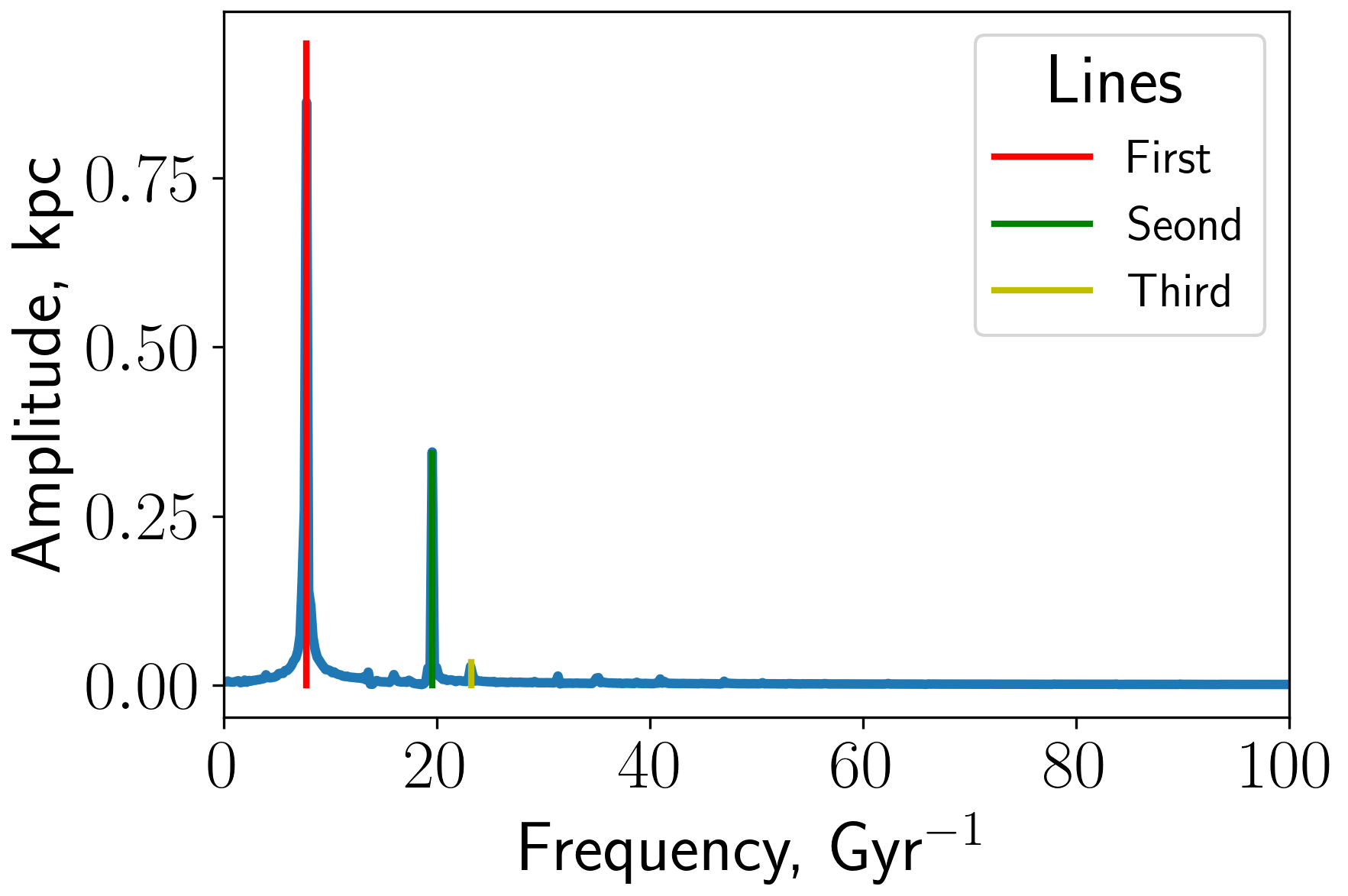} 
	\end{minipage}%
     \begin{minipage}[t]{0.49\textwidth}
	    %\vspace*{-0.1cm}
		\includegraphics[width=0.85\textwidth]{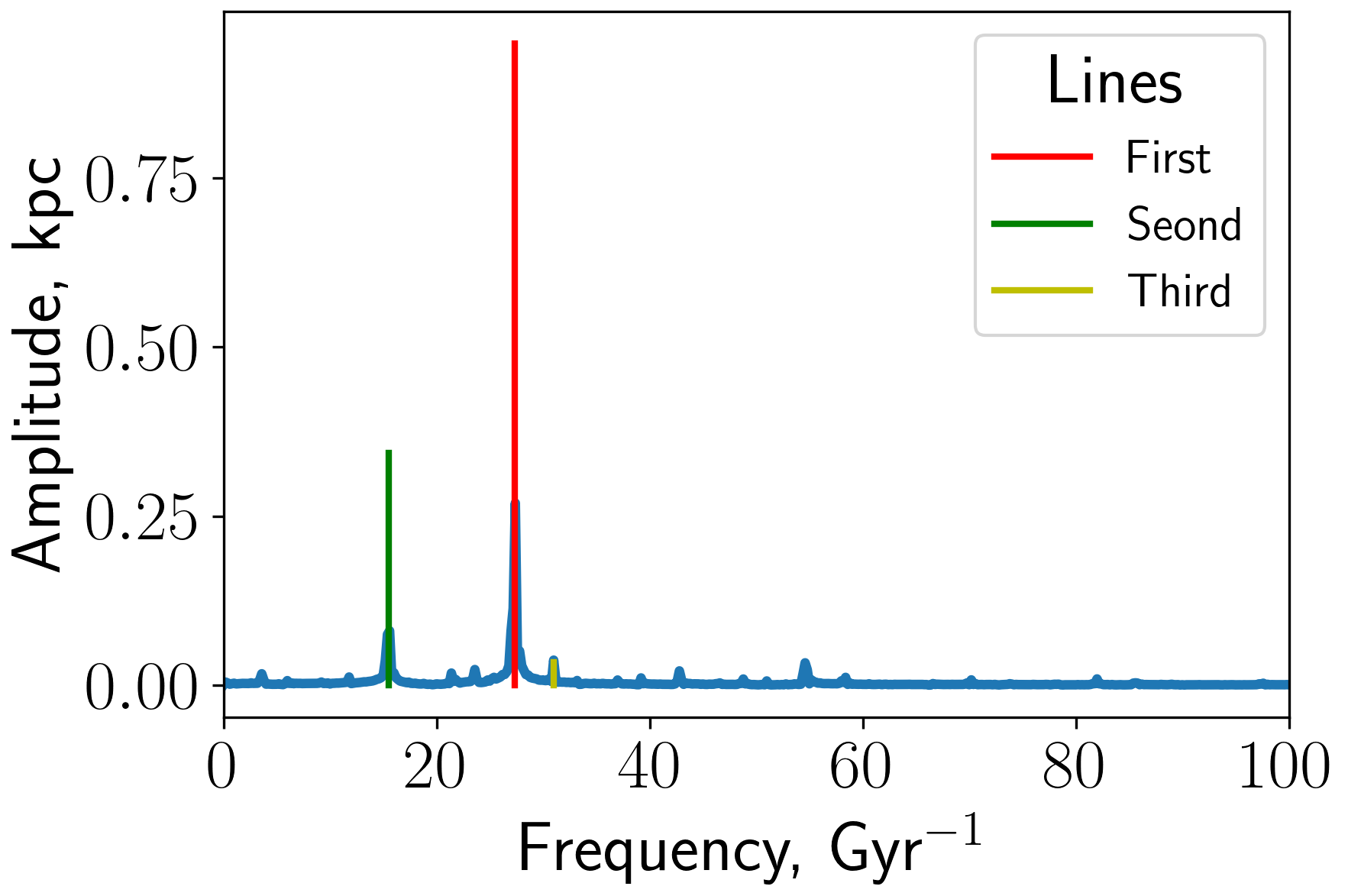} 
	\end{minipage}%
	    
	\end{minipage}% \\
	   \\
	\caption{An example of the orbit of NGC 6266 in case of $\Omega_\mathrm{p}=45$ km/s/kpc, $M_\mathrm{bar}/M_\mathrm{b}=0.95$, $a_\mathrm{bar}=5$ kpc, $p=2.0$, and $q=3.0$ (\textit{left}, shown in the bar rotating reference frame). The red dot marks the initial position of the cluster. \textit{Second} and \textit{third} columns: coordinate time series (\textit{top} row) and corresponding spectra (\textit{bottom} row).}
    \label{fig:spectra}
\end{figure*}

To classify orbits, we apply the  methods of spectral dynamics pioneered by~\cite{Binney_Spergel1982}. In this approach, one calculates the coordinate spectra of the orbit, i.e. Fourier transforms of the time series $x$, $y$, $z$, and $R$ taken in a bar rotating frame, and then finds dominant frequencies $f_x, f_y, f_z$, and $f_\mathrm{R}$ corresponding to the highest spectral lines. 
The spectra are calculated as follows
%%%%%%%%%%%%%%%%%%%%%%%%%%%%%%%%%%%%%%%%%%%%%%%%%%%%%%%%%%%%%%%%%%%%%%%
%%%%%%%%%%%%%%%%%%%%%%%%%%%%%%%%%%%%%%%%%%%%%%%%%%%%%%%%%%%%%%%%%%%%%%%
\begin{equation}
P_j =  \frac{1}{N_t} \left|\sum_{k=0}^{N_t-1} x_k  
\exp (-2\pi i f_{j} t_k) \right|, 
\label{eq:FT}
\end{equation}
%%%%%%%%%%%%%%%%%%%%%%%%%%%%%%%%%%%%%%%%%%%%%%%%%%%%%%%%%%%%%%%%%%%%%%%
%%%%%%%%%%%%%%%%%%%%%%%%%%%%%%%%%%%%%%%%%%%%%%%%%%%%%%%%%%%%%%%%%%%%%%%
where  $f_{j} = j/\Delta T$, $\Delta T = 5$ Gyr, $t_k = k \Delta t$, $\Delta t =1$ Myr, $0\leq j \leq (N_t-1)/2$, and $N_t$ is the length of the time series. To improve the resolution of the peaks, we use a subroutine similar to zero-padding (see details in~\citealt{Parul_etal2020}, where a similar analysis was applied to the study the orbital families of B/PS bulges).
\par 
For regular orbits, the spectra consist of discrete lines, these lines can be distinguished, and the corresponding frequencies can be studied to understand which orbital group or family the orbit belongs to~\citep{Binney_Spergel1982}. This approach made it possible to obtain many fruitful results on the orbital composition of the bar and the importance of various resonances for the structure of the bar in a number of studies~\citep{Gajda_etal2016, Wang_etal2016, Portail2017,Lokas2019,Parul_etal2020, Tikhonenko_etal2021,Smirnov_etal2021}. \cite{PV2020} also calculated the orbital frequencies to determine whether a particular GC follows the bar or not. Here, we use the same approach and assume that if $f_\mathrm{R}/f_x=2.0 \pm 0.1$, then the GC with such a ratio of frequencies is the bar supporting one, i.e. follows the bar. 
\par
A typical example of the orbit of NGC 6266, along with its time series of coordinates and their spectra, is presented in Fig.~\ref{fig:spectra}. \textcolor{black}{Hereinafter, all orbits presented in the figures are shown in the bar rotating frame if not specified otherwise.} Bar parameters are $\Omega_\mathrm{p}=45$ km/s/kpc, $M_\mathrm{bar}/M_\mathrm{b}=0.95$, $a_\mathrm{bar}=5$ kpc, $p=2.0$, and $q=3.0$ in this case. Integration is carried out in the potential of BB2016. We note that, although the orbit has a nice-looking regular profile, it does not actually follow the bar, since $f_\mathrm{R}/f_x\approx3.5$.
%Our second approach \textcolor{black}{LATER} MaySoS, Maybe Ej.

\section{Orbit type depending on the bar parameters}
\label{sec:systematics}

    \begin{figure*}
    \centering
    	\begin{minipage}[t]{0.33\textwidth}
	    %\vspace*{-0.1cm}
		\includegraphics[width=0.99\textwidth]{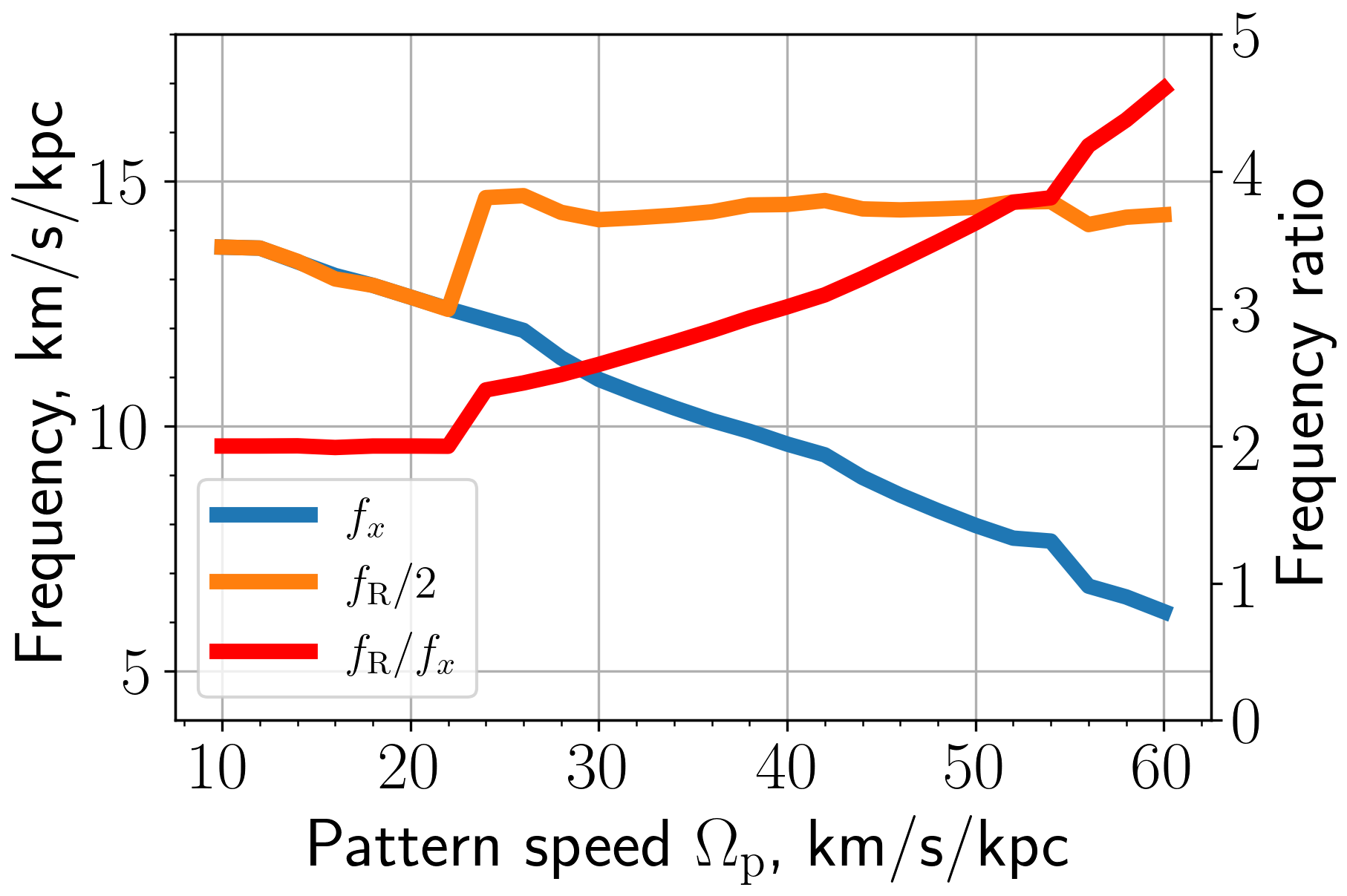} 
	\end{minipage}% 
    \begin{minipage}[t]{0.33\textwidth}
	    %\vspace*{-0.2cm}
		\includegraphics[width=0.99\textwidth]{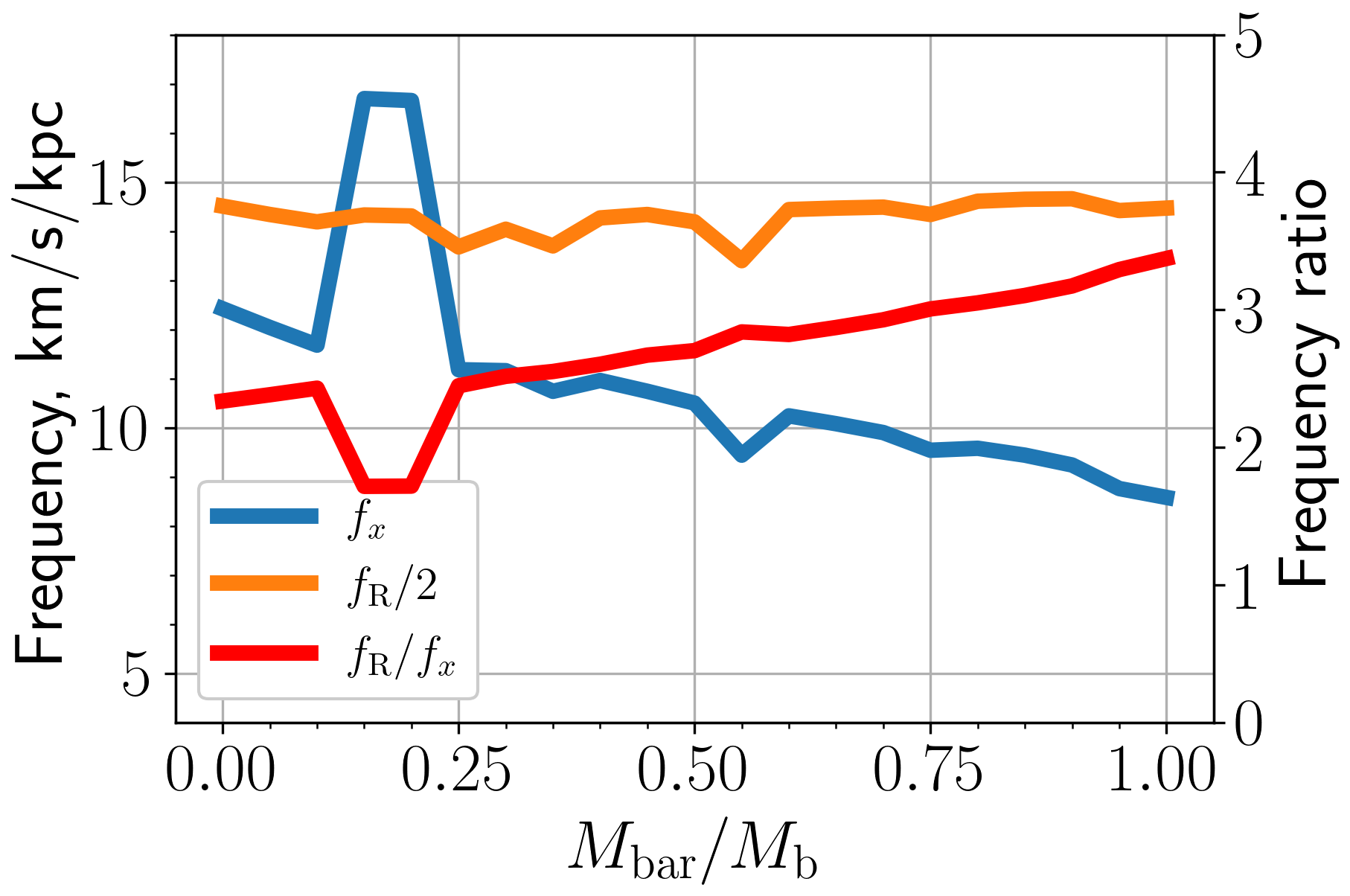} 
	\end{minipage}%
	 \begin{minipage}[t]{0.33\textwidth}
	    %\vspace*{-0.2cm}
		\includegraphics[width=0.99\textwidth]{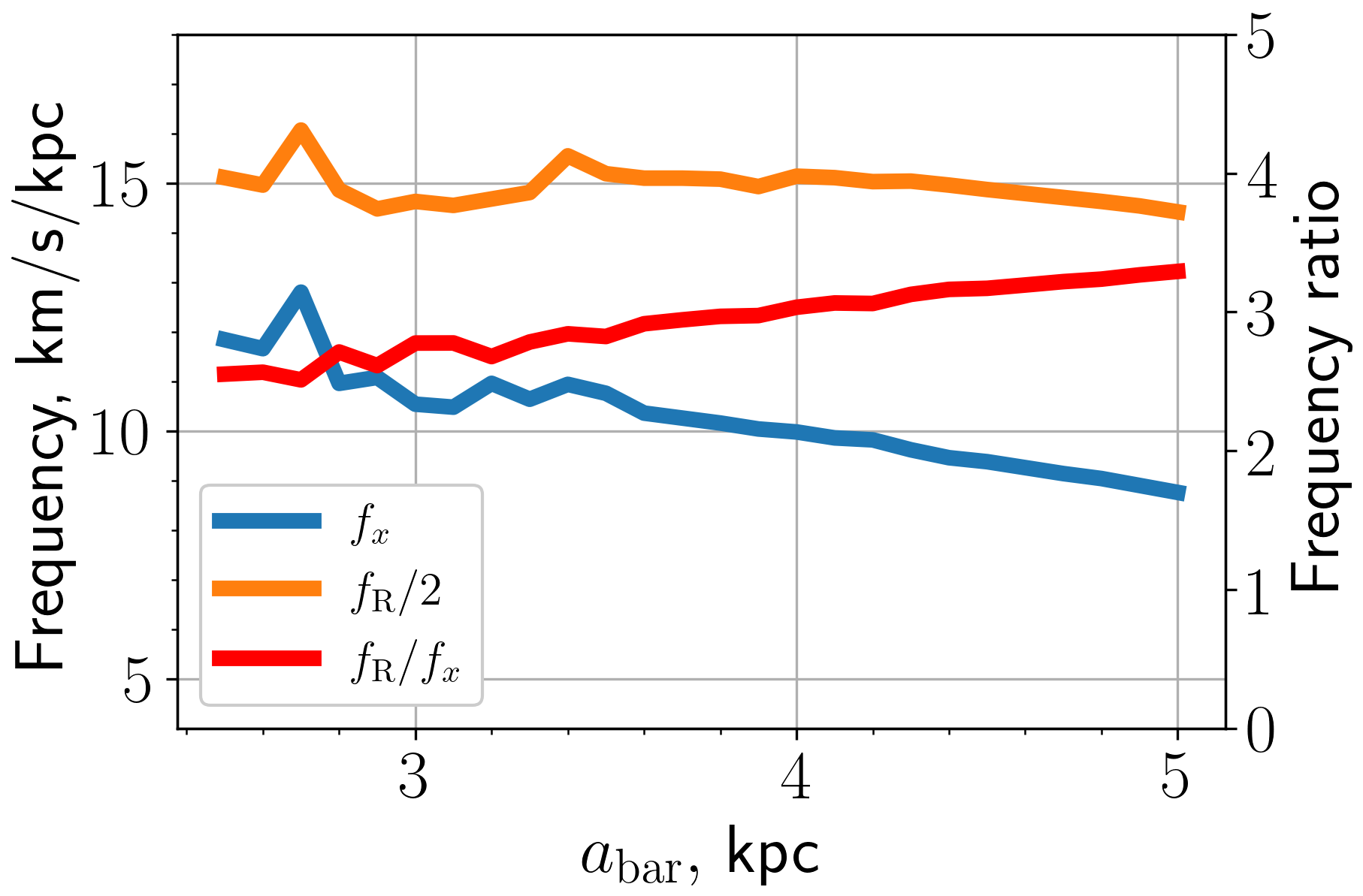} 
	\end{minipage}%
	\caption{Dependence of the orbital frequencies $f_x$, $f_\mathrm{R}$, and the ratio $f_\mathrm{R}/f_x$ for NGC 6266 on the bar parameters for the potential of BB2016.}
    \label{fig:NGC6266_BB}
\end{figure*}

    \begin{figure*}
    \centering
    	\begin{minipage}[t]{0.31\textwidth}
	    %\vspace*{-0.1cm}
		\includegraphics[width=0.85\textwidth]{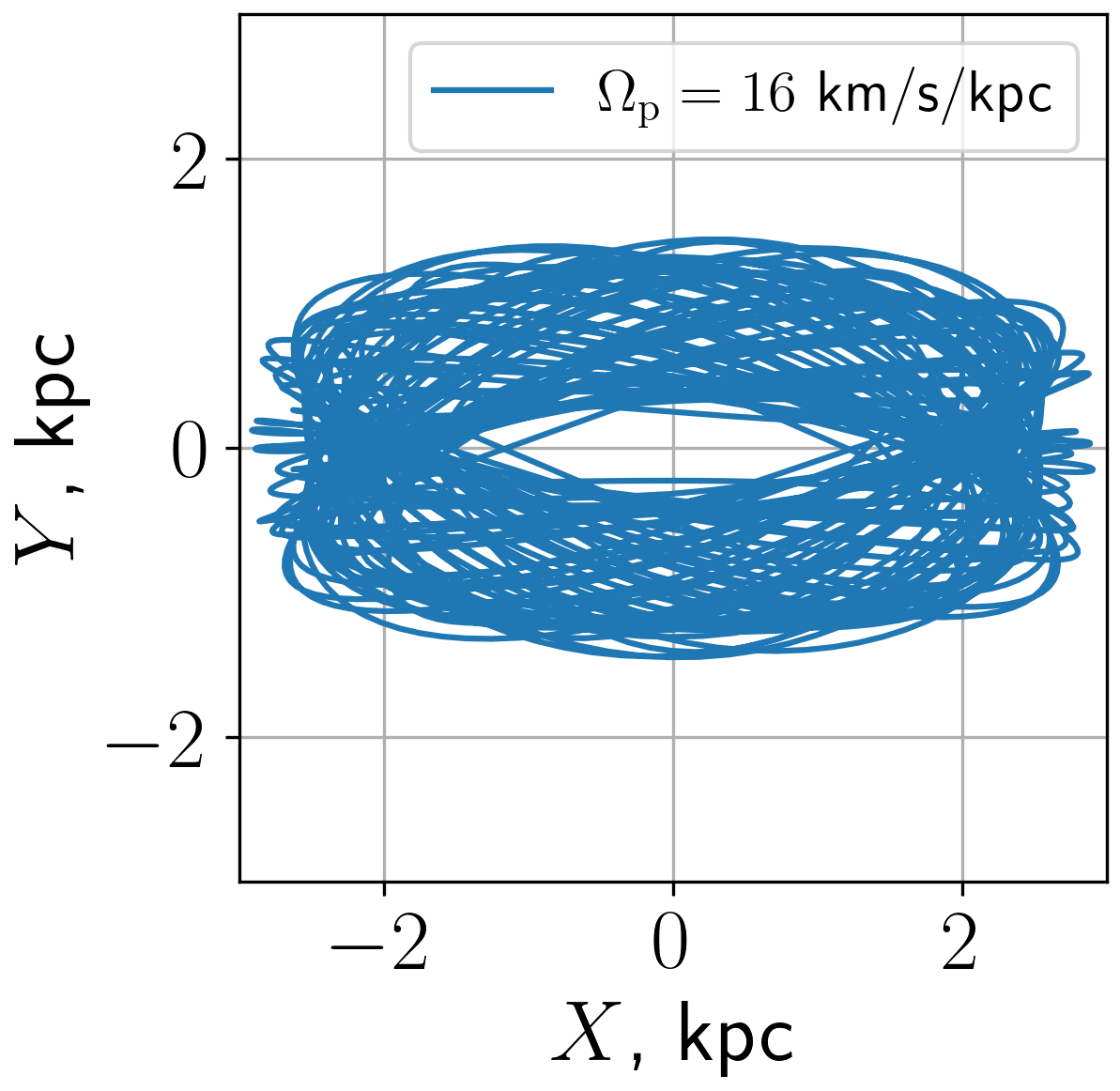} 
	\end{minipage}% 
	    \begin{minipage}[t]{0.31\textwidth}
	    %\vspace*{-0.2cm}
		\includegraphics[width=0.85\textwidth]{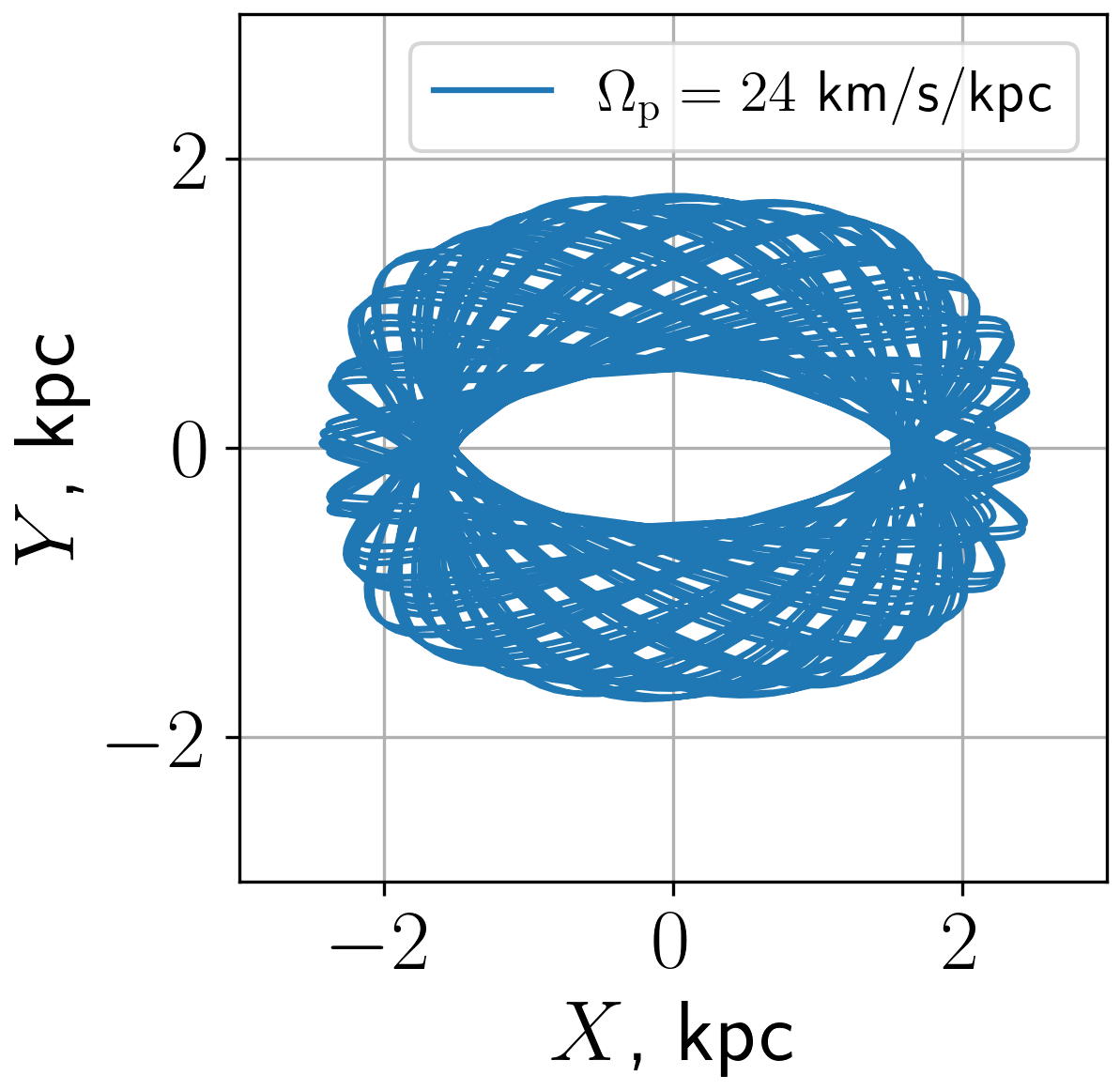}% 
	\end{minipage}%
	\begin{minipage}[t]{0.31\textwidth}
	    %\vspace*{-0.2cm}
		\includegraphics[width=0.85\textwidth]{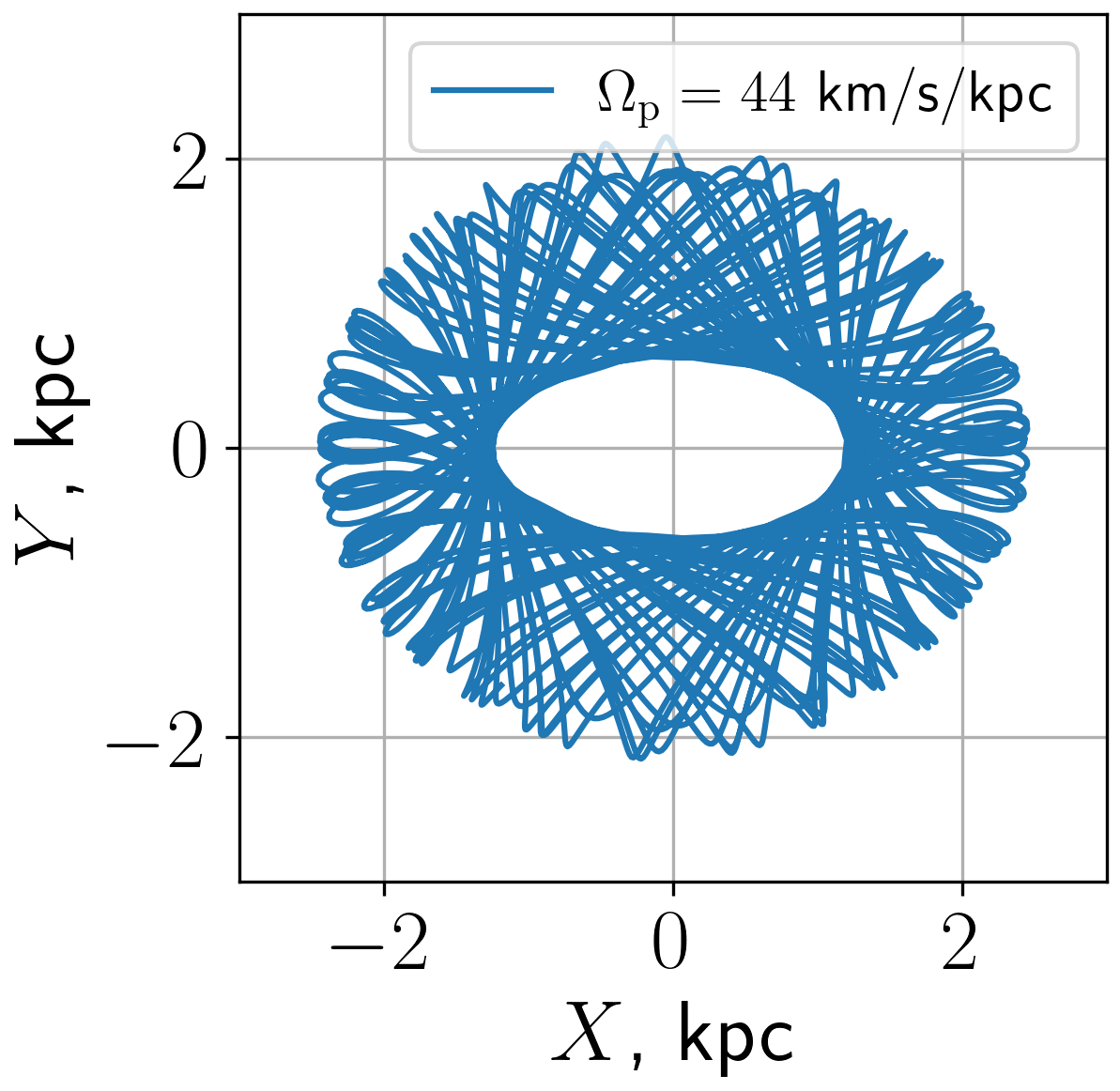}% 
	\end{minipage}% \\
	 \\
    \begin{minipage}[t]{0.31\textwidth}
	    %\vspace*{-0.2cm}
		\includegraphics[width=0.85\textwidth]{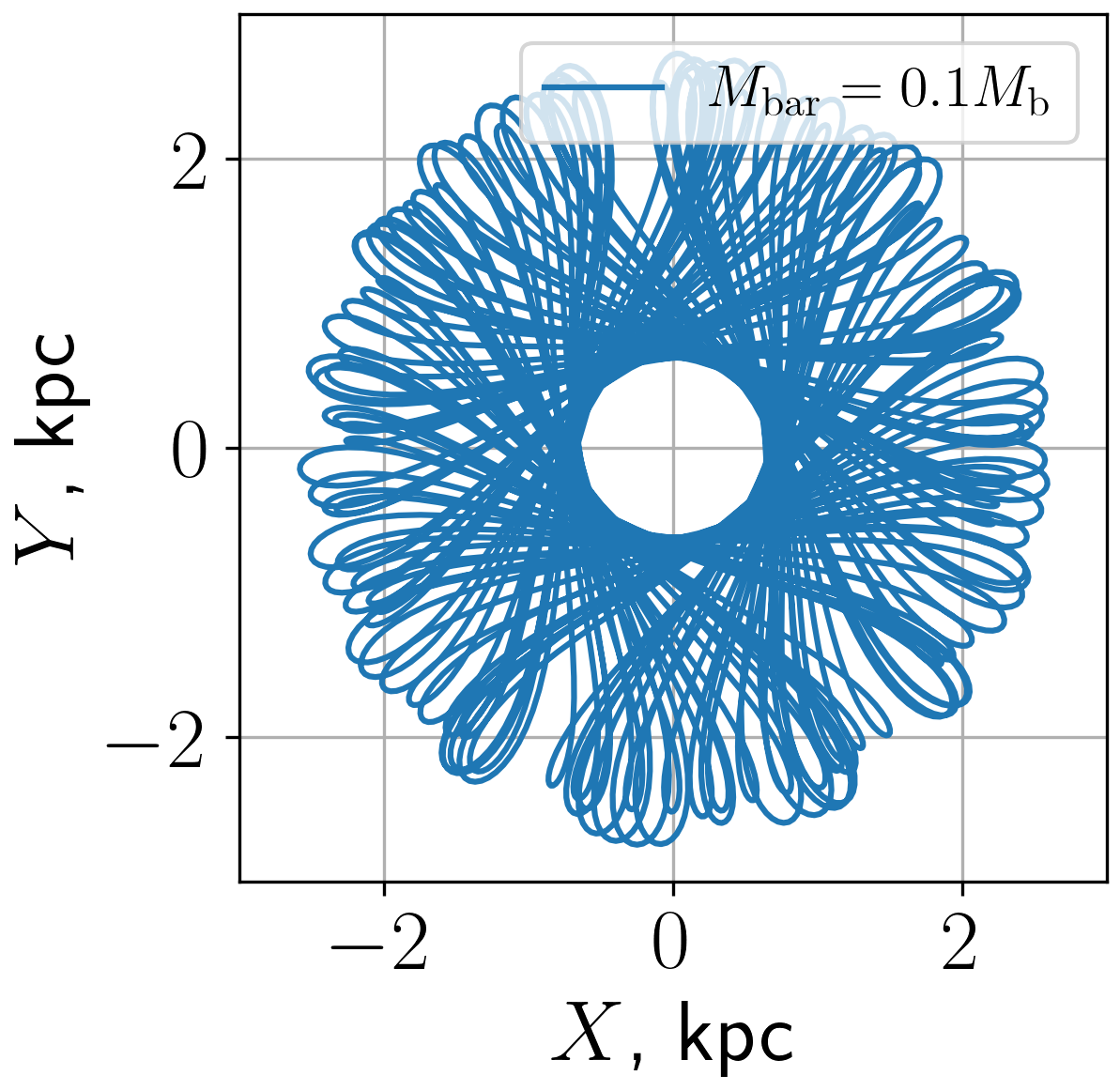} 
	\end{minipage}%
	    	\begin{minipage}[t]{0.31\textwidth}
	    %\vspace*{-0.2cm}
		\includegraphics[width=0.85\textwidth]{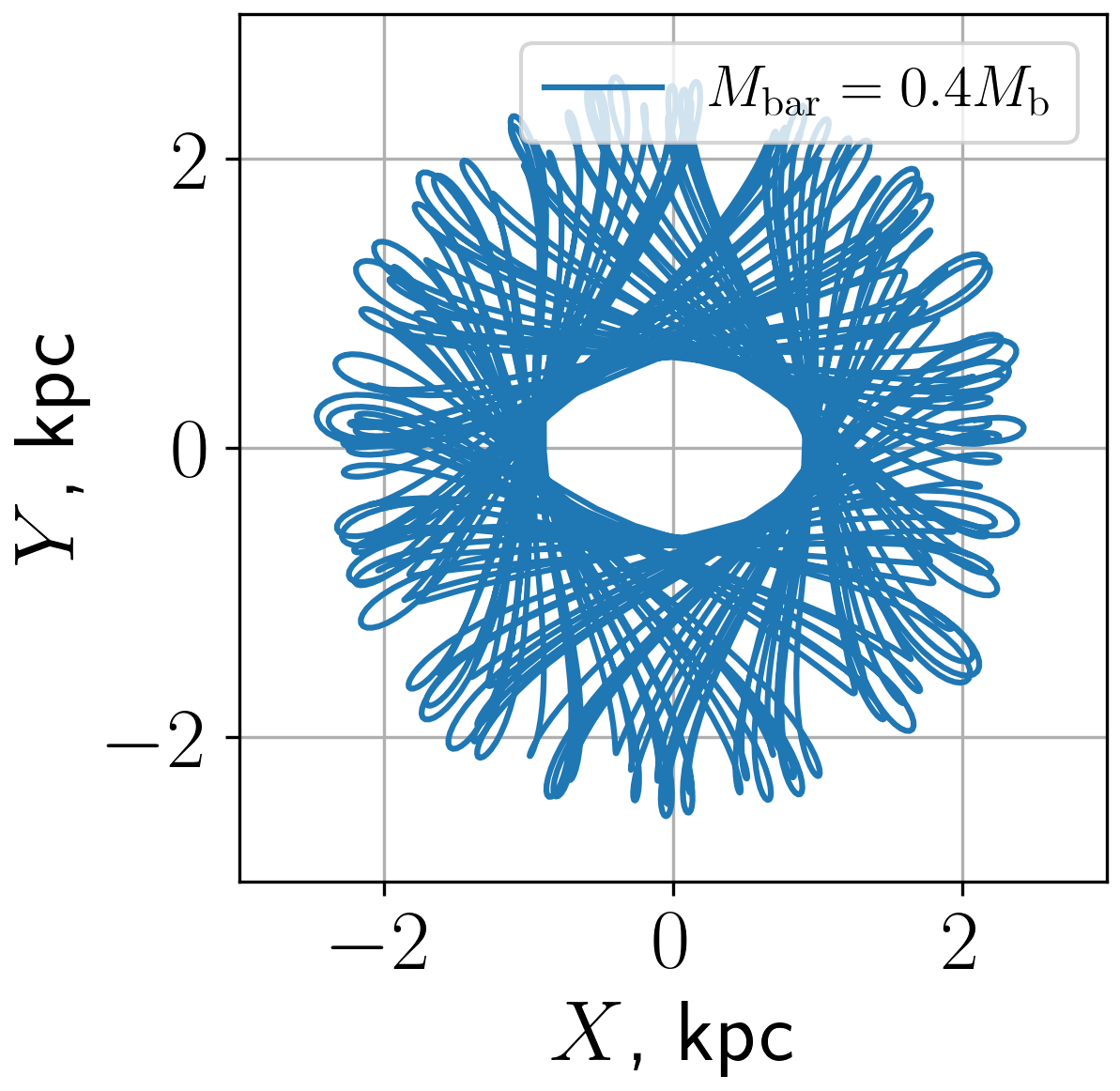}% 
	\end{minipage}%
	\begin{minipage}[t]{0.31\textwidth}
	    %\vspace*{-0.2cm}
		\includegraphics[width=0.85\textwidth]{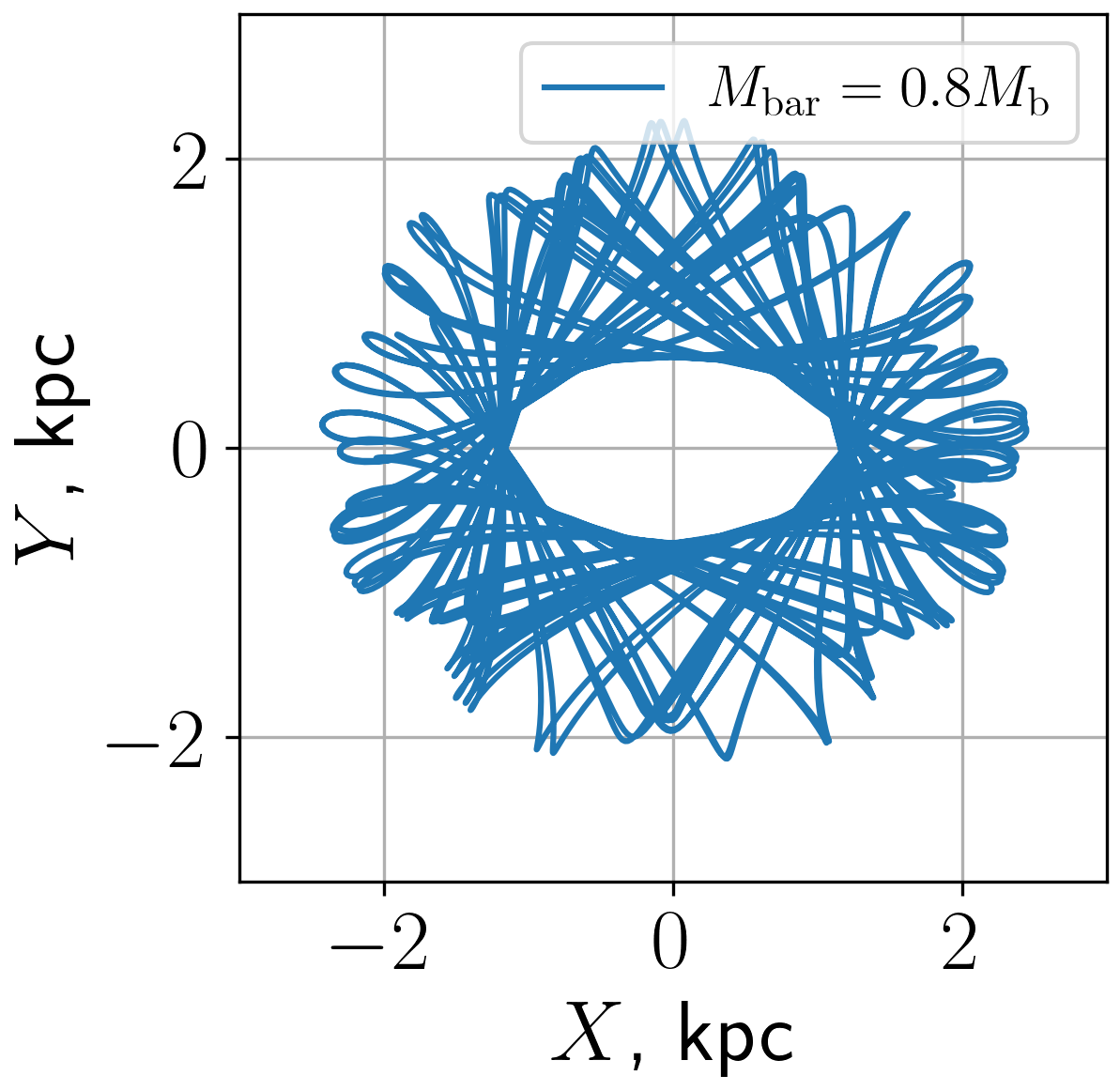}% 
	\end{minipage}%
	\\
	 \begin{minipage}[t]{0.31\textwidth}
	    %\vspace*{-0.2cm}
		\includegraphics[width=0.85\textwidth]{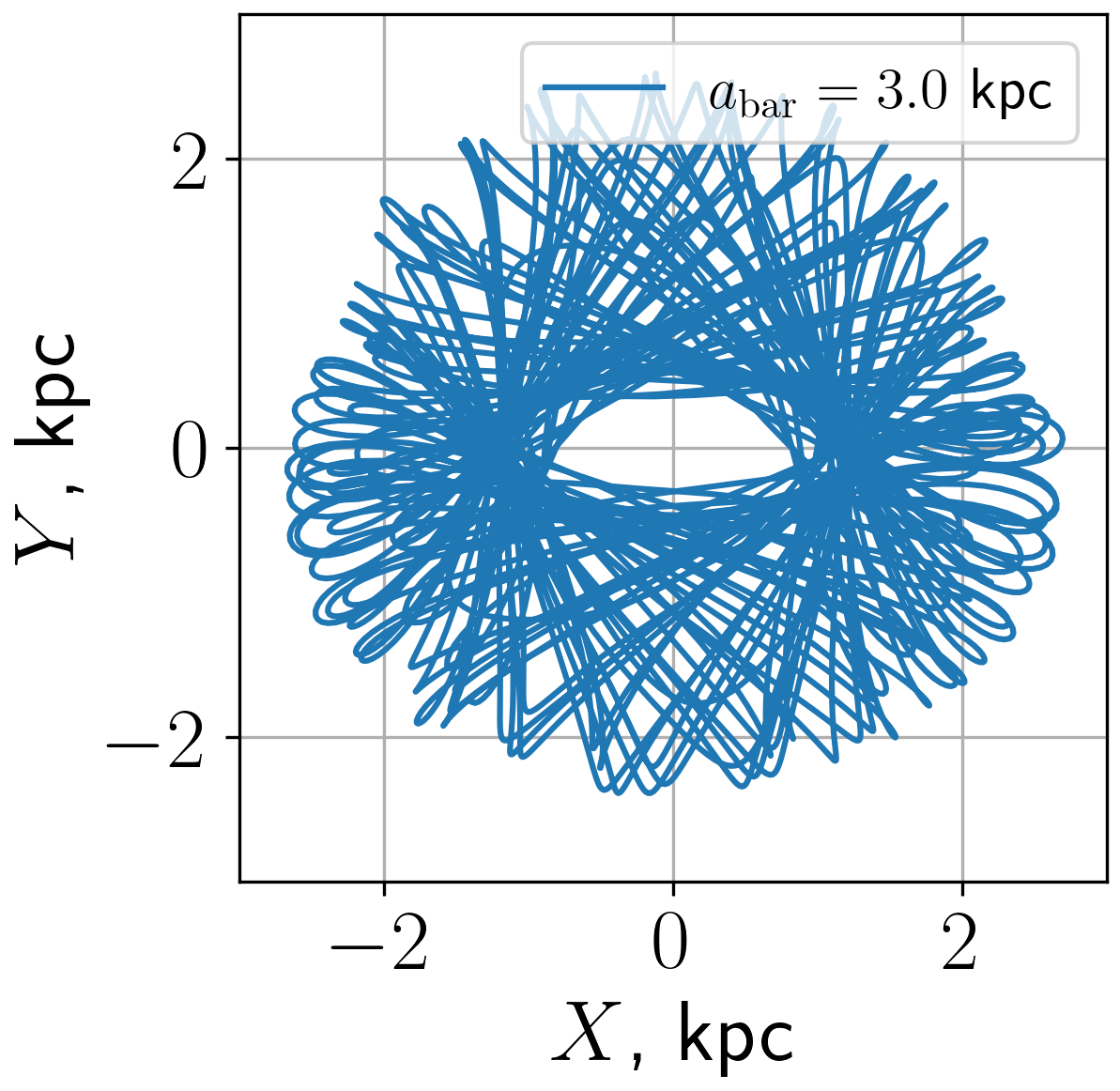} 
	\end{minipage}%
	    	\begin{minipage}[t]{0.31\textwidth}
	    %\vspace*{-0.2cm}
		\includegraphics[width=0.85\textwidth]{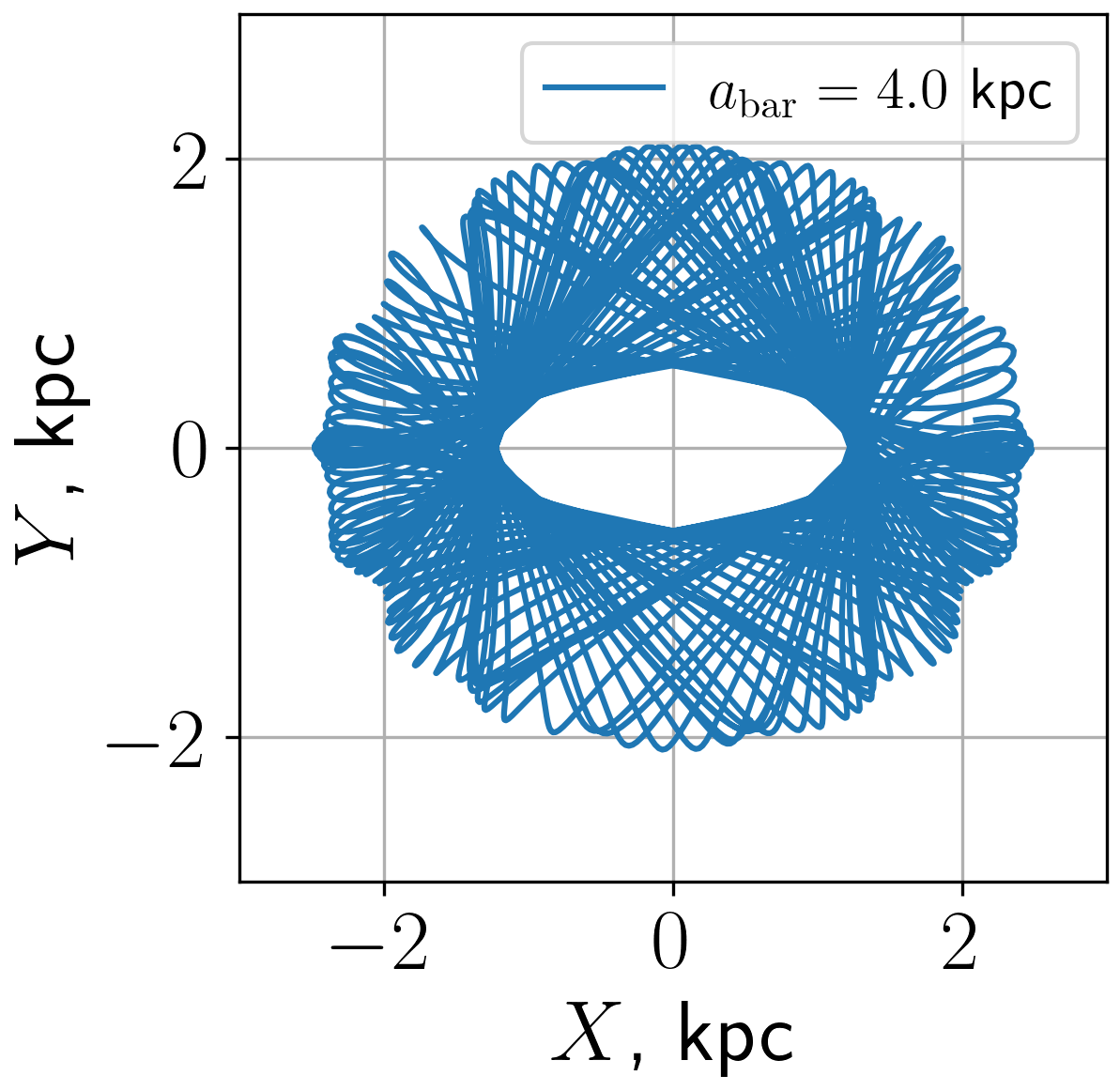}% 
	\end{minipage}%
	\begin{minipage}[t]{0.31\textwidth}
	    %\vspace*{-0.2cm}
		\includegraphics[width=0.85\textwidth]{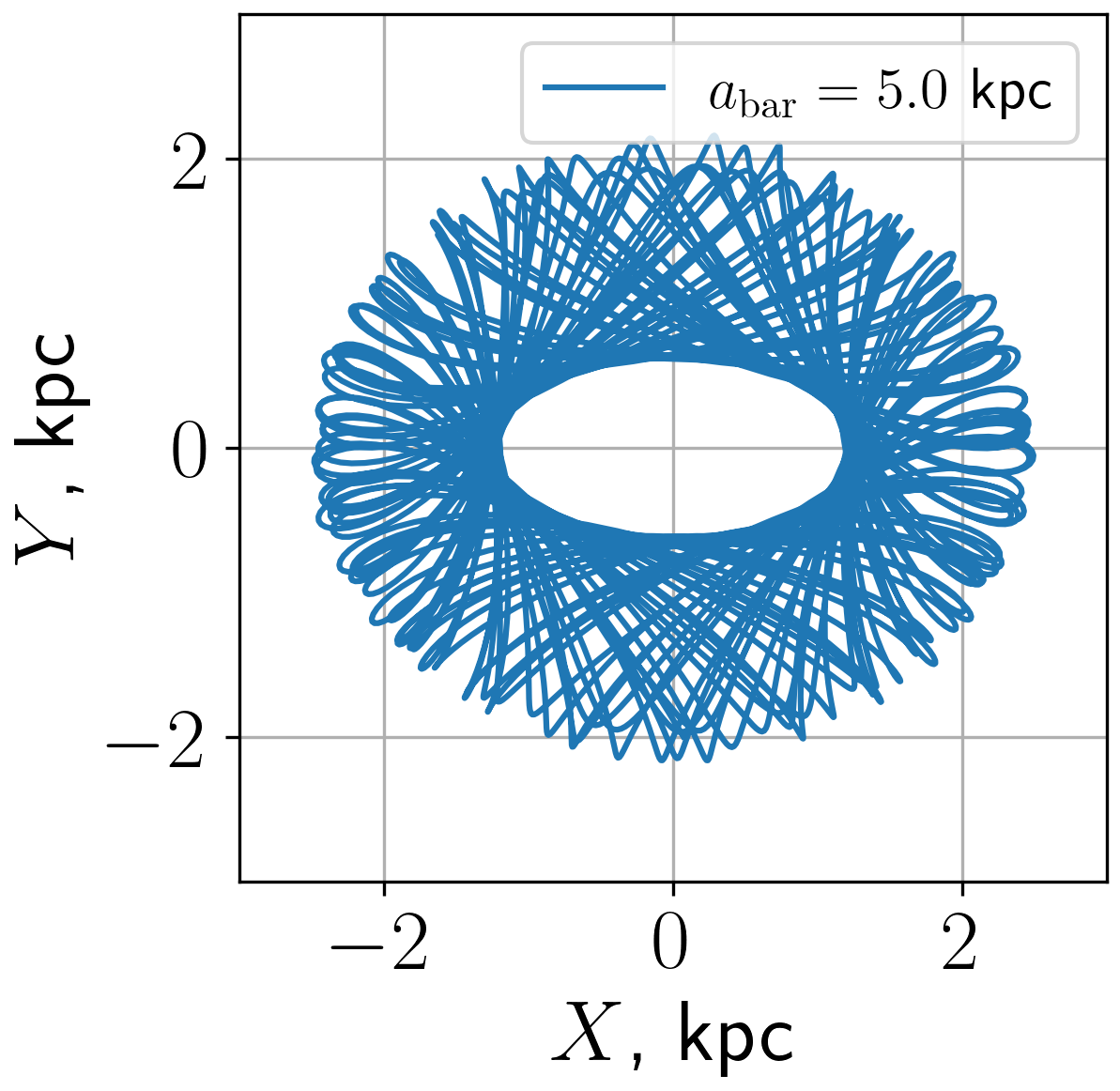}% 
	\end{minipage}%
	\caption{Evolution of the orbital profiles of NGC 6266 depending on the bar parameters for the potential of BB2016. From the first to the third row: dependencies on the pattern speed, \textcolor{black}{bar-to-bulge mass ratio}, and size of the bar, respectively.}
    \label{fig:NGC6266_BB_orbits}
\end{figure*}
First of all, we would like to explore how the choice of bar parameters affects the type of orbit. %Doing so, we will outline some important dependencies concerning the classification of the orbit and the bar pattern speed and mass.
We begin this Section by considering only one GC, namely NGC 6626. There is no particular reason for this choice, except that this example is illustrative. By a detailed analysis of one orbit, we outline the systematics in the classification of orbits that arise due to changes in the bar parameters. 
\par 
Fig.~\ref{fig:NGC6266_BB} shows how frequencies $f_x$ and $f_\mathrm{R}$ and their ratio $f_\mathrm{R}/f_x$ change with the bar pattern speed, mass, and size for the potential of BB2016. To study dependencies, we first consider the one-dimensional case, where one parameter changes, while the rest are fixed. Unless  otherwise specified, all bar parameters are fixed at the following values: $\Omega_\mathrm{p}=45$ km/s/kpc $M_\mathrm{bar}/M_\mathrm{b}=0.95$, $a_\mathrm{bar}=5.0$ kpc, $p=2.0$, and $q=3.0$. We present orbital profiles in Fig.~\ref{fig:NGC6266_BB_orbits} to illustrate how they change when the corresponding parameter is changed. As can be seen from the individual subpanels, there are clear systematic shifts in frequencies and, accordingly, frequency ratios:
\begin{enumerate}
    \item With an increase of the pattern speed, frequency of radial oscillations decreases. This continues up to the point at about $24$ km/s/kpc, then the frequency increases  abruptly, after which it remains constant.
    \item Frequency $f_x$ decreases monotonically with an increase in the pattern speed.
    \item The frequency ratio $f_\mathrm{R}/f_x$ shows an interesting behaviour as a result of changes in individual frequencies. Initially, $f_\mathrm{R}/f_x=2$ (a typical ratio for orbits following the bar), but at $\Omega_\mathrm{p}\approx24$ km/s/kpc and after that, it deviates more and more from this value.
\end{enumerate}
The described changes of frequencies are reflected in the orbit profile. In the case of $f_\mathrm{R}/f_x\approx2$, one observes a very regular orbit captured by the bar. For $f_\mathrm{R}/f_x\gtrsim2$, the orbit becomes more ``windy'' and now oscillates around the bar.
\par 
For the \textcolor{black}{bar-to-bulge mass ratio} and size (\textit{second} and \textit{third} rows of Fig.~\ref{fig:NGC6266_BB}), one can see that changing these parameters affects the orbit profile and the corresponding frequency ratios, but their influence is not so strong compared to the pattern speed. An increase in the bar mass and size leads to a slight decrease in $f_x$, which leads to small changes in the frequency ratios, from $f_\mathrm{R}/f_x\approx2.2-2.4$ at the left boundary of the interval to about $f_\mathrm{R}/f_x\approx 3.0$ on the right.
\par
\begin{figure*}
    \centering
    	\begin{minipage}[t]{0.33\textwidth}
	    %\vspace*{-0.2cm}
		\includegraphics[width=0.9\textwidth]{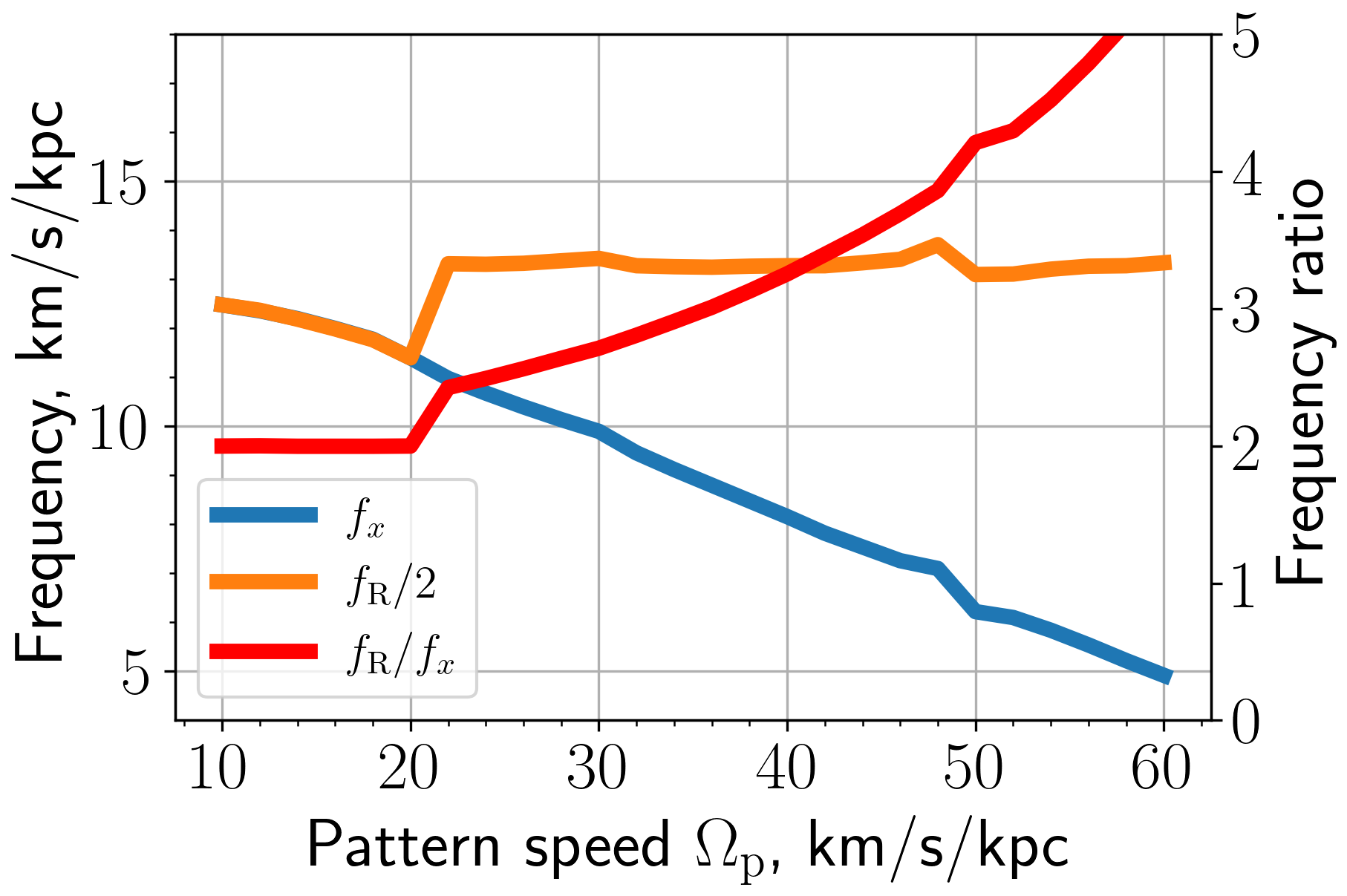} 
	\end{minipage}% 
	    	\begin{minipage}[t]{0.33\textwidth}
	    %\vspace*{-0.2cm}
		\includegraphics[width=0.9\textwidth]{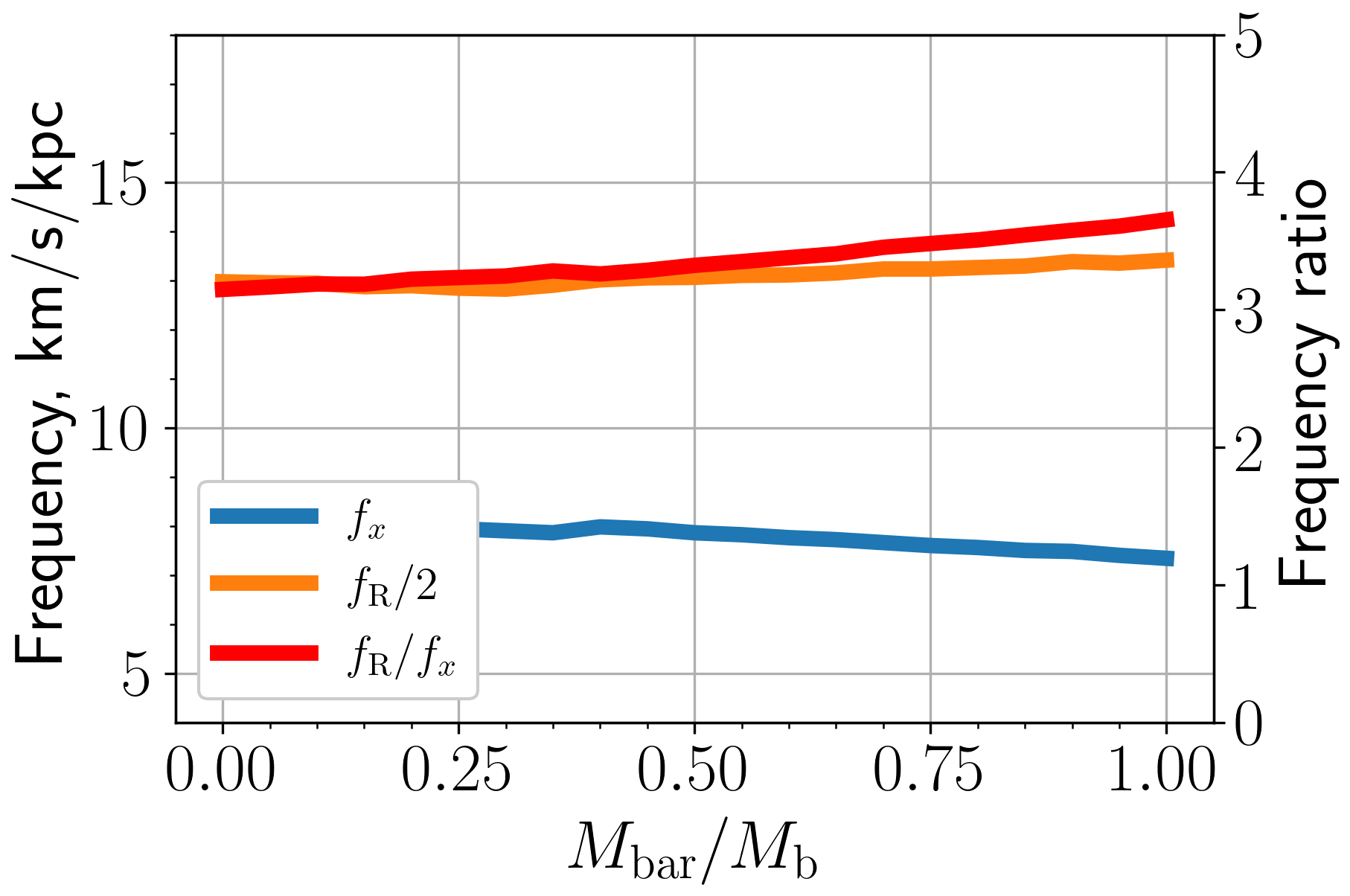}% 
	\end{minipage}%
	\begin{minipage}[t]{0.33\textwidth}
	    %\vspace*{-0.2cm}
		\includegraphics[width=0.9\textwidth]{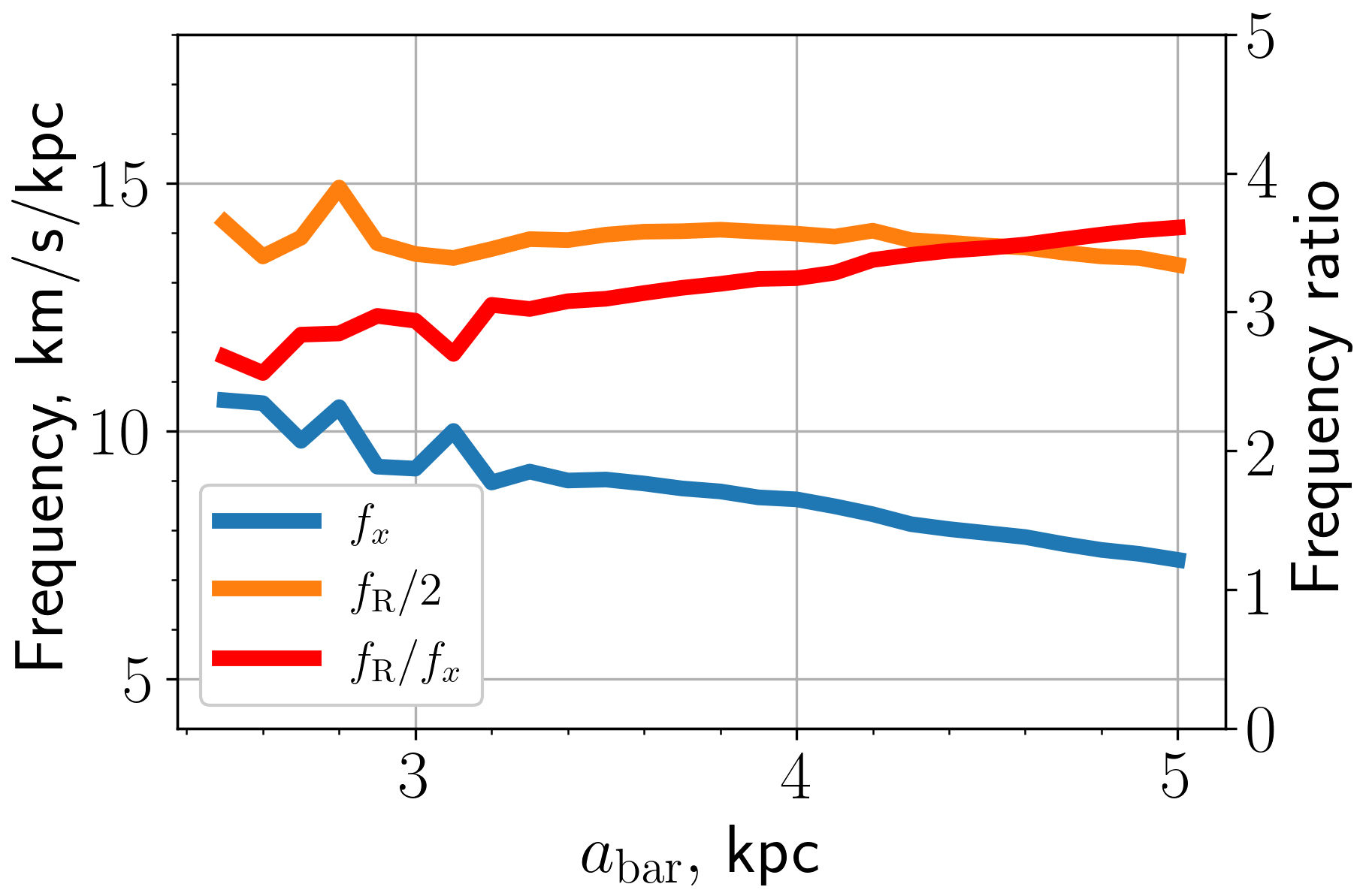}% 
	\end{minipage}% \\
	\caption{Dependence of the orbital frequencies $f_x$, $f_\mathrm{R}$, and the ratio $f_\mathrm{R}/f_x$ on the bar parameters for NGC 6266 in potential of MC2017.}
    \label{fig:NGC6266_MC}
\end{figure*}

 
Comparing the results for BB2016 (Fig.~\ref{fig:NGC6266_BB}) and MC2017 (Fig.~\ref{fig:NGC6266_MC}), one can see that the trends in changes of frequencies between them are similar, i.e. there is a sudden change in the frequency ratio at a particular value of the bar pattern speed. For the MC2017 potential, this change occurs at a somewhat smaller value of $\Omega\approx20$ km/s/kpc. In the case of MC2017, changing the \textcolor{black}{bar-to-bulge mass ratio} has almost no effect on the frequency ratio. This can be explained by the fact that the bulge in the model of MC2017 already has a certain degree of flatness (along the vertical direction) and its transformation into an elongated component does not significantly affect the potential.
\par
\begin{figure*}
    \centering
	    %\vspace*{-0.2cm}
		\includegraphics[width=0.85\textwidth]{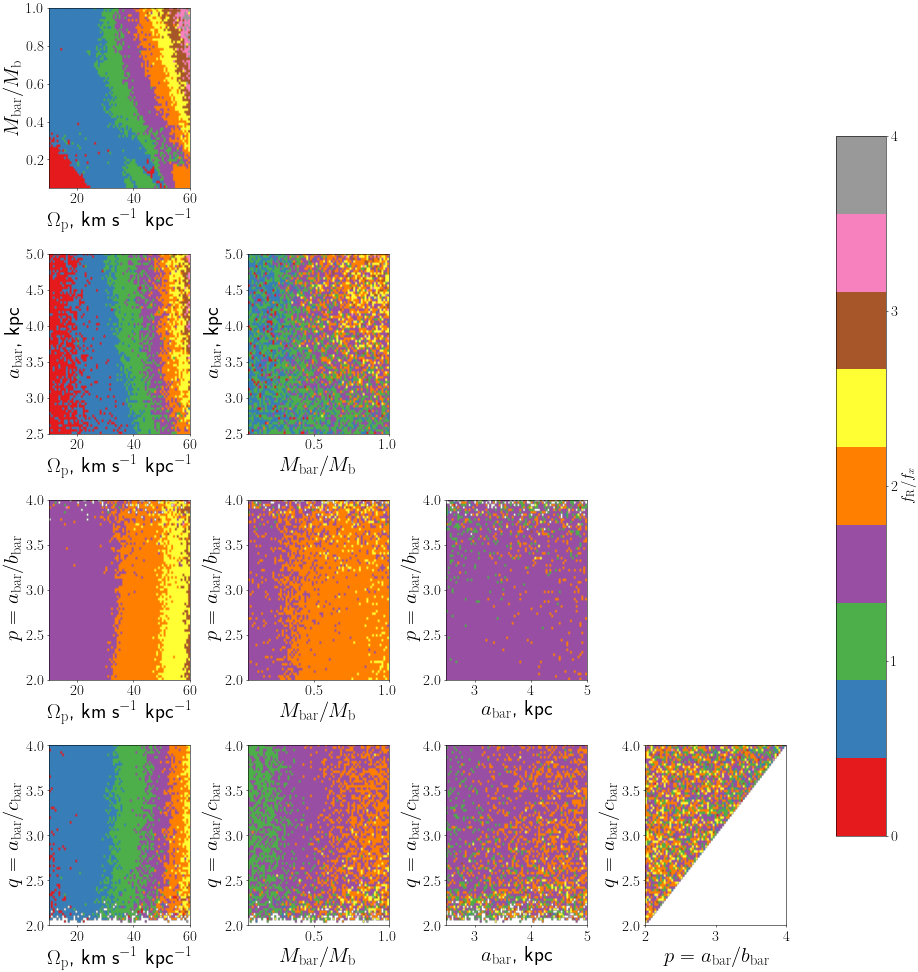} 
    \caption{Colour coded frequency ratios $f_\mathrm{R}/f_x$ for NGC 6266 depending on the bar parameters (pattern speed $\Omega_\mathrm{p}$, \textcolor{black}{bar-to-bulge} mass ratio $M_\mathrm{bar}/M_\mathrm{b}$, size $a_\mathrm{bar}$, and  the ratios of the axes in the disc plane $p$ and vertical direction $q$) for the potential of BB2016.}
    \label{fig:MC_bar}
\end{figure*}

In Fig.~\ref{fig:NGC6266_BB} and Fig.~\ref{fig:NGC6266_MC}, we fixed all bar parameters, except for one, which then varied. However, doing so, we did not take into account the possibility that, with a different combination of bar parameters, the observed dependencies may well change or simply disappear. To explore such a behaviour in more detail, we conduct a following suit of simulations. We run Monte-Carlo simulations, choosing a set of bar parameters from the intervals specified in Table~\ref{tab:models1_pars} uniformly, then we calculate the orbit and the corresponding ratio of its frequencies. We performed $10^5$ of such iterations. Fig.~\ref{fig:MC_bar} show the results in a form of a matrix plot for all parameters, with the   average value of frequency ratio for a given pixel highlighted in different colours. \textcolor{black}{Each subplot presents a 2D histogram obtained by averaging the values within 100 bins from minimum to maximum values for each axis.} The subplots show qualitatively similar results compared to those presented in the 1D plots (Fig.~\ref{fig:NGC6266_BB} and Fig.~\ref{fig:NGC6266_MC}). Again, the pattern speed is the most important  parameter, i.~e. in each subpanel in the first column there is a gradual progression of colours. For other parameters, there is no such correlations, except for a weak correlation of frequency rations with $M_\mathrm{bar}$. Thus, changing all other parameters does not strongly affect the frequency ratio. This means that the pattern speed may be very well the most important factor when one is trying to asses orbit families and check whether a particular orbit follows a bar or not.
\\
\par
\begin{figure*}
    \centering
    \begin{minipage}[t]{0.49\textwidth}
	    %\vspace*{-0.2cm}
		\includegraphics[width=0.85\textwidth]{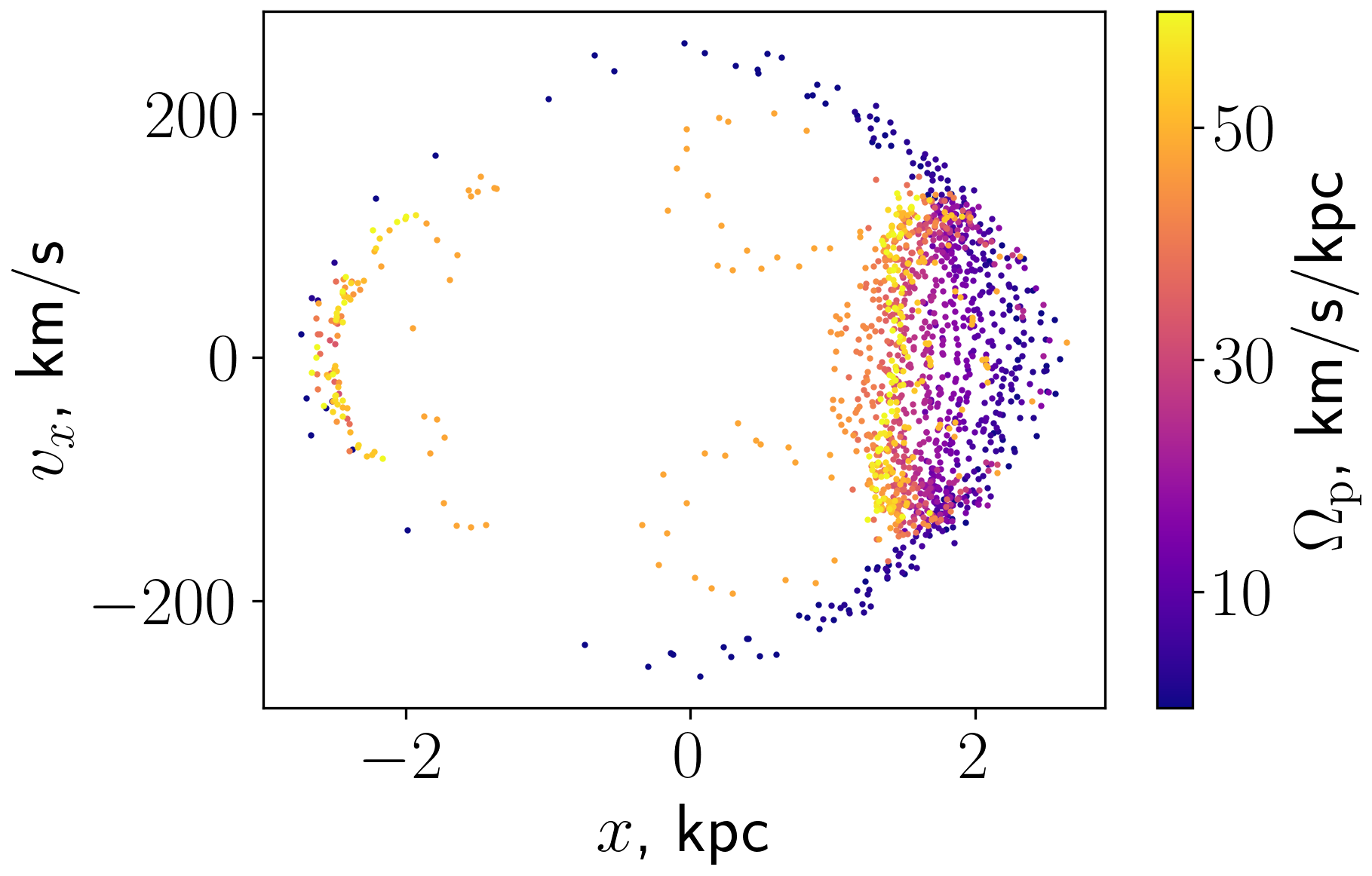}% 
	\end{minipage}%
     \begin{minipage}[t]{0.49\textwidth}
	    %\vspace*{-0.2cm}
		\includegraphics[width=0.85\textwidth]{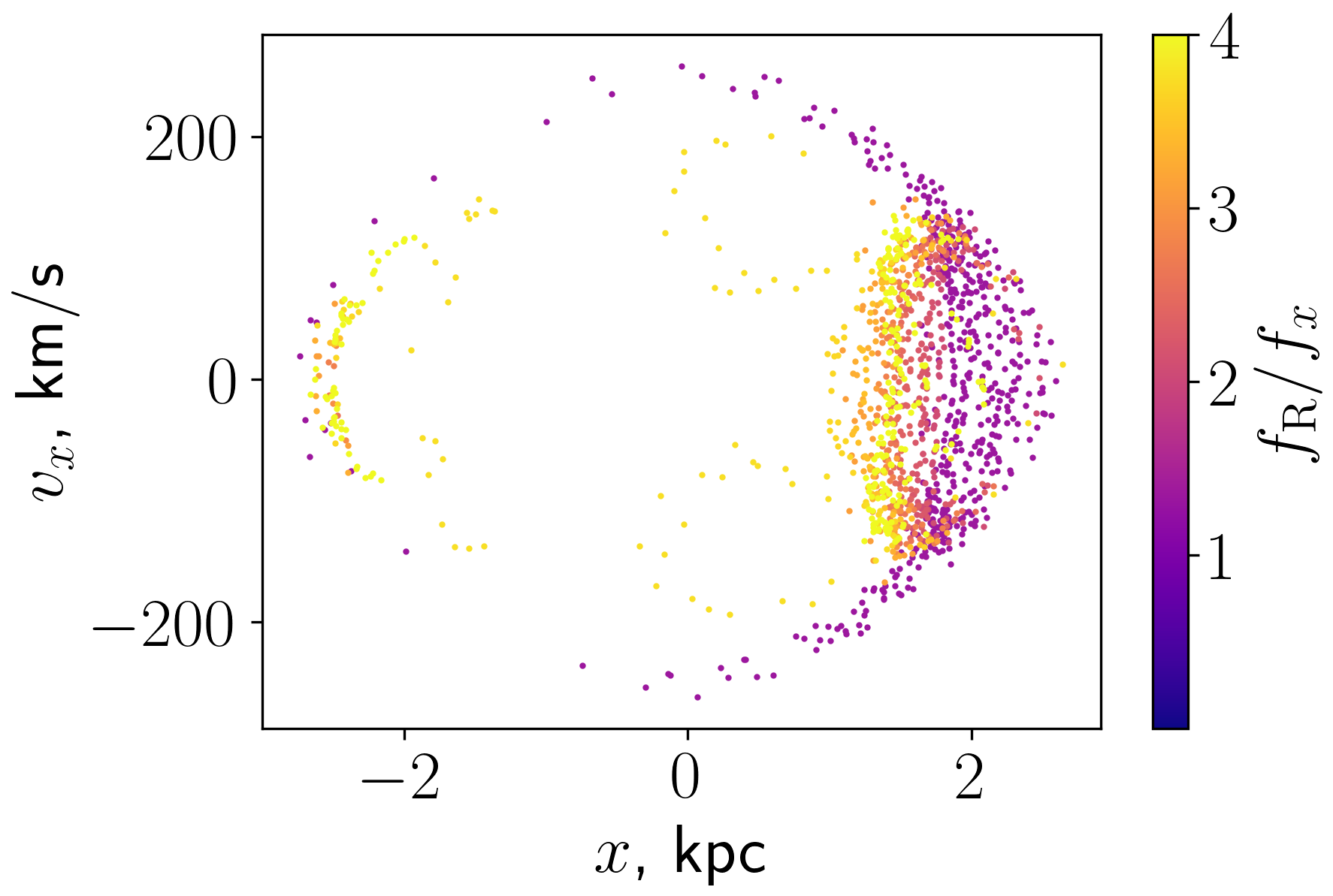}% 
	\end{minipage}%

    \caption{Surface of sections for the orbit of NGC 6266 with a colorbar indicating the values of the pattern speed $\Omega_\mathrm{p}$ (\textit{left}) and the frequency ratio $f_\mathrm{R}/f_x$ (\textit{right}).}
    \label{fig:SoS}
\end{figure*}

 
To understand why frequencies abruptly change with the pattern speed, we calculated the Poincare surface of sections (SoSs) for the range of pattern speeds. %(the rest of the bar parameters are fixed at $M_\mathrm{bar}=0.95 M_\mathrm{b}, a_\mathrm{bar}=5$ kpc, $a/b=2$, $a/c=3$). 
\textcolor{black}{For 3D orbits, SoSs are four dimensional objects, i.e. $(x,z,V_x, V_z)$ taken at $y=0$ and $V_y>0$ (or any other similar combinations).  Here, we plot SoS projections on $(x, V_x)$ plane taken at $y=0$ and $V_y>0$. A similar approach was used in~\cite{2004A&A...428..905K,2007MNRAS.381..757V}, where 3D $N$-body orbits were studied.}
Fig.~\ref{fig:SoS} demonstrate how the SoSs change either with $\Omega_\mathrm{p}$ or with the corresponding frequency ratio $f_\mathrm{R}/f_x$. Note that the SoSs presented are not exactly typical.  Usually, the Jacobi energy is a fixed variable and one investigates various orbits for a chosen energy value. In Fig.~\ref{fig:SoS}, the pattern speed (and, thus, the corresponding Jacobi energy) changes from orbit to orbit, not the initial velocity or position. Nevertheless, as can be seen from the figures, the family to  which the orbit belongs gradually changes with the pattern speed. The orbit starts on the island close to $x1$ family (they reside in rightmost corner of the plot), then gradually expands to the left side of the diagram. At some point (after $\Omega_\mathrm{p}\gtrsim30$), new islands appear. The orbit clearly ceases to be a member of the $x1$ family, as indicated by its increasing frequency ratio $f_\mathrm{R}/f_x$. As for the question of which family an orbit ends up, this question is not easy to answer, since a bar can be populated by orbits with multiplicity greater than 2:1, see a recent work by~\cite{Wang2022}. From frequency ratios, it follows that the orbit considered here gradually changes its multiplicity with an increase of the pattern speed, becoming a 3:1 orbit, then a 4:1 orbit, and so on. %to the left leaving the associated island, and ends up as a typical $x3$ and $x4$ orbit. Such orbits are usually associated with the co-rotation resonance~\citep{Contopoulos_Papayannopoulos1980}. Add something...

%Thus, we observe here  inner Lindbland resonance, where $f_\mathrm{R}/f_x=2$

\section{Frequency ratios for the sample of globular clusters}
\label{sec:all_orbits}

\begin{table*}
\begin{tabular}{ c | c | c | c | c | c | c |}
\hline
%ID & \Omega_\mathrm{p}$$ & $V_\mathrm{R}$ & $V_\mathrm{T}$ & $z$ & $V_z$& $\phi$\\
ID & $f_\mathrm{R}/f_x ({\Omega=30})$& $f_\mathrm{R}/f_x ({\Omega=45})$& $f_\mathrm{R}/f_x ({\Omega=60})$& $f_\mathrm{R}/f_x^{*} ({\Omega=30})$& $f_\mathrm{R}/f_x^{*}  ({\Omega=45})$& $f_\mathrm{R}/f_x^{*}  ({\Omega=60})$\\
\hline 
\hline
BH 229 &2.2${\displaystyle \pm}0.6$ &3.2${\displaystyle \pm}0.8$ &3.9${\displaystyle \pm}1.3$ &2.6${\displaystyle \pm}0.6$ &3.5${\displaystyle \pm}0.8$ &5.0${\displaystyle \pm}1.1$ \\ 
ESO 452-11 &2.6${\displaystyle \pm}0.1$ &3.5${\displaystyle \pm}0.6$ &4.5${\displaystyle \pm}0.4$ &2.8${\displaystyle \pm}0.2$ &3.9${\displaystyle \pm}0.6$ &5.4${\displaystyle \pm}0.4$ \\ 
Liller 1 &1.5${\displaystyle \pm}0.2$ &1.4${\displaystyle \pm}0.1$ &1.3${\displaystyle \pm}0.1$ &1.6${\displaystyle \pm}0.2$ &1.4${\displaystyle \pm}0.1$ &1.3${\displaystyle \pm}0.1$ \\ 
NGC 6144 &3.2${\displaystyle \pm}0.4$ &4.9${\displaystyle \pm}1.3$ &11.2${\displaystyle \pm}3.7$ &3.6${\displaystyle \pm}0.5$ &9.2${\displaystyle \pm}1.5$ &4.9${\displaystyle \pm}13.8$ \\ 
NGC 6266 &2.1${\displaystyle \pm}0.2$ &3.1${\displaystyle \pm}0.3$ &4.5${\displaystyle \pm}0.4$ &2.2${\displaystyle \pm}0.3$ &3.5${\displaystyle \pm}0.2$ &5.2${\displaystyle \pm}0.7$ \\ 
NGC 6273 &3.4${\displaystyle \pm}0.7$ &6.0${\displaystyle \pm}1.7$ &17.9${\displaystyle \pm}18.6$ &4.4${\displaystyle \pm}0.6$ &17.6${\displaystyle \pm}8.1$ &11.3${\displaystyle \pm}4.5$ \\ 
NGC 6293 &2.7${\displaystyle \pm}0.2$ &3.9${\displaystyle \pm}0.7$ &6.1${\displaystyle \pm}1.4$ &3.0${\displaystyle \pm}0.3$ &4.6${\displaystyle \pm}1.0$ &9.6${\displaystyle \pm}2.4$ \\ 
NGC 6342 &1.8${\displaystyle \pm}0.7$ &2.4${\displaystyle \pm}1.4$ &3.6${\displaystyle \pm}2.4$ &2.2${\displaystyle \pm}0.9$ &1.9${\displaystyle \pm}1.3$ &3.5${\displaystyle \pm}2.8$ \\ 
NGC 6355 &2.7${\displaystyle \pm}0.2$ &3.7${\displaystyle \pm}0.5$ &5.6${\displaystyle \pm}1.4$ &3.0${\displaystyle \pm}0.4$ &4.5${\displaystyle \pm}0.9$ &10.2${\displaystyle \pm}2.1$ \\ 
NGC 6380 &2.7${\displaystyle \pm}0.1$ &3.2${\displaystyle \pm}0.1$ &4.3${\displaystyle \pm}0.3$ &2.8${\displaystyle \pm}0.1$ &3.5${\displaystyle \pm}0.2$ &4.9${\displaystyle \pm}0.4$ \\ 
NGC 6401 &2.7${\displaystyle \pm}0.1$ &3.5${\displaystyle \pm}0.2$ &5.3${\displaystyle \pm}0.6$ &2.9${\displaystyle \pm}0.1$ &4.3${\displaystyle \pm}0.2$ &9.0${\displaystyle \pm}2.1$ \\ 
NGC 6440 &2.5${\displaystyle \pm}0.1$ &2.8${\displaystyle \pm}0.1$ &3.4${\displaystyle \pm}0.1$ &2.6${\displaystyle \pm}0.1$ &3.1${\displaystyle \pm}0.1$ &3.9${\displaystyle \pm}0.4$ \\ 
NGC 6453 &2.8${\displaystyle \pm}0.4$ &3.6${\displaystyle \pm}0.7$ &5.2${\displaystyle \pm}1.6$ &3.0${\displaystyle \pm}0.2$ &4.1${\displaystyle \pm}0.7$ &7.2${\displaystyle \pm}3.2$ \\ 
NGC 6522 &2.4${\displaystyle \pm}0.3$ &2.4${\displaystyle \pm}0.5$ &3.1${\displaystyle \pm}0.9$ &2.2${\displaystyle \pm}0.3$ &2.4${\displaystyle \pm}0.6$ &3.8${\displaystyle \pm}1.0$ \\ 
NGC 6528 &2.7${\displaystyle \pm}0.2$ &3.3${\displaystyle \pm}0.7$ &4.3${\displaystyle \pm}0.8$ &3.0${\displaystyle \pm}0.1$ &4.0${\displaystyle \pm}0.3$ &5.1${\displaystyle \pm}1.5$ \\ 
NGC 6558 &2.6${\displaystyle \pm}0.1$ &3.2${\displaystyle \pm}0.1$ &4.3${\displaystyle \pm}0.5$ &2.7${\displaystyle \pm}0.1$ &3.7${\displaystyle \pm}0.2$ &5.6${\displaystyle \pm}0.7$ \\ 
NGC 6624 &2.3${\displaystyle \pm}0.5$ &2.7${\displaystyle \pm}0.7$ &3.7${\displaystyle \pm}0.7$ &2.7${\displaystyle \pm}0.4$ &2.7${\displaystyle \pm}1.0$ &4.3${\displaystyle \pm}0.6$ \\ 
NGC 6626 &2.9${\displaystyle \pm}0.1$ &4.1${\displaystyle \pm}0.5$ &6.4${\displaystyle \pm}1.0$ &3.0${\displaystyle \pm}0.1$ &4.4${\displaystyle \pm}0.4$ &8.3${\displaystyle \pm}1.9$ \\ 
NGC 6638 &2.7${\displaystyle \pm}0.2$ &3.3${\displaystyle \pm}0.2$ &4.5${\displaystyle \pm}0.8$ &2.8${\displaystyle \pm}0.1$ &3.6${\displaystyle \pm}0.2$ &5.4${\displaystyle \pm}0.6$ \\ 
NGC 6637 &1.5${\displaystyle \pm}0.4$ &1.6${\displaystyle \pm}0.6$ &1.5${\displaystyle \pm}0.8$ &1.5${\displaystyle \pm}0.4$ &1.6${\displaystyle \pm}0.8$ &1.4${\displaystyle \pm}0.7$ \\ 
NGC 6642 &2.5${\displaystyle \pm}0.1$ &3.0${\displaystyle \pm}0.2$ &3.7${\displaystyle \pm}0.7$ &2.6${\displaystyle \pm}0.2$ &3.1${\displaystyle \pm}0.5$ &3.5${\displaystyle \pm}1.3$ \\ 
NGC 6717 &2.8${\displaystyle \pm}0.4$ &3.7${\displaystyle \pm}0.4$ &5.4${\displaystyle \pm}1.0$ &3.0${\displaystyle \pm}0.2$ &4.2${\displaystyle \pm}0.5$ &7.1${\displaystyle \pm}0.8$ \\ 
NGC 6723 &3.1${\displaystyle \pm}0.8$ &4.6${\displaystyle \pm}1.9$ &10.1${\displaystyle \pm}5.5$ &3.5${\displaystyle \pm}0.8$ &6.4${\displaystyle \pm}3.0$ &7.1${\displaystyle \pm}4.3$ \\ 
Pal 6 &2.6${\displaystyle \pm}0.8$ &4.0${\displaystyle \pm}1.4$ &5.5${\displaystyle \pm}5.9$ &2.6${\displaystyle \pm}1.1$ &4.8${\displaystyle \pm}2.2$ &12.4${\displaystyle \pm}18.1$ \\ 
Terzan 1 &2.5${\displaystyle \pm}0.1$ &3.2${\displaystyle \pm}0.1$ &4.5${\displaystyle \pm}0.4$ &2.6${\displaystyle \pm}0.1$ &3.5${\displaystyle \pm}0.2$ &5.0${\displaystyle \pm}0.5$ \\ 
Terzan 2 &2.4${\displaystyle \pm}0.1$ &2.8${\displaystyle \pm}0.2$ &3.2${\displaystyle \pm}0.3$ &2.6${\displaystyle \pm}0.1$ &3.1${\displaystyle \pm}0.2$ &3.9${\displaystyle \pm}0.3$ \\ 
Terzan 4 &2.1${\displaystyle \pm}0.1$ &2.6${\displaystyle \pm}0.2$ &2.9${\displaystyle \pm}0.2$ &2.3${\displaystyle \pm}0.3$ &2.9${\displaystyle \pm}0.1$ &3.5${\displaystyle \pm}0.2$ \\ 
Terzan 5 &2.5${\displaystyle \pm}0.1$ &3.0${\displaystyle \pm}0.1$ &3.7${\displaystyle \pm}0.4$ &2.6${\displaystyle \pm}0.2$ &3.2${\displaystyle \pm}0.2$ &4.1${\displaystyle \pm}0.7$ \\ 
Terzan 6 &2.6${\displaystyle \pm}0.3$ &3.0${\displaystyle \pm}0.8$ &3.5${\displaystyle \pm}1.2$ &2.8${\displaystyle \pm}0.2$ &3.5${\displaystyle \pm}0.5$ &3.6${\displaystyle \pm}1.6$ \\ 
Terzan 9 &2.7${\displaystyle \pm}0.1$ &3.4${\displaystyle \pm}0.2$ &4.6${\displaystyle \pm}0.5$ &2.8${\displaystyle \pm}0.1$ &3.8${\displaystyle \pm}0.5$ &5.5${\displaystyle \pm}0.7$ \\ 
\hline
\end{tabular}
    \caption{Frequency ratios for different pattern speeds of the bar for the potential of BB2016 and MC2017. The latter are marked with an asterisk.}
    \label{tab:CG_freqs}
\end{table*}
%%%%%%%%%%%%%%%%%%%%%%%%%%%%%%%%%%%%%%%%%%%%%%%%%%%%%%%
%%%%%%%%%%%%%%%%%%%%%%%%%%%%%%%%%%%%%%%%%%%%%%%%%%%%%%%

%%%%%%%%%%%%%%%%%%%%%%%%%%%%%%%%%%%%%%%%%%%%%%%%%%%%%%%%%%%%%%%%%%%%%%%%%%%%%%%%%%%%%%%%%%%%%%%%%%%%%%%
\begin{table}
\centering
\caption{List of GCs following the bar for $\Omega_\mathrm{p}=39$ km/s/kpc. Orbit types are obtained by visual classification. Note that there are no strictly periodic orbits, and we denote here the family of orbits around which an orbit with a regular profile oscillates.}
\begin{tabular}{ c | c | c |}
ID & Potential & Orbital family\\
\hline \hline
NGC 6266 & $N$-body & x1\\
NGC 6380 & $N$-body & x1\\
NGC 6440 & $N$-body & x2\\
NGC 6522 & BB2016, MC2017 & x1\\
NGC 6522 & $N$-body & x2\\
NGC 6642 & $N$-body & x2\\
Terzan 1 & $N$-body & x1\\
Terzan 2 & $N$-body & x1\\
Terzan 4 & $N$-body & x1\\
Terzan 5 & $N$-body & x1\\
%%%%%%%%%%%%%%%%%%%%%%%%%%%%%%%%%%%%%%%%%%%%%%%%%%%%%%%%%%%%%%%%%%%%%%%%%%%%%%%%%%%%%%%%%%%%%%%%%%%%%%%
%\multicolumn{3}{p{0.4\textwidth}}
%{\footnotesize{\textit{Notes}: }}
\end{tabular}
\label{tab:bar_captured}
\end{table} 
%%%%%%%%%%%%%%%%%%%%%%%%%%%%%%%%%%%%%%%%%%%%%%%%%%%%%%%%%%%%%%%%%%%%%%%%%%%%%%%%%%%%%%%%%%%%%%%%%%%%%%%

\begin{table*}
\begin{tabular}{ c | c | c | c }
\hline
%ID & \Omega_\mathrm{p}$$ & $V_\mathrm{R}$ & $V_\mathrm{T}$ & $z$ & $V_z$& $\phi$\\
ID & $f_\mathrm{R}/f_x$ (BB2016) & $f_\mathrm{R}/f_x$ (MC2017) & $f_\mathrm{R}/f_x$ ($N$-body)  \\ 
\hline 
\hline
BH 229 &2.8${\displaystyle \pm}0.8$ &3.1${\displaystyle \pm}0.7$ &1.9${\displaystyle \pm}0.8$ \\ 
ESO 452-11 &3.0${\displaystyle \pm}0.7$ &3.4${\displaystyle \pm}0.3$ &2.3${\displaystyle \pm}0.5$ \\ 
Liller 1 &1.4${\displaystyle \pm}0.1$ &1.5${\displaystyle \pm}0.2$ &1.5${\displaystyle \pm}0.2$ \\ 
NGC 6144 &4.0${\displaystyle \pm}1.0$ &5.2${\displaystyle \pm}1.0$ &2.1${\displaystyle \pm}1.1$ \\ 
NGC 6266 &2.7${\displaystyle \pm}0.4$ &3.0${\displaystyle \pm}0.3$ &2.0${\displaystyle \pm}0.2$ \\ 
NGC 6273 &4.6${\displaystyle \pm}0.9$ &7.7${\displaystyle \pm}1.6$ &2.2${\displaystyle \pm}1.1$ \\ 
NGC 6293 &3.3${\displaystyle \pm}0.4$ &3.8${\displaystyle \pm}0.5$ &2.3${\displaystyle \pm}0.6$ \\ 
NGC 6342 &2.3${\displaystyle \pm}1.1$ &2.5${\displaystyle \pm}1.3$ &1.4${\displaystyle \pm}0.8$ \\ 
NGC 6355 &3.2${\displaystyle \pm}0.3$ &3.9${\displaystyle \pm}0.4$ &2.0${\displaystyle \pm}0.7$ \\ 
NGC 6380 &3.0${\displaystyle \pm}0.1$ &3.1${\displaystyle \pm}0.1$ &2.3${\displaystyle \pm}0.3$ \\ 
NGC 6401 &3.1${\displaystyle \pm}0.2$ &3.6${\displaystyle \pm}0.2$ &2.2${\displaystyle \pm}0.3$ \\ 
NGC 6440 &2.7${\displaystyle \pm}0.1$ &2.9${\displaystyle \pm}0.1$ &2.0${\displaystyle \pm}0.2$ \\ 
NGC 6453 &3.2${\displaystyle \pm}0.5$ &3.5${\displaystyle \pm}0.4$ &1.9${\displaystyle \pm}1.1$ \\ 
NGC 6522 &2.3${\displaystyle \pm}0.4$ &2.3${\displaystyle \pm}0.5$ &1.9${\displaystyle \pm}0.3$ \\ 
NGC 6528 &2.8${\displaystyle \pm}0.7$ &3.3${\displaystyle \pm}0.7$ &1.9${\displaystyle \pm}0.6$ \\ 
NGC 6558 &2.9${\displaystyle \pm}0.1$ &3.2${\displaystyle \pm}0.1$ &2.2${\displaystyle \pm}0.3$ \\ 
NGC 6624 &2.3${\displaystyle \pm}0.8$ &2.6${\displaystyle \pm}0.8$ &2.2${\displaystyle \pm}0.3$ \\ 
NGC 6626 &3.4${\displaystyle \pm}0.2$ &3.7${\displaystyle \pm}0.3$ &2.5${\displaystyle \pm}0.6$ \\ 
NGC 6638 &3.0${\displaystyle \pm}0.1$ &3.2${\displaystyle \pm}0.1$ &2.2${\displaystyle \pm}0.3$ \\ 
NGC 6637 &1.6${\displaystyle \pm}0.6$ &1.6${\displaystyle \pm}0.6$ &1.4${\displaystyle \pm}0.8$ \\ 
NGC 6642 &2.8${\displaystyle \pm}0.2$ &2.8${\displaystyle \pm}0.4$ &2.1${\displaystyle \pm}0.2$ \\ 
NGC 6717 &3.2${\displaystyle \pm}0.2$ &3.5${\displaystyle \pm}0.2$ &2.4${\displaystyle \pm}0.6$ \\ 
NGC 6723 &3.8${\displaystyle \pm}1.4$ &4.8${\displaystyle \pm}1.6$ &2.8${\displaystyle \pm}1.7$ \\ 
Pal 6 &3.2${\displaystyle \pm}1.0$ &3.8${\displaystyle \pm}1.3$ &2.5${\displaystyle \pm}2.0$ \\ 
Terzan 1 &2.9${\displaystyle \pm}0.1$ &3.1${\displaystyle \pm}0.1$ &2.2${\displaystyle \pm}0.2$ \\ 
Terzan 2 &2.6${\displaystyle \pm}0.2$ &2.9${\displaystyle \pm}0.1$ &2.0${\displaystyle \pm}0.2$ \\ 
Terzan 4 &2.4${\displaystyle \pm}0.2$ &2.7${\displaystyle \pm}0.2$ &2.0${\displaystyle \pm}0.1$ \\ 
Terzan 5 &2.7${\displaystyle \pm}0.1$ &2.9${\displaystyle \pm}0.2$ &2.2${\displaystyle \pm}0.2$ \\ 
Terzan 6 &2.7${\displaystyle \pm}0.6$ &3.0${\displaystyle \pm}0.6$ &2.2${\displaystyle \pm}0.5$ \\ 
Terzan 9 &3.1${\displaystyle \pm}0.1$ &3.3${\displaystyle \pm}0.1$ &2.4${\displaystyle \pm}0.3$ \\    
\hline
\end{tabular}
    \caption{Comparison of frequencies for three different potentials considered in the present work. All frequency ratios are calculated for $\Omega_\mathrm{p}=39$ km/s/kpc.}
    \label{tab:CG_comare}
\end{table*}
%%%%%%%%%%%%%%%%%%%%%%%%%%%%%%%%%%%%%%%%%%%%%%%%%%%%%%%
%%%%%%%%%%%%%%%%%%%%%%%%%%%%%%%%%%%%%%%%%%%%%%%%%%%%%%%

\begin{figure*}
    \centering
    \begin{minipage}[t]{0.33\textwidth}
	    %\vspace*{-0.2cm}
		\includegraphics[width=0.85\textwidth]{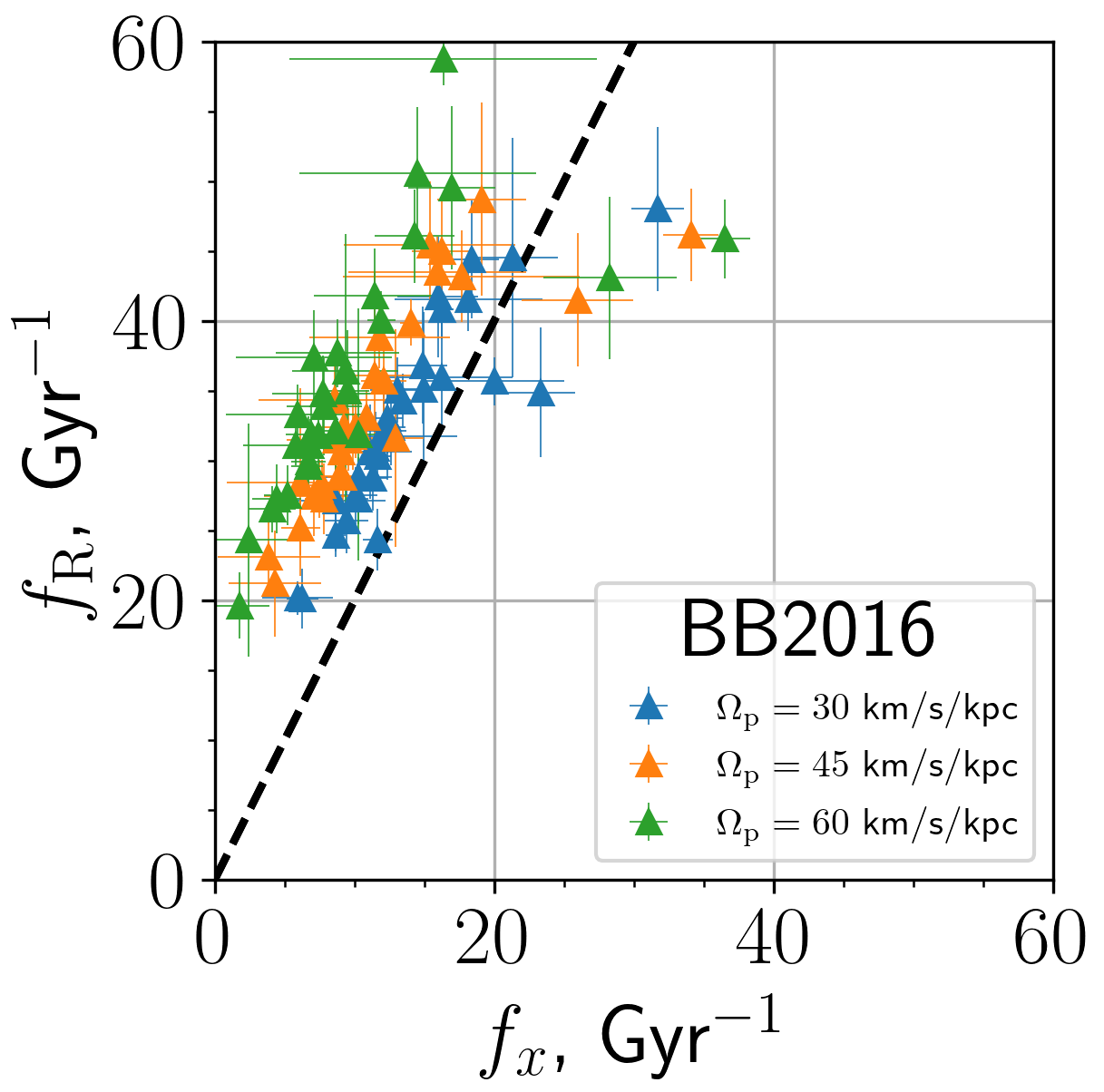}% 
	\end{minipage}%
     \begin{minipage}[t]{0.33\textwidth}
	    %\vspace*{-0.2cm}
		\includegraphics[width=0.85\textwidth]{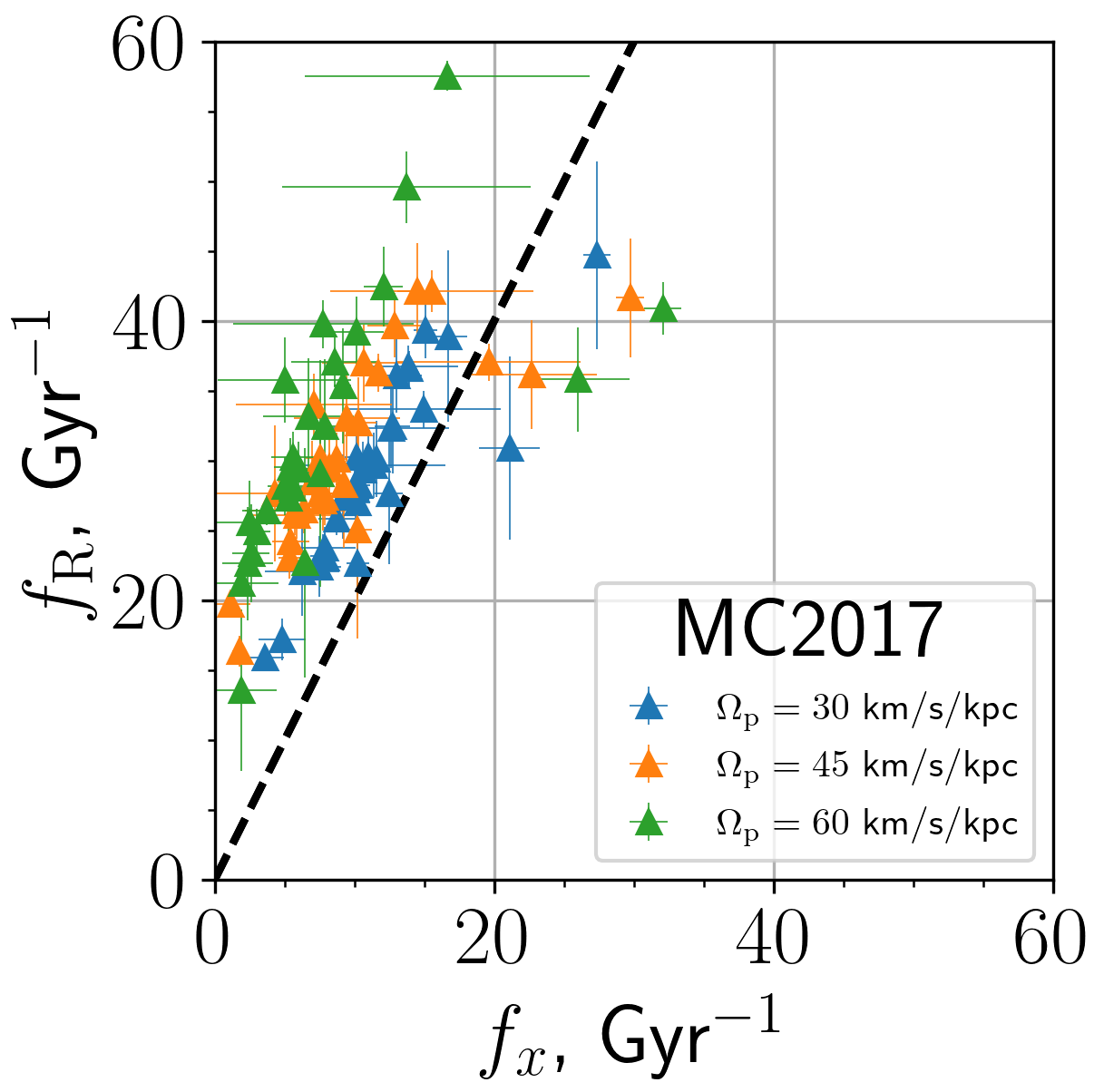}% 
	\end{minipage}%
     \begin{minipage}[t]{0.33\textwidth}
	    %\vspace*{-0.2cm}
		\includegraphics[width=0.85\textwidth]{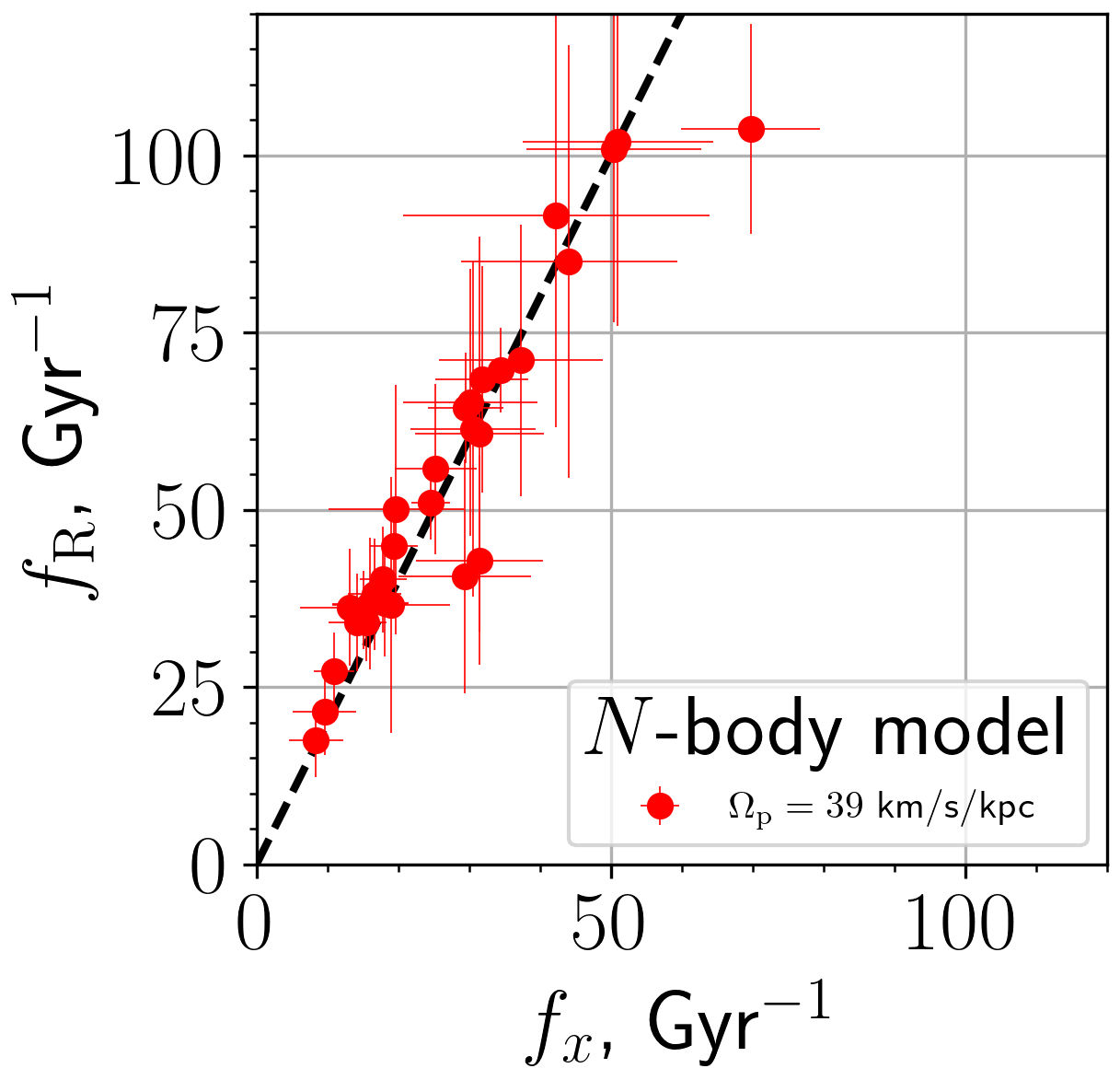}% 
	\end{minipage}%
       
    \caption{Frequency ratios depending on bar pattern speed for BB2016 and MC2017 potentials (\textit{left} and \textit{middle} subpanels, respectively), and frequency ratios in the $N$-body model.}
    \label{fig:fxfR_fyfR_all}
\end{figure*}

Here we consider in detail how the frequencies change with the pattern speed for all GCs in our sample. Fig.~\ref{fig:fxfR_fyfR_all} shows the frequencies $f_x$ and $f_\mathrm{R}$ for all three potentials, and Table~\ref{tab:CG_freqs} lists the exact values. For clarity, we investigate only three values of the pattern speed, $\Omega_\mathrm{p}=(30,45,60)$ km/s/kpc, while the rest of the bar parameters are fixed at $M_\mathrm{bar}/M_\mathrm{b}=0.95, a_\mathrm{bar}=5.0$ kpc, $p=2.0$, $q=3.0$. For each orbit, we use Monte-Carlo simulations ($10^3$ iterations) to estimate frequency errors due to uncertainty in GCs' positions and velocities. It can be seen from the figure that the orbital frequencies of almost all GCs behave in the same way as it was shown earlier for NGC 6266. As the pattern speed increases, the frequencies ratio $f_\mathrm{R}/f_x$ begins to deviate more and more from the resonance line 2:1. %There are also some peculiar GCs, which do not behave this way, i. e. they have resonance ratio 2:1 at rather high value of $\Omega_\mathrm{p}=60$ km/s/kpc. 
%\textcolor{black}{discuss these orbits separately}
\par 
In general, Fig.~\ref{fig:fxfR_fyfR_all} demonstrates that, in analytical potentials (both in BB2016 and MC2017), most GCs are do not follow the bar for all the considered values of the pattern speed. We should also note that one cannot overstep the limits of the pattern speed considered here, since they are motivated by observations. The rightmost panel of Fig.~\ref{fig:fxfR_fyfR_all} shows orbital frequencies obtained for the same GCs in the $N$-body potential. For the $N$-body model, we do not consider different pattern speeds, since in this case its value $39$ km/s/kpc follows from direct and precise measurements of the bar properties in the model~\citep{2021arXiv211105466T}. As can be seen, there is much more orbits with the resonance frequency ratio of 2:1 in such a potential. We have compiled a list of them in Table~\ref{tab:bar_captured}. \textcolor{black}{Based on the orbital profiles, we distinguish the orbits into two types, the well-known $x1$ family consisting of orbits elongated along the bar and supporting its structure and $x2$ orbits which are elongated in the direction perpendicular to the bar major axis and observed in the most central regions~(\citealt{1980A&A....92...33C}, see also reviews by~\citealt{1989A&ARv...1..261C} and a more recent one by~\citealt{2014RvMP...86....1S}).}

%We also note that orbits oscillate much faster in the $N$-body potential, i. e. frequencies are much greater there (note the change of scales along both axes for the subplot, depicting $N$-body results). 
\par
\begin{figure*}
    \centering
    \begin{minipage}[t]{0.45\textwidth}
	    %\vspace*{-0.2cm}
		\includegraphics[width=0.85\textwidth]{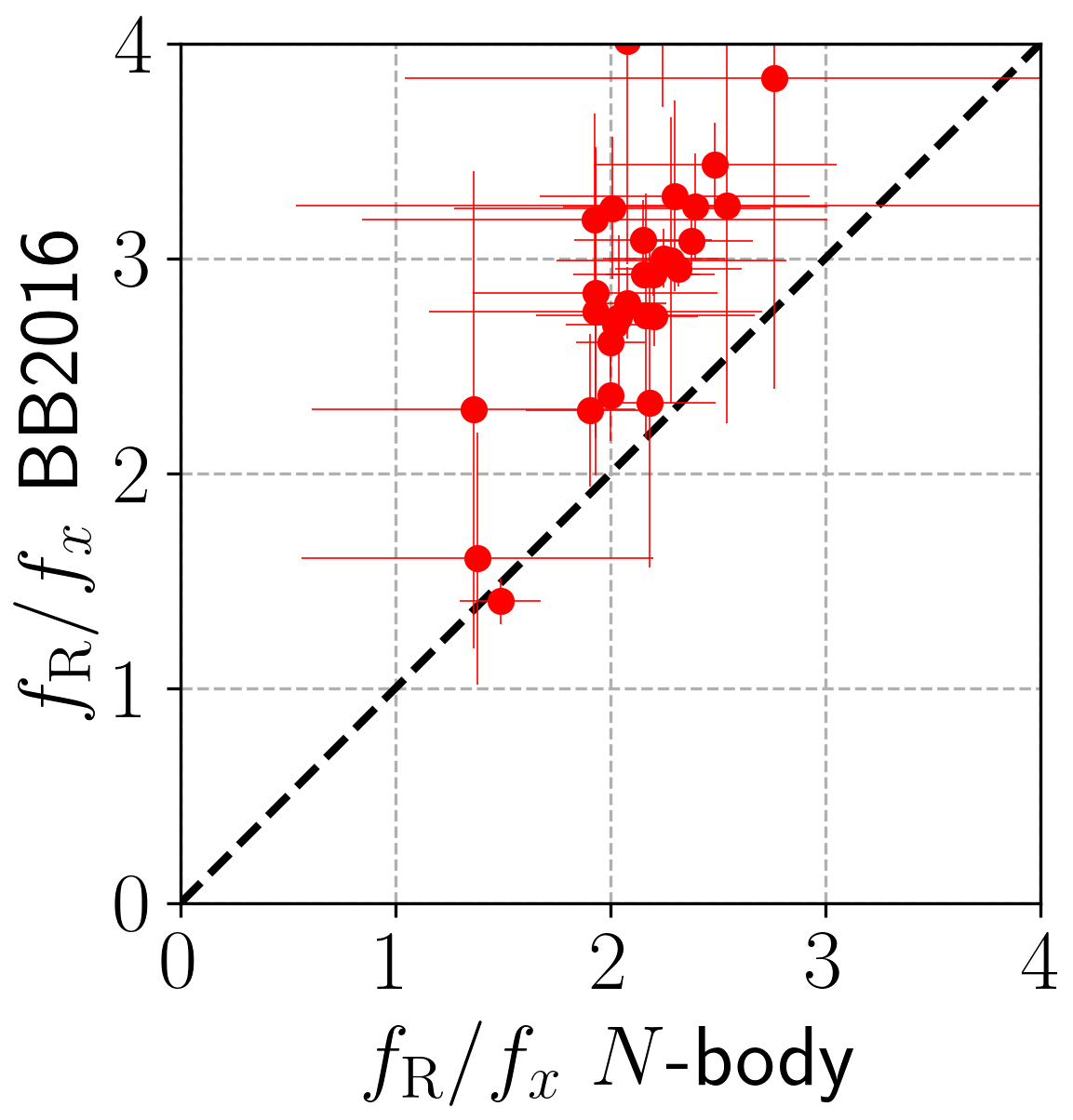}% 
	\end{minipage}%
     \begin{minipage}[t]{0.45\textwidth}
	    %\vspace*{-0.2cm}
		\includegraphics[width=0.85\textwidth]{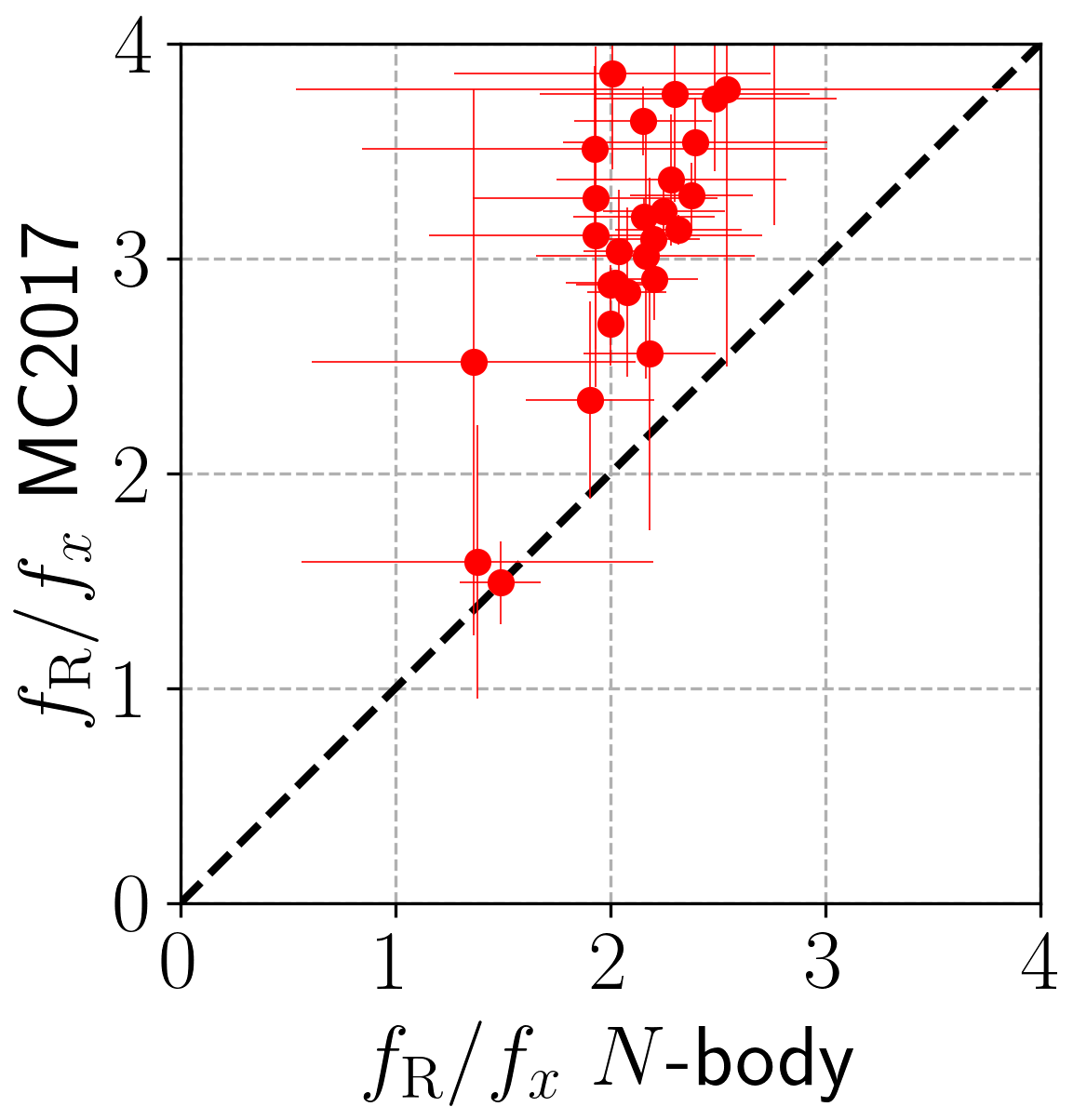}% 
	\end{minipage}%
       
    \caption{Comparison of frequency ratios obtained for different potentials, BB2016 and $N$-body (\textit{left}) and MC2017 and $N$-body (\textit{right}).}
    \label{fig:fRfx_comp_between_potentials}
\end{figure*}

\begin{figure*}
    \centering
		\includegraphics[width=0.95\textwidth]{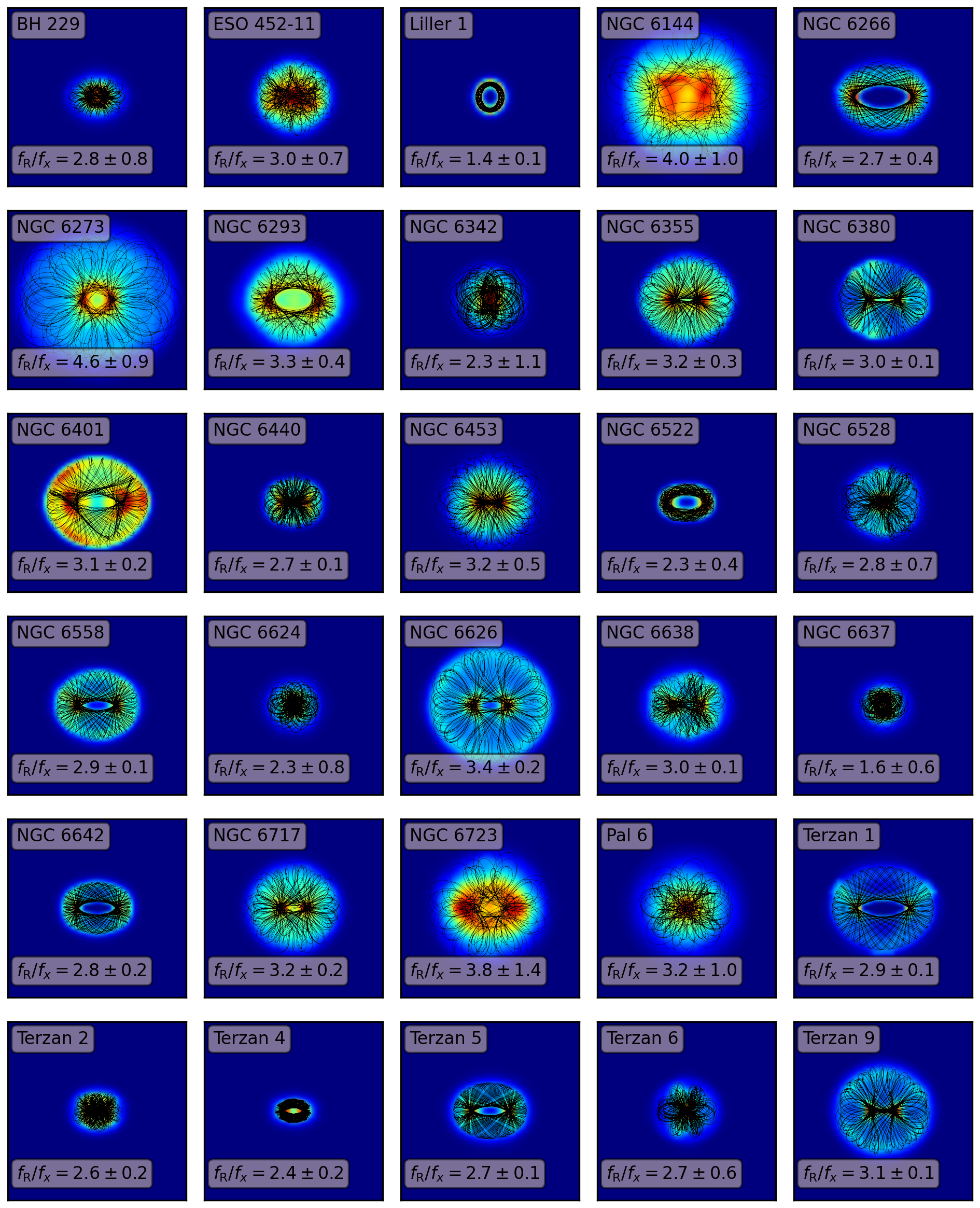}%
    \caption{Orbits of globular clusters for $\Omega_\mathrm{p}=39$, $M_\mathrm{bar}/M_\mathrm{b}=0.95$, $a_\mathrm{bar}=5$ kpc, $p=2.0$, $q=3.0$ for the potential of BB2016. \textcolor{black}{For all orbits, $(xy)$ projection in the bar rotating frame is shown in the square area of 5 kpc $\times$ 5 kpc.} The black line shows the orbit for the middle values of the parameters from Table~\ref{tab:CG_prop}. The colour map depicts the probably of finding an orbit according to Monte-Carlo simulations.}
    \label{fig:orbitsBB}
\end{figure*}

\begin{figure*}
    \centering
		\includegraphics[width=0.95\textwidth]{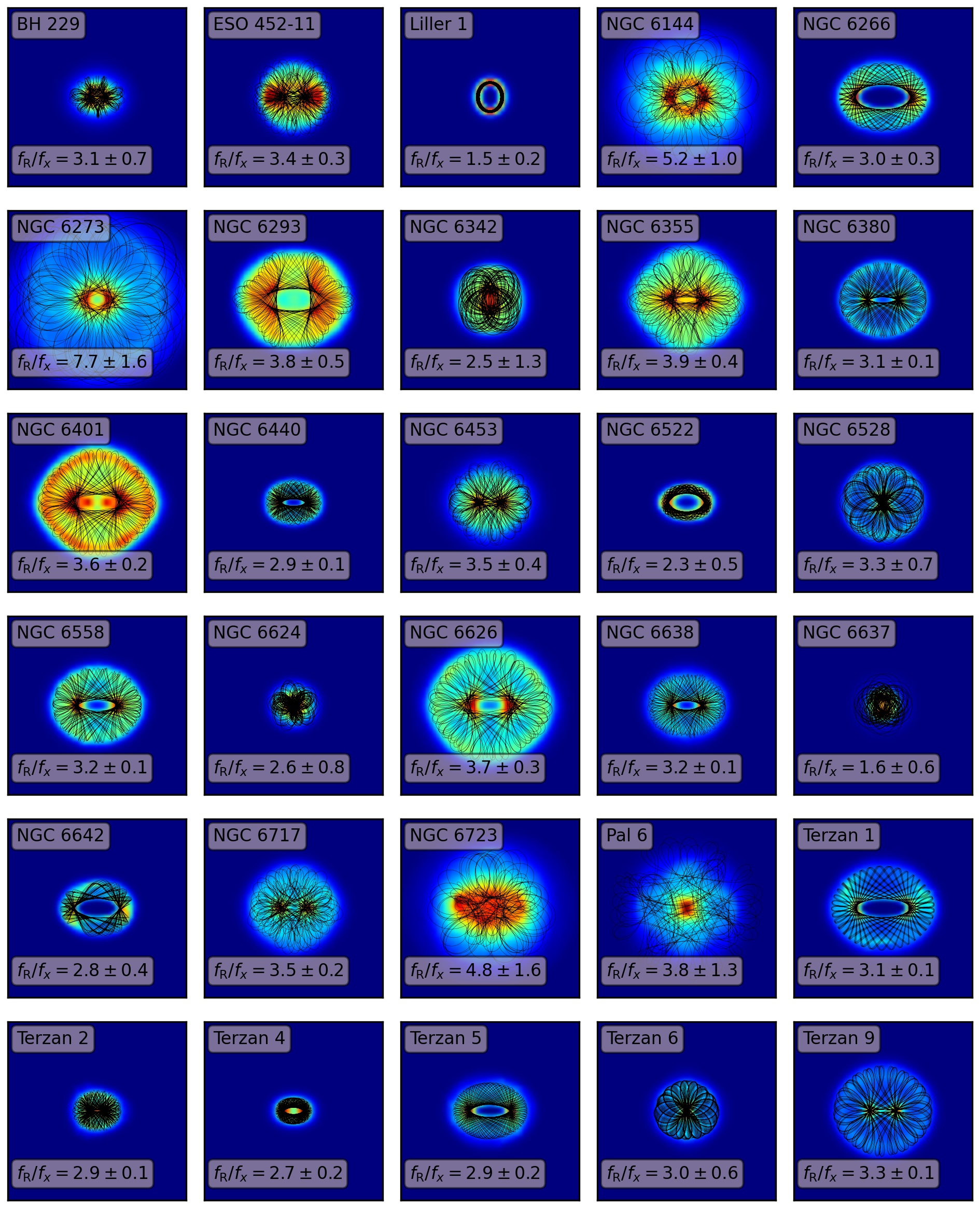}% 
      \caption{Same as Fig.~\ref{fig:orbitsBB}, but for the potential of MC2017.}
    \label{fig:orbitsMC}
\end{figure*}

\begin{figure*}
    \centering
		\includegraphics[width=0.95\textwidth]{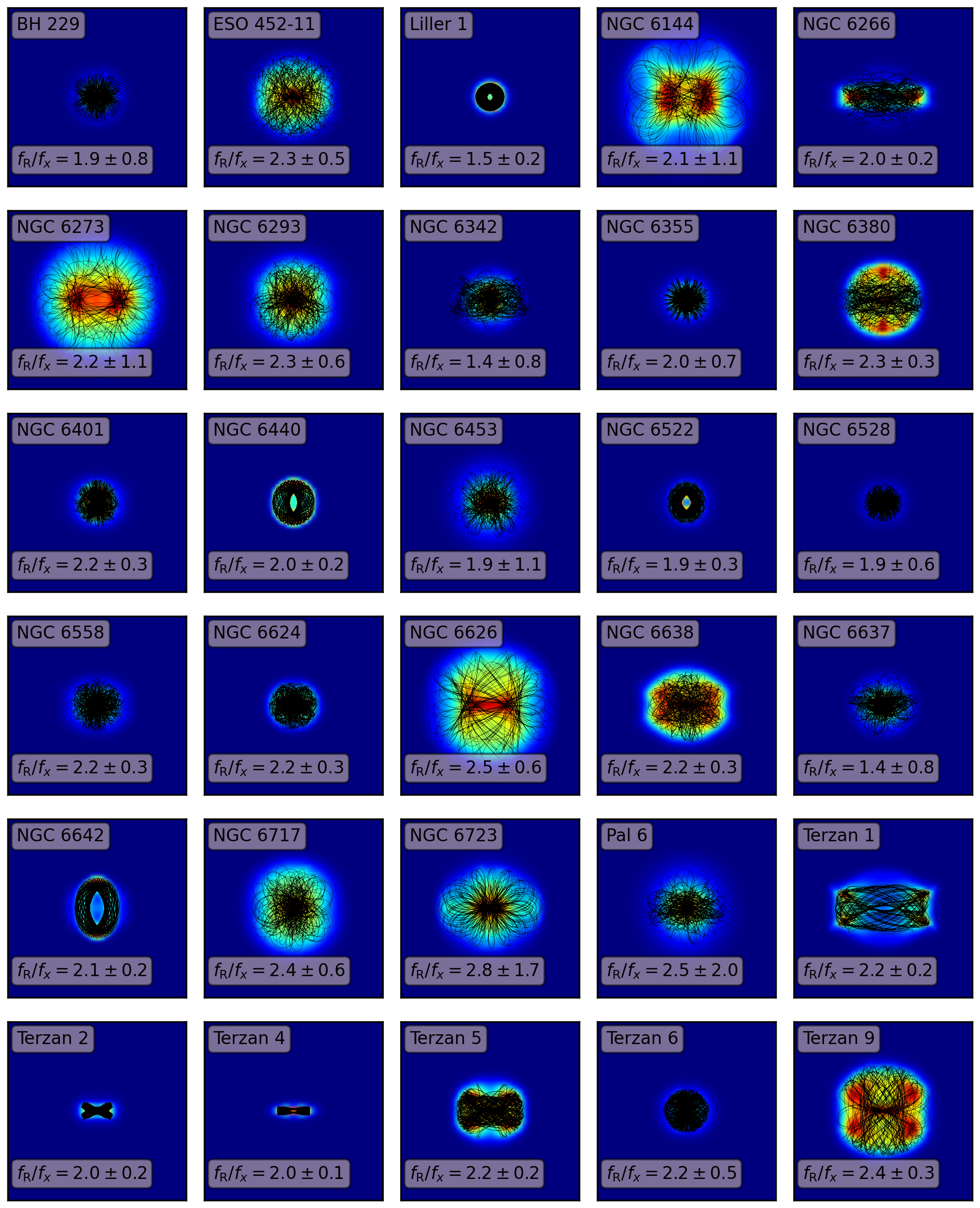}%
    \caption{Orbits of globular clusters for the $N$-body potential of TG2021.}
    \label{fig:orbitsNbody}
\end{figure*}

 
In Fig.~\ref{fig:fRfx_comp_between_potentials} and Table~\ref{tab:CG_comare}, $f_\mathrm{R}/f_x$ values are compared for all considered potentials. The orbits themselves are presented in Fig.~\ref{fig:orbitsBB}, Fig.~\ref{fig:orbitsMC}, and Fig.~\ref{fig:orbitsNbody}. For a better comparison, we fixed the bar pattern speed in analytical potentials at the value of the pattern speed in the $N$-body simulation, i.~e. $\Omega_\mathrm{p}=39$ km/s/kpc for all cases. \textcolor{black}{The rest bar parameters are the same as in the previous section: $M_\mathrm{bar}/M_\mathrm{b}=0.95$, $a_\mathrm{bar}=5.0$ kpc, $p=2.0$, $q=3.0$. We should note that one can try to change the bar-to-bulge mass ratio somewhat to make the BB2016 and MC2017 pontentials resemble the $N$-body potential more, but, in practice, it is hard to estimate the ratio in the $N$-body model itself. For example, if one considers the ratio of masses of the classical bulge and the bar plus the said bulge  in the $N$-body model, it is about 0.6. However, at the same time, the total mass of the bar plus the bulge is about the half of the original disc mass~(see table~1 in~\citealt{Smirnov_etal2021}), while, for the BB2016 potential, the bulge plus bar is about 20\% of the disc. The root of the problem is that, for the $N$-body model, the bar is formed from the disc material, and the disc itself does not go all the way towards the centre~(see~\citealt{2020MNRAS.499..462S}, figure 1 there). This is clearly not the case for the potentials of BB2016 and MC2017 obtained from the velocity curve fitting, where the disc goes all the way towards the centre and, thus, has high contribution in terms of mass there. One can possibly alleviate this issue by reducing the disc mass in the center or by initially considering the disc with the hole in the centre. We leave the solution of this problem for future studies. Here, we stick to the appoach by~\cite{Ortolani2019a, Ortolani2019b, PV2020}, where the whole or almost the whole bulge is thought to be a bar.}
\par
%the value of the bar/classic bulge mass ratio in the $N$-body model. 
As can be seen, $f_\mathrm{R}/f_x$ in the analytical potentials are shifted towards larger values compared to those in the $N$-body potential, for both BB2016 and MC2017. Although, we should note that the difference between BB2016 and $N$-body is a bit smaller on average than between MC2017 and $N$-body.
\par 
In addition to the orbits following the bar, we want to mention some of the interesting ones with  frequency ratios above or below 2:1. Liller 1 in all three potentials has a frequency ratio of about 3:2, the orbit itself looks regular, but has circle-like profile, and clearly does not follow the bar. \textcolor{black}{NGC 6380 in BB2016 has $f_\mathrm{R}/f_x$ close to 3, which is reflected in its overall trefoil-like shape.} We should also mention, that, while most of orbits do not follow the bar in BB2017 and MC2017, some of their profiles look regular and resemble those previously shown for NGC 6266 in Fig.~\ref{fig:NGC6266_BB_orbits}. \textcolor{black}{These are NGC 6642, NGC 6558, Terzan 1, Terzan 5 for BB2016 and NGC 6380, NCC 6440, NGC 6522, NGC 6642, Terzan 1, Terzan 4, and Terzan 5 for MC 2017.} It is interesting to note that most of these orbits have a rather small error in their frequency ratios ($\Delta f_\mathrm{R}/f_x = 0.1-0.2$)

\section{Discussion}
\label{sec:comp}
The change in the ratio of frequencies with the pattern speed has been indirectly observed in some of the previous works. In particular, ~\cite{PV2020} found that the percentage of orbits following a bar decreases with bar rotation, except for NGC 6304, NGC 6342, and NGC 6637, which are not considered in the present work. If we assume the percentage of orbits following the bar should increase as the frequency ratio gets closer to 2:1, which is reasonable, then results of~\cite{PV2020} support the idea that decreasing the pattern speed causes the frequency ratios $f_\mathrm{R}/f_x$ get closer to the bar frequency ratio 2:1. 
%The latter was clearly shown in  for NGC 6266 and in Section~\ref{sec:all_orbits} for most of GCs from our sample. 
\par 
A decrease in  the frequency $f_x$ with the patter speed, which is one of the reasons why the frequency ratio deviates from 2:1, was also observed by~\cite{2020MNRAS.495.3175S} for the self-consistent $N$-body model. Strictly speaking, what was observed is an increase of $f_x$ with a decrease in $\Omega_\mathrm{p}$ (i.e. bar slow downing), which is essentially coincides as our result. We should also note that attributing the effect to a change in the pattern speed only in the case of~\cite{2020MNRAS.495.3175S} may be somewhat biased, since other properties of the bar (mass and size), were also changing there in accordance with the self-consistent evolution of the model.  
\par 
As for the particular GCs, \cite{PV2020} found that for Liller 1, NGC 6304, NGC 6522, NGC 6528, NGC 6540, NGC 6553, Terzan 5, and Terzan 9, more then 20 percent of orbits follow the bar.
Comparing our results to \cite{PV2020}, we find that, for all potentials, Liller 1 and Terzan 9 do not follow the bar, while Terzan 5 follow the bar in the $N$-body model. \textcolor{black}{NGC 6522 follows the bar in the potentials of BB2016 and MC2017, but perpendicular to it in the $N$-body potential ($x2$ family).} For NGC 6528, the frequency ratio is close to 2:1, but the orbits themselves have an irregular profile, therefore, it is hard to say that this orbit can support a bar.
\par

%Terzan 5, and Terzan 9 found for their potential that the changes in the pattern speed has a negligible value on the bulge CGs' orbits.   %we should also note the following. \cite{PV2020} studied the bulge GCs using the potential consisting of Miyamoto-Nagai disc and NFW halo. The bar component was modelled by Ferrer's triaxial ellipsoid. 
%At the same time, the authors noted that, with an increase of the pattern speed, the fraction of orbits that follow the bar, decreases (their table 7).
%In this study, we find that changing  the bar pattern speed directly lead to systematic changes in the orbits' frequency ratio $f_\mathrm{R}/f_x$.%, which is important if one classify orbits based on $f_\mathrm{R}/f_x$.%, i.~e. with an increase in the pattern speed $f_\mathrm{R}/f_x$ ratio becomes farther and farther from the resonance ratio 2:1. 
%The problem is that~\cite{PV2020} considered a rather small pattern speed interval, $\Omega_\mathrm{p}=(40-50)$ km/s/kpc. While in this work, we consider a larger interval, $\Omega_\mathrm{p}=(30-60)$ km/s/kpc, according to the range of estimates from observational data. Since the interval is greater than that considered by ~\cite{PV2020}, systematic changes are more prominent and easy to notice.
%\par 

\section{Conclusions}
\label{sec:conc}
\begin{enumerate}
    \item We calculated the evolution of 30 globular clusters located in the inner area of the Galaxy ($R\lesssim5$) backwards in time for 5 Gyr in a non-axisymmetric galaxy potential using \textcolor{black}{Gaia DR2 data for line-of sight velocities~\citep{Vasiliev2019} and the newest Gaia DR3 data for proper motions and distances~\citep{Vasiliev2021, Baumgardt2021}}. Throughout this work, we have compared the results for three potentials, two of which are analytical, obtained by fitting the rotation curve from~\cite{Bajkova_2016,Bajkova_2017} and ~\cite{McMillan_2017}, and one is taken directly from $N$-body simulations recently preparedby~\cite{2021arXiv211105466T} (``surrogate Milky Way''). 
    \item For all orbits, we calculated their coordinate spectra and determined the corresponding main frequencies, $f_x$ and $f_\mathrm{R}$, for a range of bar parameters (pattern speed, mass, size, shape) in analytical potentials and for a fixed pattern speed for the $N$-body model. 
    \item We distinguish orbits by their frequency ratio $f_\mathrm{R}/f_{x}$ to test whether a particular orbit follows the bar. Most of orbits in both considered analytical potentials do not support the bar in the "usual" sense~(either $f_\mathrm{R}/f_x \gtrsim 2.1$ or $f_\mathrm{R}/f_x \lesssim 1.9$) for a physically reasonable values of the pattern speed, while, for the case of the $N$-body potential, 10 GCs follow the bar ($f_\mathrm{R}/f_\mathrm{x}\approx2.0$).
    \item On the example of one orbit (NGC 6266), we verified how the the frequency ratio changes depending on the pattern speed, the mass and size of the bar tracking the changes in a wide range of parameters using a small relative step. We found the the frequency ratio does not depend much on the mass \textcolor{black}{ratio of the bar and the spherical bulge (``classic'' one)}, bar size, or its shape parameters. Most of the changes occur due to the changes in the pattern speed.  For $\Omega_\mathrm{p}\lesssim20$ km/s/kpc, the orbit perfectly follows the bar $(f_\mathrm{R}/f_\mathrm{x}\approx2.0)$ for all values of the pattern speed and has a typical ``bar''-like profile. Then, at a certain value of the pattern speed depending on the potential, the frequency ratio changes abruptly, becoming either greater or smaller than $f_\mathrm{R}/f_\mathrm{x}\approx2.0$.
    The orbit then begins to oscillate around the bar and no longer supports it.

    %\par To understand the nature of changes, we investigated Poincare surface of sections. We found that the frequency changes along with the family of orbits a particular orbit belongs too. Specifically, considered orbit changes from $x1$ type to $x3$ (or $x4$) usually associated with the co-rotation resonance~\citep{Contopoulos_Papayannopoulos1980}.

\end{enumerate}
%%%%%%%%%%%%%%%%%%%%%%%%%%%%%%%%%%%%%%%%%%%%%%%%%%%%%%%
%%%%%%%%%%%%%%%%%%%%%%%%%%%%%%%%%%%%%%%%%%%%%%%%%%%%%%%
Overall, our results show that comparing orbital classifications between different potentials is indeed valuable, as the results turn out to be vastly different between them. An interesting question that we could not find an answer to in the present work, is why the $N$-body model demonstrates a lot more bar following orbits compared to the analytic potentials. This can possibly indicate that the self-consistency of the potential should play an important factor in orbital studies of GCs.

%%%%%%%%%%%%%%%%%%%%%%%%%%%%%%%%%%%%%%%%%%%%%%%%%%%%%%%
%%%%%%%%%%%%%%%%%%%%%%%%%%%%%%%%%%%%%%%%%%%%%%%%%%%%%%%
\section*{Data availability}
The data underlying this article will be shared on reasonable request to the corresponding author.
%%%%%%%%%%%%%%%%%%%%%%%%%%%%%%%%%%%%%%%%%%%%%%%%%%%%%%%
%%%%%%%%%%%%%%%%%%%%%%%%%%%%%%%%%%%%%%%%%%%%%%%%%%%%%%%

%%%%%%%%%%%%%%%%%%%%%%%%%%%%%%%%%%%%%%%%%%%%%%%%%%%%%%%
%%%%%%%%%%%%%%%%%%%%%%%%%%%%%%%%%%%%%%%%%%%%%%%%%%%%%%%
\section*{Acknowledgements}
We acknowledge the use of the~\texttt{AGAMA}~\citep{agama} \texttt{python} package without which the present work would not be possible. We are grateful to the anonymous reviewer for their valuable comments that contributed to a improvement of the scientific quality of the manuscript and a clearer presentation of our results.

%%%%%%%%%%%%%%%%%%%%%%%%%%%%%%%%%%%%%%%%%%%%%%%%%%%%%%%
%%%%%%%%%%%%%%%%%%%%%%%%%%%%%%%%%%%%%%%%%%%%%%%%%%%%%%%
\bibliographystyle{mnras}
\bibliography{main.bib} % if your bibtex file is called example.bib

\begin{thebibliography}{}
\makeatletter
\relax
\def\mn@urlcharsother{\let\do\@makeother \do\$\do\&\do\#\do\^\do\_\do\%\do\~}
\def\mn@doi{\begingroup\mn@urlcharsother \@ifnextchar [ {\mn@doi@}
  {\mn@doi@[]}}
\def\mn@doi@[#1]#2{\def\@tempa{#1}\ifx\@tempa\@empty \href
  {http://dx.doi.org/#2} {doi:#2}\else \href {http://dx.doi.org/#2} {#1}\fi
  \endgroup}
\def\mn@eprint#1#2{\mn@eprint@#1:#2::\@nil}
\def\mn@eprint@arXiv#1{\href {http://arxiv.org/abs/#1} {{\tt arXiv:#1}}}
\def\mn@eprint@dblp#1{\href {http://dblp.uni-trier.de/rec/bibtex/#1.xml}
  {dblp:#1}}
\def\mn@eprint@#1:#2:#3:#4\@nil{\def\@tempa {#1}\def\@tempb {#2}\def\@tempc
  {#3}\ifx \@tempc \@empty \let \@tempc \@tempb \let \@tempb \@tempa \fi \ifx
  \@tempb \@empty \def\@tempb {arXiv}\fi \@ifundefined
  {mn@eprint@\@tempb}{\@tempb:\@tempc}{\expandafter \expandafter \csname
  mn@eprint@\@tempb\endcsname \expandafter{\@tempc}}}

\bibitem[\protect\citeauthoryear{{Antoja} et~al.,}{{Antoja}
  et~al.}{2014}]{Antoja2014}
{Antoja} T.,  et~al., 2014, \mn@doi [\aap] {10.1051/0004-6361/201322623}, \href
  {https://ui.adsabs.harvard.edu/abs/2014A&A...563A..60A} {563, A60}

\bibitem[\protect\citeauthoryear{{Asano}, {Fujii}, {Baba}, {B{\'e}dorf},
  {Sellentin}  \& {Portegies Zwart}}{{Asano} et~al.}{2020}]{Asano2020}
{Asano} T.,  {Fujii} M.~S.,  {Baba} J.,  {B{\'e}dorf} J.,  {Sellentin} E.,
  {Portegies Zwart} S.,  2020, \mn@doi [\mnras] {10.1093/mnras/staa2849}, \href
  {https://ui.adsabs.harvard.edu/abs/2020MNRAS.499.2416A} {499, 2416}

\bibitem[\protect\citeauthoryear{{Bajkova} \& {Bobylev}}{{Bajkova} \&
  {Bobylev}}{2016}]{Bajkova_2016}
{Bajkova} A.~T.,  {Bobylev} V.~V.,  2016, \mn@doi [Astronomy Letters]
  {10.1134/S1063773716090012}, \href
  {https://ui.adsabs.harvard.edu/abs/2016AstL...42..567B} {42, 567}

\bibitem[\protect\citeauthoryear{{Bajkova} \& {Bobylev}}{{Bajkova} \&
  {Bobylev}}{2017}]{Bajkova_2017}
{Bajkova} A.,  {Bobylev} V.,  2017, \mn@doi [Open Astronomy]
  {10.1515/astro-2017-0016}, \href
  {https://ui.adsabs.harvard.edu/abs/2017OAst...26...72B} {26, 72}

\bibitem[\protect\citeauthoryear{{Bajkova} \& {Bobylev}}{{Bajkova} \&
  {Bobylev}}{2019}]{2019MNRAS.488.3474B}
{Bajkova} A.~T.,  {Bobylev} V.~V.,  2019, \mn@doi [\mnras]
  {10.1093/mnras/stz2061}, \href
  {https://ui.adsabs.harvard.edu/abs/2019MNRAS.488.3474B} {488, 3474}

\bibitem[\protect\citeauthoryear{{Bajkova} \& {Bobylev}}{{Bajkova} \&
  {Bobylev}}{2021}]{BB2021}
{Bajkova} A.~T.,  {Bobylev} V.~V.,  2021, \mn@doi [Research in Astronomy and
  Astrophysics] {10.1088/1674-4527/21/7/173}, \href
  {https://ui.adsabs.harvard.edu/abs/2021RAA....21..173B} {21, 173}

\bibitem[\protect\citeauthoryear{{Bajkova}, {Carraro}, {Korchagin}, {Budanova}
  \& {Bobylev}}{{Bajkova} et~al.}{2020a}]{BB2020}
{Bajkova} A.~T.,  {Carraro} G.,  {Korchagin} V.~I.,  {Budanova} N.~O.,
  {Bobylev} V.~V.,  2020a, \mn@doi [\apj] {10.3847/1538-4357/ab8ea7}, \href
  {https://ui.adsabs.harvard.edu/abs/2020ApJ...895...69B} {895, 69}

\bibitem[\protect\citeauthoryear{{Bajkova}, {Carraro}, {Korchagin}, {Budanova}
  \& {Bobylev}}{{Bajkova} et~al.}{2020b}]{2020ApJ...895...69B}
{Bajkova} A.~T.,  {Carraro} G.,  {Korchagin} V.~I.,  {Budanova} N.~O.,
  {Bobylev} V.~V.,  2020b, \mn@doi [\apj] {10.3847/1538-4357/ab8ea7}, \href
  {https://ui.adsabs.harvard.edu/abs/2020ApJ...895...69B} {895, 69}

\bibitem[\protect\citeauthoryear{{Baumgardt} \& {Vasiliev}}{{Baumgardt} \&
  {Vasiliev}}{2021}]{Baumgardt2021}
{Baumgardt} H.,  {Vasiliev} E.,  2021, \mn@doi [\mnras]
  {10.1093/mnras/stab1474}, \href
  {https://ui.adsabs.harvard.edu/abs/2021MNRAS.505.5957B} {505, 5957}

\bibitem[\protect\citeauthoryear{{Becklin} \& {Neugebauer}}{{Becklin} \&
  {Neugebauer}}{1968}]{Becklin1968}
{Becklin} E.~E.,  {Neugebauer} G.,  1968, \mn@doi [\apj] {10.1086/149425},
  \href {https://ui.adsabs.harvard.edu/abs/1968ApJ...151..145B} {151, 145}

\bibitem[\protect\citeauthoryear{{Bhattacharjee}, {Chaudhury}  \&
  {Kundu}}{{Bhattacharjee} et~al.}{2014}]{2014ApJ...785...63B}
{Bhattacharjee} P.,  {Chaudhury} S.,   {Kundu} S.,  2014, \mn@doi [\apj]
  {10.1088/0004-637X/785/1/63}, \href
  {https://ui.adsabs.harvard.edu/abs/2014ApJ...785...63B} {785, 63}

\bibitem[\protect\citeauthoryear{{Bica}, {Ortolani}  \& {Barbuy}}{{Bica}
  et~al.}{2016}]{Bica2016}
{Bica} E.,  {Ortolani} S.,   {Barbuy} B.,  2016, \mn@doi [\pasa]
  {10.1017/pasa.2015.47}, \href
  {https://ui.adsabs.harvard.edu/abs/2016PASA...33...28B} {33, e028}

\bibitem[\protect\citeauthoryear{{Binney}}{{Binney}}{2020}]{Binney2020}
{Binney} J.,  2020, \mn@doi [\mnras] {10.1093/mnras/staa1103}, \href
  {https://ui.adsabs.harvard.edu/abs/2020MNRAS.495..895B} {495, 895}

\bibitem[\protect\citeauthoryear{{Binney} \& {Spergel}}{{Binney} \&
  {Spergel}}{1982}]{Binney_Spergel1982}
{Binney} J.,  {Spergel} D.,  1982, \mn@doi [\apj] {10.1086/159559}, 252, 308

\bibitem[\protect\citeauthoryear{{Binney} \& {Tremaine}}{{Binney} \&
  {Tremaine}}{2008}]{Binney_Tremaine2008}
{Binney} J.,  {Tremaine} S.,  2008, {Galactic Dynamics: Second Edition}.
Princeton University Press

\bibitem[\protect\citeauthoryear{{Bland-Hawthorn} \&
  {Gerhard}}{{Bland-Hawthorn} \& {Gerhard}}{2016}]{BG2016}
{Bland-Hawthorn} J.,  {Gerhard} O.,  2016, \mn@doi [\araa]
  {10.1146/annurev-astro-081915-023441}, \href
  {https://ui.adsabs.harvard.edu/abs/2016ARA&A..54..529B} {54, 529}

\bibitem[\protect\citeauthoryear{{Bobylev} \& {Bajkova}}{{Bobylev} \&
  {Bajkova}}{2016}]{2016AstL...42....1B}
{Bobylev} V.~V.,  {Bajkova} A.~T.,  2016, \mn@doi [Astronomy Letters]
  {10.1134/S1063773716010023}, \href
  {https://ui.adsabs.harvard.edu/abs/2016AstL...42....1B} {42, 1}

\bibitem[\protect\citeauthoryear{{Bovy}, {Leung}, {Hunt}, {Mackereth},
  {Garc{\'\i}a-Hern{\'a}ndez}  \& {Roman-Lopes}}{{Bovy}
  et~al.}{2019}]{Bovy2019}
{Bovy} J.,  {Leung} H.~W.,  {Hunt} J. A.~S.,  {Mackereth} J.~T.,
  {Garc{\'\i}a-Hern{\'a}ndez} D.~A.,   {Roman-Lopes} A.,  2019, \mn@doi
  [\mnras] {10.1093/mnras/stz2891}, \href
  {https://ui.adsabs.harvard.edu/abs/2019MNRAS.490.4740B} {490, 4740}

\bibitem[\protect\citeauthoryear{{Buta} \& {Zhang}}{{Buta} \&
  {Zhang}}{2009}]{2009ApJS..182..559B}
{Buta} R.~J.,  {Zhang} X.,  2009, \mn@doi [\apjs]
  {10.1088/0067-0049/182/2/559}, \href
  {https://ui.adsabs.harvard.edu/abs/2009ApJS..182..559B} {182, 559}

\bibitem[\protect\citeauthoryear{{Chiba} \& {Sch{\"o}nrich}}{{Chiba} \&
  {Sch{\"o}nrich}}{2021}]{Chiba2021}
{Chiba} R.,  {Sch{\"o}nrich} R.,  2021, \mn@doi [\mnras]
  {10.1093/mnras/stab1094}, \href
  {https://ui.adsabs.harvard.edu/abs/2021MNRAS.505.2412C} {505, 2412}

\bibitem[\protect\citeauthoryear{{Clarke} \& {Gerhard}}{{Clarke} \&
  {Gerhard}}{2022}]{Clarke2022}
{Clarke} J.~P.,  {Gerhard} O.,  2022, \mn@doi [\mnras] {10.1093/mnras/stac603},
  \href {https://ui.adsabs.harvard.edu/abs/2022MNRAS.512.2171C} {512, 2171}

\bibitem[\protect\citeauthoryear{{Contopoulos} \& {Grosbol}}{{Contopoulos} \&
  {Grosbol}}{1989}]{1989A&ARv...1..261C}
{Contopoulos} G.,  {Grosbol} P.,  1989, \mn@doi [\aapr] {10.1007/BF00873080},
  \href {https://ui.adsabs.harvard.edu/abs/1989A&ARv...1..261C} {1, 261}

\bibitem[\protect\citeauthoryear{{Contopoulos} \&
  {Papayannopoulos}}{{Contopoulos} \&
  {Papayannopoulos}}{1980}]{1980A&A....92...33C}
{Contopoulos} G.,  {Papayannopoulos} T.,  1980, \aap, \href
  {https://ui.adsabs.harvard.edu/abs/1980A&A....92...33C} {92, 33}

\bibitem[\protect\citeauthoryear{{C{\^o}t{\'e}}}{{C{\^o}t{\'e}}}{1999}]{Cote1999}
{C{\^o}t{\'e}} P.,  1999, \mn@doi [\aj] {10.1086/300930}, \href
  {https://ui.adsabs.harvard.edu/abs/1999AJ....118..406C} {118, 406}

\bibitem[\protect\citeauthoryear{{Cuomo}, {Lopez Aguerri}, {Corsini},
  {Debattista}, {M{\'e}ndez-Abreu}  \& {Pizzella}}{{Cuomo}
  et~al.}{2019}]{2019A&A...632A..51C}
{Cuomo} V.,  {Lopez Aguerri} J.~A.,  {Corsini} E.~M.,  {Debattista} V.~P.,
  {M{\'e}ndez-Abreu} J.,   {Pizzella} A.,  2019, \mn@doi [\aap]
  {10.1051/0004-6361/201936415}, \href
  {https://ui.adsabs.harvard.edu/abs/2019A&A...632A..51C} {632, A51}

\bibitem[\protect\citeauthoryear{{Gajda}, {{\L}okas}  \&
  {Athanassoula}}{{Gajda} et~al.}{2016}]{Gajda_etal2016}
{Gajda} G.,  {{\L}okas} E.~L.,   {Athanassoula} E.,  2016, \mn@doi [\apj]
  {10.3847/0004-637X/830/2/108}, 830, 108

\bibitem[\protect\citeauthoryear{{Garma-Oehmichen}, {Cano-D{\'\i}az},
  {Hern{\'a}ndez-Toledo}, {Aquino-Ort{\'\i}z}, {Valenzuela}, {Aguerri},
  {S{\'a}nchez}  \& {Merrifield}}{{Garma-Oehmichen}
  et~al.}{2020}]{2020MNRAS.491.3655G}
{Garma-Oehmichen} L.,  {Cano-D{\'\i}az} M.,  {Hern{\'a}ndez-Toledo} H.,
  {Aquino-Ort{\'\i}z} E.,  {Valenzuela} O.,  {Aguerri} J.~A.~L.,  {S{\'a}nchez}
  S.~F.,   {Merrifield} M.,  2020, \mn@doi [\mnras] {10.1093/mnras/stz3101},
  \href {https://ui.adsabs.harvard.edu/abs/2020MNRAS.491.3655G} {491, 3655}

\bibitem[\protect\citeauthoryear{{Garma-Oehmichen} et~al.,}{{Garma-Oehmichen}
  et~al.}{2022}]{2022MNRAS.517.5660G}
{Garma-Oehmichen} L.,  et~al., 2022, \mn@doi [\mnras] {10.1093/mnras/stac3069},
  \href {https://ui.adsabs.harvard.edu/abs/2022MNRAS.517.5660G} {517, 5660}

\bibitem[\protect\citeauthoryear{{Guo}, {Mao}, {Athanassoula}, {Li}, {Ge},
  {Long}, {Merrifield}  \& {Masters}}{{Guo} et~al.}{2019}]{2019MNRAS.482.1733G}
{Guo} R.,  {Mao} S.,  {Athanassoula} E.,  {Li} H.,  {Ge} J.,  {Long} R.~J.,
  {Merrifield} M.,   {Masters} K.,  2019, \mn@doi [\mnras]
  {10.1093/mnras/sty2715}, \href
  {https://ui.adsabs.harvard.edu/abs/2019MNRAS.482.1733G} {482, 1733}

\bibitem[\protect\citeauthoryear{{Hernquist}}{{Hernquist}}{1990}]{Hernquist1990}
{Hernquist} L.,  1990, \mn@doi [\apj] {10.1086/168845}, 356, 359

\bibitem[\protect\citeauthoryear{{Kalapotharakos}, {Voglis}  \&
  {Contopoulos}}{{Kalapotharakos} et~al.}{2004}]{2004A&A...428..905K}
{Kalapotharakos} C.,  {Voglis} N.,   {Contopoulos} G.,  2004, \mn@doi [\aap]
  {10.1051/0004-6361:20041492}, \href
  {https://ui.adsabs.harvard.edu/abs/2004A&A...428..905K} {428, 905}

\bibitem[\protect\citeauthoryear{{Kawata}, {Baba}, {Hunt}, {Sch{\"o}nrich},
  {Ciuc{\u{a}}}, {Friske}, {Seabroke}  \& {Cropper}}{{Kawata}
  et~al.}{2021}]{Kawata2021}
{Kawata} D.,  {Baba} J.,  {Hunt} J. A.~S.,  {Sch{\"o}nrich} R.,  {Ciuc{\u{a}}}
  I.,  {Friske} J.,  {Seabroke} G.,   {Cropper} M.,  2021, \mn@doi [\mnras]
  {10.1093/mnras/stab2582}, \href
  {https://ui.adsabs.harvard.edu/abs/2021MNRAS.508..728K} {508, 728}

\bibitem[\protect\citeauthoryear{{Kunder} et~al.,}{{Kunder}
  et~al.}{2012}]{2012AJ....143...57K}
{Kunder} A.,  et~al., 2012, \mn@doi [\aj] {10.1088/0004-6256/143/3/57}, \href
  {https://ui.adsabs.harvard.edu/abs/2012AJ....143...57K} {143, 57}

\bibitem[\protect\citeauthoryear{{Kunder} et~al.,}{{Kunder}
  et~al.}{2016}]{2016ApJ...821L..25K}
{Kunder} A.,  et~al., 2016, \mn@doi [\apjl] {10.3847/2041-8205/821/2/L25},
  \href {https://ui.adsabs.harvard.edu/abs/2016ApJ...821L..25K} {821, L25}

\bibitem[\protect\citeauthoryear{{Li}, {Shen}, {Gerhard}  \& {Clarke}}{{Li}
  et~al.}{2022}]{Li2022}
{Li} Z.,  {Shen} J.,  {Gerhard} O.,   {Clarke} J.~P.,  2022, \mn@doi [\apj]
  {10.3847/1538-4357/ac3823}, \href
  {https://ui.adsabs.harvard.edu/abs/2022ApJ...925...71L} {925, 71}

\bibitem[\protect\citeauthoryear{{{\L}okas}}{{{\L}okas}}{2019}]{Lokas2019}
{{\L}okas} E.~L.,  2019, \mn@doi [\aap] {10.1051/0004-6361/201936056}, 629, A52

\bibitem[\protect\citeauthoryear{{Massari}, {Koppelman}  \& {Helmi}}{{Massari}
  et~al.}{2019}]{2019A&A...630L...4M}
{Massari} D.,  {Koppelman} H.~H.,   {Helmi} A.,  2019, \mn@doi [\aap]
  {10.1051/0004-6361/201936135}, \href
  {https://ui.adsabs.harvard.edu/abs/2019A&A...630L...4M} {630, L4}

\bibitem[\protect\citeauthoryear{{McMillan}}{{McMillan}}{2017}]{McMillan_2017}
{McMillan} P.~J.,  2017, \mn@doi [\mnras] {10.1093/mnras/stw2759}, \href
  {https://ui.adsabs.harvard.edu/abs/2017MNRAS.465...76M} {465, 76}

\bibitem[\protect\citeauthoryear{{McWilliam} \& {Zoccali}}{{McWilliam} \&
  {Zoccali}}{2010}]{McWilliam2010}
{McWilliam} A.,  {Zoccali} M.,  2010, \mn@doi [\apj]
  {10.1088/0004-637X/724/2/1491}, \href
  {https://ui.adsabs.harvard.edu/abs/2010ApJ...724.1491M} {724, 1491}

\bibitem[\protect\citeauthoryear{{Minchev}, {Nordhaus}  \& {Quillen}}{{Minchev}
  et~al.}{2007}]{Minchev2007}
{Minchev} I.,  {Nordhaus} J.,   {Quillen} A.~C.,  2007, \mn@doi [\apjl]
  {10.1086/520578}, \href
  {https://ui.adsabs.harvard.edu/abs/2007ApJ...664L..31M} {664, L31}

\bibitem[\protect\citeauthoryear{{Miyamoto} \& {Nagai}}{{Miyamoto} \&
  {Nagai}}{1975}]{1975PASJ...27..533M}
{Miyamoto} M.,  {Nagai} R.,  1975, \pasj, \href
  {https://ui.adsabs.harvard.edu/abs/1975PASJ...27..533M} {27, 533}

\bibitem[\protect\citeauthoryear{{Mosenkov}, {Savchenko}, {Smirnov}  \&
  {Camps}}{{Mosenkov} et~al.}{2021}]{Mosenkov2021}
{Mosenkov} A.~V.,  {Savchenko} S.~S.,  {Smirnov} A.~A.,   {Camps} P.,  2021,
  \mn@doi [\mnras] {10.1093/mnras/stab2445}, \href
  {https://ui.adsabs.harvard.edu/abs/2021MNRAS.507.5246M} {507, 5246}

\bibitem[\protect\citeauthoryear{{Nataf}, {Udalski}, {Gould}, {Fouqu{\'e}}  \&
  {Stanek}}{{Nataf} et~al.}{2010}]{Nataf2010}
{Nataf} D.~M.,  {Udalski} A.,  {Gould} A.,  {Fouqu{\'e}} P.,   {Stanek} K.~Z.,
  2010, \mn@doi [\apjl] {10.1088/2041-8205/721/1/L28}, \href
  {https://ui.adsabs.harvard.edu/abs/2010ApJ...721L..28N} {721, L28}

\bibitem[\protect\citeauthoryear{{Navarro}, {Frenk}  \& {White}}{{Navarro}
  et~al.}{1996}]{NFW}
{Navarro} J.~F.,  {Frenk} C.~S.,   {White} S.~D.~M.,  1996, \mn@doi [\apj]
  {10.1086/177173}, 462, 563

\bibitem[\protect\citeauthoryear{{Ness} et~al.,}{{Ness}
  et~al.}{2013}]{2013MNRAS.430..836N}
{Ness} M.,  et~al., 2013, \mn@doi [\mnras] {10.1093/mnras/sts629}, \href
  {https://ui.adsabs.harvard.edu/abs/2013MNRAS.430..836N} {430, 836}

\bibitem[\protect\citeauthoryear{{Ness} et~al.,}{{Ness}
  et~al.}{2016}]{2016ApJ...819....2N}
{Ness} M.,  et~al., 2016, \mn@doi [\apj] {10.3847/0004-637X/819/1/2}, \href
  {https://ui.adsabs.harvard.edu/abs/2016ApJ...819....2N} {819, 2}

\bibitem[\protect\citeauthoryear{{Ortolani}, {Nardiello}, {P{\'e}rez-Villegas},
  {Bica}  \& {Barbuy}}{{Ortolani} et~al.}{2019a}]{Ortolani2019a}
{Ortolani} S.,  {Nardiello} D.,  {P{\'e}rez-Villegas} A.,  {Bica} E.,
  {Barbuy} B.,  2019a, \mn@doi [\aap] {10.1051/0004-6361/201834477}, \href
  {https://ui.adsabs.harvard.edu/abs/2019A&A...622A..94O} {622, A94}

\bibitem[\protect\citeauthoryear{{Ortolani} et~al.,}{{Ortolani}
  et~al.}{2019b}]{Ortolani2019b}
{Ortolani} S.,  et~al., 2019b, \mn@doi [\aap] {10.1051/0004-6361/201935726},
  \href {https://ui.adsabs.harvard.edu/abs/2019A&A...627A.145O} {627, A145}

\bibitem[\protect\citeauthoryear{{Parul}, {Smirnov}  \& {Sotnikova}}{{Parul}
  et~al.}{2020}]{Parul_etal2020}
{Parul} H.~D.,  {Smirnov} A.~A.,   {Sotnikova} N.~Y.,  2020, \mn@doi [\apj]
  {10.3847/1538-4357/ab76ce}, \href
  {https://ui.adsabs.harvard.edu/abs/2020ApJ...895...12P} {895, 12}

\bibitem[\protect\citeauthoryear{{Pasquato} \& {Chung}}{{Pasquato} \&
  {Chung}}{2019}]{Pasquato2019}
{Pasquato} M.,  {Chung} C.,  2019, \mn@doi [\mnras] {10.1093/mnras/stz2766},
  \href {https://ui.adsabs.harvard.edu/abs/2019MNRAS.490.3392P} {490, 3392}

\bibitem[\protect\citeauthoryear{{P{\'e}rez-Villegas}, {Barbuy}, {Kerber},
  {Ortolani}, {Souza}  \& {Bica}}{{P{\'e}rez-Villegas} et~al.}{2020}]{PV2020}
{P{\'e}rez-Villegas} A.,  {Barbuy} B.,  {Kerber} L.~O.,  {Ortolani} S.,
  {Souza} S.~O.,   {Bica} E.,  2020, \mn@doi [\mnras] {10.1093/mnras/stz3162},
  \href {https://ui.adsabs.harvard.edu/abs/2020MNRAS.491.3251P} {491, 3251}

\bibitem[\protect\citeauthoryear{{Plummer}}{{Plummer}}{1911}]{1911MNRAS..71..460P}
{Plummer} H.~C.,  1911, \mn@doi [\mnras] {10.1093/mnras/71.5.460}, \href
  {https://ui.adsabs.harvard.edu/abs/1911MNRAS..71..460P} {71, 460}

\bibitem[\protect\citeauthoryear{{Portail}, {Gerhard}, {Wegg}  \&
  {Ness}}{{Portail} et~al.}{2017}]{Portail2017}
{Portail} M.,  {Gerhard} O.,  {Wegg} C.,   {Ness} M.,  2017, \mn@doi [\mnras]
  {10.1093/mnras/stw2819}, \href
  {https://ui.adsabs.harvard.edu/abs/2017MNRAS.465.1621P} {465, 1621}

\bibitem[\protect\citeauthoryear{{Salo} \& {Laurikainen}}{{Salo} \&
  {Laurikainen}}{2017}]{Salo_Laurikainen2017_v2}
{Salo} H.,  {Laurikainen} E.,  2017, \mn@doi [\apj]
  {10.3847/1538-4357/835/2/252}, 835, 252

\bibitem[\protect\citeauthoryear{{Sanders}, {Smith}  \& {Evans}}{{Sanders}
  et~al.}{2019}]{Sanders2019}
{Sanders} J.~L.,  {Smith} L.,   {Evans} N.~W.,  2019, \mn@doi [\mnras]
  {10.1093/mnras/stz1827}, \href
  {https://ui.adsabs.harvard.edu/abs/2019MNRAS.488.4552S} {488, 4552}

\bibitem[\protect\citeauthoryear{{Sch{\"o}nrich}, {Binney}  \&
  {Dehnen}}{{Sch{\"o}nrich} et~al.}{2010}]{Sch2010}
{Sch{\"o}nrich} R.,  {Binney} J.,   {Dehnen} W.,  2010, \mn@doi [\mnras]
  {10.1111/j.1365-2966.2010.16253.x}, \href
  {https://ui.adsabs.harvard.edu/abs/2010MNRAS.403.1829S} {403, 1829}

\bibitem[\protect\citeauthoryear{{Sellwood}}{{Sellwood}}{2014}]{2014RvMP...86....1S}
{Sellwood} J.~A.,  2014, \mn@doi [Reviews of Modern Physics]
  {10.1103/RevModPhys.86.1}, \href
  {https://ui.adsabs.harvard.edu/abs/2014RvMP...86....1S} {86, 1}

\bibitem[\protect\citeauthoryear{{Sellwood} \& {Gerhard}}{{Sellwood} \&
  {Gerhard}}{2020}]{2020MNRAS.495.3175S}
{Sellwood} J.~A.,  {Gerhard} O.,  2020, \mn@doi [\mnras]
  {10.1093/mnras/staa1336}, \href
  {https://ui.adsabs.harvard.edu/abs/2020MNRAS.495.3175S} {495, 3175}

\bibitem[\protect\citeauthoryear{{Smirnov} \& {Savchenko}}{{Smirnov} \&
  {Savchenko}}{2020}]{2020MNRAS.499..462S}
{Smirnov} A.~A.,  {Savchenko} S.~S.,  2020, \mn@doi [\mnras]
  {10.1093/mnras/staa2892}, \href
  {https://ui.adsabs.harvard.edu/abs/2020MNRAS.499..462S} {499, 462}

\bibitem[\protect\citeauthoryear{{Smirnov} \& {Sotnikova}}{{Smirnov} \&
  {Sotnikova}}{2018}]{Smirnov_Sotnikova2018}
{Smirnov} A.~A.,  {Sotnikova} N.~Y.,  2018, \mn@doi [\mnras]
  {10.1093/mnras/sty2423}, 481, 4058

\bibitem[\protect\citeauthoryear{{Smirnov}, {Tikhonenko}  \&
  {Sotnikova}}{{Smirnov} et~al.}{2021}]{Smirnov_etal2021}
{Smirnov} A.~A.,  {Tikhonenko} I.~S.,   {Sotnikova} N.~Y.,  2021, \mn@doi
  [\mnras] {10.1093/mnras/stab327}, 502, 4689

\bibitem[\protect\citeauthoryear{{Sun}, {Wang}, {Liu}, {Long}, {Chen}  \&
  {Gao}}{{Sun} et~al.}{2023}]{Sun2022}
{Sun} G.,  {Wang} Y.,  {Liu} C.,  {Long} R.~J.,  {Chen} X.,   {Gao} Q.,  2023,
  \mn@doi [Research in Astronomy and Astrophysics] {10.1088/1674-4527/ac9e91},
  \href {https://ui.adsabs.harvard.edu/abs/2023RAA....23a5013S} {23, 015013}

\bibitem[\protect\citeauthoryear{{Tepper-Garcia} et~al.,}{{Tepper-Garcia}
  et~al.}{2021}]{2021arXiv211105466T}
{Tepper-Garcia} T.,  et~al., 2021, arXiv e-prints, \href
  {https://ui.adsabs.harvard.edu/abs/2021arXiv211105466T} {p. arXiv:2111.05466}

\bibitem[\protect\citeauthoryear{{Tikhonenko}, {Smirnov}  \&
  {Sotnikova}}{{Tikhonenko} et~al.}{2021}]{Tikhonenko_etal2021}
{Tikhonenko} I.~S.,  {Smirnov} A.~A.,   {Sotnikova} N.~Y.,  2021, \mn@doi
  [\aap] {10.1051/0004-6361/202140703}, 648, L4

\bibitem[\protect\citeauthoryear{{Vasiliev}}{{Vasiliev}}{2019a}]{agama}
{Vasiliev} E.,  2019a, \mn@doi [\mnras] {10.1093/mnras/sty2672}, \href
  {https://ui.adsabs.harvard.edu/abs/2019MNRAS.482.1525V} {482, 1525}

\bibitem[\protect\citeauthoryear{{Vasiliev}}{{Vasiliev}}{2019b}]{Vasiliev2019}
{Vasiliev} E.,  2019b, \mn@doi [\mnras] {10.1093/mnras/stz171}, \href
  {https://ui.adsabs.harvard.edu/abs/2019MNRAS.484.2832V} {484, 2832}

\bibitem[\protect\citeauthoryear{{Vasiliev} \& {Baumgardt}}{{Vasiliev} \&
  {Baumgardt}}{2021}]{Vasiliev2021}
{Vasiliev} E.,  {Baumgardt} H.,  2021, \mn@doi [\mnras]
  {10.1093/mnras/stab1475}, \href
  {https://ui.adsabs.harvard.edu/abs/2021MNRAS.505.5978V} {505, 5978}

\bibitem[\protect\citeauthoryear{{Voglis}, {Harsoula}  \&
  {Contopoulos}}{{Voglis} et~al.}{2007}]{2007MNRAS.381..757V}
{Voglis} N.,  {Harsoula} M.,   {Contopoulos} G.,  2007, \mn@doi [\mnras]
  {10.1111/j.1365-2966.2007.12263.x}, \href
  {https://ui.adsabs.harvard.edu/abs/2007MNRAS.381..757V} {381, 757}

\bibitem[\protect\citeauthoryear{{Wang}, {Athanassoula}  \& {Mao}}{{Wang}
  et~al.}{2016}]{Wang_etal2016}
{Wang} Y.,  {Athanassoula} E.,   {Mao} S.,  2016, \mn@doi [\mnras]
  {10.1093/mnras/stw2301}, 463, 3499

\bibitem[\protect\citeauthoryear{{Wang}, {Athanassoula}, {Patsis}  \&
  {Mao}}{{Wang} et~al.}{2022}]{Wang2022}
{Wang} Y.,  {Athanassoula} E.,  {Patsis} P.,   {Mao} S.,  2022, \mn@doi [\aap]
  {10.1051/0004-6361/202243699}, \href
  {https://ui.adsabs.harvard.edu/abs/2022A&A...668A..55W} {668, A55}

\bibitem[\protect\citeauthoryear{{Wegg} \& {Gerhard}}{{Wegg} \&
  {Gerhard}}{2013}]{Wegg2013}
{Wegg} C.,  {Gerhard} O.,  2013, \mn@doi [\mnras] {10.1093/mnras/stt1376},
  \href {https://ui.adsabs.harvard.edu/abs/2013MNRAS.435.1874W} {435, 1874}

\makeatother
\end{thebibliography}
%%%%%%%%%%%%%%%%%%%%%%%%%%%%%%%%%%%%%%%%%%%%%%%%%%%%%%%
%%%%%%%%%%%%%%%%%%%%%%%%%%%%%%%%%%%%%%%%%%%%%%%%%%%%%%%

%%%%%%%%%%%%%%%%%%%%%%%%%%%%%%%%%%%%%%%%%%%%%%%%%%%%%%%
%%%%%%%%%%%%%%%%%%%%%% APPENDICES %%%%%%%%%%%%%%%%%%%%%
%\appendix

%%%%%%%%%%%%%%%%%%%%%%%%%%%%%%%%%%%%%%%%%%%%%%%%%%%%%%%
%%%%%%%%%%%%%%%%%%%%%%%%%%%%%%%%%%%%%%%%%%%%%%%%%%%%%%%
%\section{Whatever}
%%%%%%%%%%%%%%%%%%%%%%%%%%%%%%%%%%%%%%%%%%%%%%%%%%%%%%%
%%%%%%%%%%%%%%%%%%%%%%%%%%%%%%%%%%%%%%%%%%%%%%%%%%%%%%%

%%%%%%%%%%%%%%%%%%%%%%%%%%%%%%%%%%%%%%%%%%%%%%%%%%%%%%%
%%%%%%%%%%%%%%%%%%%%%%%%%%%%%%%%%%%%%%%%%%%%%%%%%%%%%%%

% Don't change these lines
\bsp	% typesetting comment
\label{lastpage}
\end{document}